\documentclass[twocolumn]{emulateapj}
\usepackage{apjfonts}
\usepackage{graphicx}
\newcommand{\bnabla}{\mbox{\boldmath$\nabla$}}

\newcommand{\be}{\begin{equation}}
\newcommand{\ee}{\end{equation} }
\newcommand{\ba}{\begin{eqnarray}}
\newcommand{\ea}{\end{eqnarray}}
\newcommand{\nn}{\mbox{} \nonumber \\ \mbox{} }
\newcommand{\kB}{k_{\rm B}}

\shorttitle{Radially Magnetized Protoplanetary Disk: Vertical Profile}
\shortauthors{Russo \& Thompson}	
\slugcomment{Astrophysical Journal 2015, in press}
\begin{document}
\title{Radially Magnetized Protoplanetary Disk: Vertical Profile}
\author{Matthew Russo}
\affil{Department of Physics, University of Toronto, 60 St. George St., Toronto, ON M5S 1A7, Canada.}
\author{Christopher Thompson}
\affil{Canadian Institute for Theoretical Astrophysics, 60 St. George St., Toronto, ON M5S 3H8, Canada.}


\begin{abstract}
This paper studies the response of a thin accretion disk to an external radial magnetic field.  Our focus is on protoplanetary disks (PPDs), which are exposed during their later evolution to an intense, magnetized wind from the central star.  A radial magnetic field is mixed into a thin surface layer, is wound up by the disk shear, and is pushed downward by a combination of turbulent mixing and ambipolar and Ohmic drift.  The toroidal field reaches much greater strengths than the seed vertical field that is usually invoked in PPD models, even becoming superthermal.  Linear stability analysis indicates that the disk experiences the magnetorotational instability (MRI) at a higher magnetization than a vertically magnetized disk when both the effects of ambipolar and Hall drift are taken into account.  Steady vertical profiles of density and magnetic field are obtained at several radii between 0.06 and 1 AU in response to a wind magnetic field $B_r \sim (10^{-4}$--$10^{-2})(r/{\rm AU})^{-2}$ G.  Careful attention is given to the radial and vertical ionization structure resulting from irradiation by stellar X-rays.  The disk is more strongly magnetized closer to the star, where it can support a higher rate of mass transfer.  As a result, the inner $\sim 1$ AU of a PPD is found to evolve toward lower surface density.  Mass transfer rates around $10^{-8}\,M_\odot$ yr$^{-1}$ are obtained under conservative assumptions about the MRI-generated stress. The evolution of the disk, and the implications for planet migration, are investigated in the accompanying paper.
\end{abstract}


\maketitle

\section{Introduction}

Understanding the formation and migration of planets in circumstellar disks presents several challenges, the first of which is to develop a deterministic model of the flow of gas and dust within the disk.The motivation for tackling this first step has been strengthened with the discovery of many planets of a substantial mass in close orbits around nearby sun-like stars\footnote{See www.exoplanets.org; www.exoplanet.eu; \cite{lissauer14}.}.   In most models of such protoplanetary disks (PPDs) -- variants of the `minimal-mass Solar nebula' \citep{hayashi81} -- the surface mass density of the disk increases inward.  Embedded planets are predicted to drift rapidly toward the star \citep{gt80,lp86}, on a timescale that is shorter than the observed disk lifetimes.   Although the direction of this drift may be reversed(or the migration stalled) in some circumstances \citep{ward91,masset01,masset06}, no robust mechanism has emerged for retaining massive planets in such an inwardly peaked disk over the wide range of orbits that is observed.  

The magnetorotational instability (MRI) is a promising mechanism for angular momentum transport in the inner disk (distance $r \lesssim 1$ AU from the protostar; \citealt{bb94,gammie96}).   Here the transport time is shorter than the disk lifetime, so that what matters most is not the absolute normalization of the viscous torque, but its {\it relative} normalization at different radii.  A key point is that the thickness of the MRI-active layer is determined by external ionization (except very close to the protostar), so that the local rate of mass transfer does not match at different radii.  This means that MRI-generated torques drive a secular change in gas surface density $\Sigma_g$.  One of the main goals of this paper is to determine the sign of this change.

The MRI-driven torque acting on a given patch of the disk depends on the magnetic flux threading it \citep{hawley95,sano04,pessah07}.This flux is usually assumed to be oriented vertically with respect to the disk plane.  Our poor understanding of how vertical flux would be distributed across the disk surface has presented something of a roadblock toward building deterministic global disk models.   Global numerical simulations of geometrically thin disks ($h/r \lesssim 0.03$) that resolve non-ideal effects cannot presently determine the evolution of vertical magnetic flux.  

Here we are guided by a basic property of protostars: they are strongly magnetized and are sources of intense
winds during the T-Tauri phase \cite{bally07}.  We focus on a stage of the disk evolution 
when the mass flux from the surrounding molecular
cloud onto the surface of the inner PPD is smaller than the accretion rate through the disk (stellar age $\gtrsim 10^5$ yr). 
When the disk accretion rate drops below $\dot M \sim \dot M_{\rm mag} \equiv R_\star B_\star^2/\Omega(R_\star) \sim 10^{-5}
(B_\star/{\rm kG})^2\,M_\odot$ yr$^{-1}$, where $\Omega$ is the Keplerian angular velocity,
a surface dipole magnetic field $B_\star$ anchored in a protostar of radius $R_\star \sim 2R_\odot$ is able to
sustain a magnetosphere.  The magnetosphere extends beyond the radius at which Keplerian orbits co-rotate with the 
stellar spin, leading to strong magnetocentrifugal effects, when the accretion rate drops further to $\dot M < 
(P_\star \Omega(R_\star)/2\pi)^{-7/3} \dot M_{\rm mag} \sim 6\times 10^{-8}\,(B_\star/{\rm kG})^2 
(P_\star/3~{\rm d})^{-7/3}\,M_\odot$ yr$^{-1}$.

Consider a stellar wind that ejects a fraction $\epsilon_w$ of the mass inflow through the disk
and reaches a terminal speed $V_w$.  Its radial kinetic pressure $\rho_w V_w^2 = 
\epsilon_w \dot M V_w/4\pi r^2$ (where $\rho_w$ is the wind density) can be compared with the thermal pressure $P_{\rm disk} 
\sim \Sigma_g c_g\Omega$ of a Keplerian disk (with isothermal sound speed $c_g$
and viscosity $\alpha c_g^2/\Omega$),
\ba
{\rho_w V_w^2\over P_{\rm disk}} &\sim& \epsilon_w \alpha {c_g V_w\over (\Omega r)^2} \nn
   &\sim&   0.003\epsilon_{w,-1} \alpha_{-1} \left({T\over 200\, \rm{K}}\right)^{1/2}
  \left({V_w\over 400~{\rm km~s}}\right)\left({r\over{\rm AU}}\right).\nn
\ea

\begin{figure}[!]
\epsscale{1.16}
\plotone{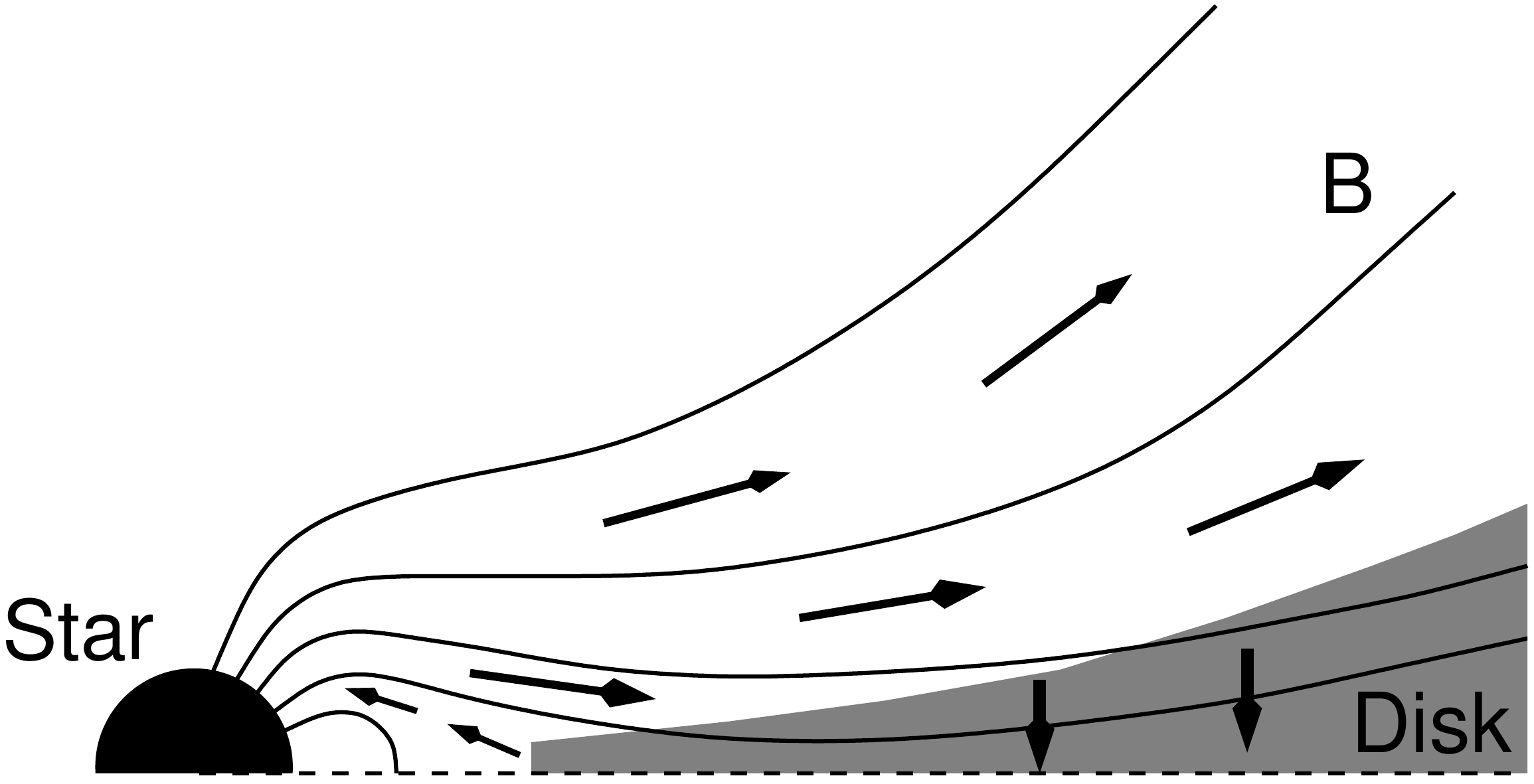}
\vskip .1in
\caption{Poloidal magnetic field of a T-Tauri star interacting with the upper layers of a geometrically thin
and weakly conducting PPD.  The kinetic pressure of the wind exceeds the pressure of any
vertical field remaining from the first stages of infall from a natal molecular cloud.}
\vskip .1in
\label{fig:Bprofile}
\end{figure}

Any poloidal magnetic field threading the disk must have a pressure larger than $\rho_w V_w^2$, or it will be swept
back.  Then the magnetic geometry differs from that postulated by \cite{bp82} and \cite{pudritz86}, and any
magnetocentrifugal wind from the disk surface \citep{bs13,lesur13} will be quenched.  Instead, the T-Tauri wind
interacts directly with the surface of the disk, and its magnetic field is entrained by a thin boundary
layer (Figure \ref{fig:Bprofile}).  

This paper is devoted to understanding how the magnetic field that is sheared out in this boundary layer
will diffuse downward into the disk, and the equilibrium structure that results.  Particular care must be
taken with regard to the radial dependence of the ionization rate in the disk, as well as the depletion of dust
in the active layer by the settling of particles from an outer gas reservoir that feeds the 
accretion flow through the inner $\sim$ AU.

Not surprisingly, strong magnetization emerges in our disk solutions within the active layer (where it is sustained as
the result of a balance between vertical diffusion and linear winding), as well as at the top in the wind-disk boundary
layer (where it reflects the strong magnetization of the T-Tauri wind).  In contrast with the strongly magnetized
Shakura-Sunyaev type disk solutions derived by \cite{pariev03} and discussed by \cite{begelman07},
a significant part of the magnetic pressure support is provided by a laminar toroidal magnetic field.  \cite{jl08} have
developed a shearing box model that is initialized with a strong toroidal (but not radial) field.  This toroidal field
is confined to the computational domain by perfectly conducting upper and lower boundaries.  Although 
the solution is found to converge in response to expansion of the vertical domain, it remains open to question 
whether such a seed magnetic flux could realistically be sustained without an external source over the 
$10^4-10^5$ orbits required for radial spreading.  The same consideration applies to an embedded radial magnetic
field.\footnote{We thank Ethan Vishniac for insightful comments about this point.}

A strong vertical seed field has been shown to suppress the MRI (e.g. \citealt{bs13}).  Hence it is
important to consider how strong magnetization and departures from ideal MHD influence 
the MRI in a toroidal field geometry.  We carry out the relevant non-axisymmetric linear stability 
analysis, which has not previously been performed in a shearing disk that is subject to ambipolar and Hall drift.
This analysis finds growth rates high enough to be consistent with sustained MRI turbulence.  Additional support
for this conclusion comes from the direction of the Poynting flux, which flows horizontally as opposed to vertically
in a horizontal background field.

Because the applied wind magnetic field decreases from the protostar in a well-defined way, we are able to 
determine the radial dependence of the torque, or equivalently of the vertically averaged
viscosity $\langle\nu\rangle$.  One finds a mild decrease in $\langle \nu\rangle \Sigma_g$ with distance 
from the protostar when the stellar outflow is nearly radial near the disk plane.

A treatment of the temporal evolution of the inner disk, and the feedback between embedded solids and the MRI,
is deferred to the accompanying paper (Russo \& Thompson 2015, hereafter Paper II).  Other details 
considered there include the implications for infrared re-emission by dust, the onset of `transition' disks, and 
planetary migration.

\subsection{Plan of the Paper}
In Section \ref{s:response} we give a simplified one-zone description of mass transfer in a radially 
magnetized disk, which is used to test the consequences of different radial magnetic field profiles.  Our handling of 
the wind-disk interaction, the effects of radiative cooling, MHD stability, X-ray ionization, and dust depletion
are all described in detail in Section \ref{s:model}.  The next Section \ref{s:model2}
presents our approach to calculating the vertical profile of the radial and toroidal magnetic field, including
the limitations on MRI turbulence that are introduced by non-ideal effects.

The numerical method is described in Section \ref{s:steady}, and results are collected in Section \ref{s:results}.
The concluding Section \ref{s:summary} outlines the implications for mass transfer in the inner $\sim$ AU
of PPDs.  The first appendix explains our calculation of X-ray ionization.   The second appendix describes
the result of a non-axisymmetric linear stability analysis of a non-ideal disk with toroidal magnetic field.

\section{Response of Disk to Radially Magnetized T-Tauri Wind}\label{s:response}
A simple one-zone model of the active layer of the disk is first developed.  This motivates the more 
detailed calculation that follows, and shows how the long-term evolution of the inner disk is insensitive
to our handling of MRI-driven turbulence.
 
\subsection{Saturated Field Strength and Viscosity}\label{s:ana}

A radial magnetic field that is imposed on a thin shearing disk can have a profound effect on angular momentum transport in the disk.
Here we briefly consider the response in the ideal MHD regime.

Simulations of the MRI starting with a weak seed field (either vertical or toroidal) suggest a non-radial
stress $T_{r\phi}/P \sim 0.5 (\langle B^2\rangle/8\pi P)$ (e.g., \citealt{hawley95}).  Here $P$ is the thermal pressure and 
$\langle B^2\rangle/8\pi$ is the mean magnetic pressure that is reached at the saturation of the instability.  

The relation between the seed and saturation magnetic fields depends on magnetic geometry, numerical resolution, and on 
the stratification of the disk.  In the largest simulations performed to date, an imposed vertical field carrying a non-vanishing flux
generates a torque that is about $10$ times stronger than a toroidal field of equal strength \citep{baistone11}, as compared with
$\sim 10^2$ times stronger in the older unstratified simulations of \citep{hawley95}.  The scaling $\langle B^2\rangle \propto 
\beta_0^{-1/2}$ is obtained in the ideal MHD limit when the plasma parameter of the seed magnetic field is $\beta_0 = 8\pi P/B_0^2 \gg 1$.  

To estimate the torque that results when a radial magnetic field $B_{r0}$ is imposed at the top of the disk, we write\footnote{Throughout the paper we use $\widetilde\alpha$ to denote the viscosity coefficient with
the dependence on background magnetic pressure factored out.}
\be\label{eq:nuana}
\nu_{\rm MRI} = \widetilde\alpha_{\rm MRI} \left({B_{\phi 0}^2\over 8\pi P}\right)^\delta {c_g^2\over\Omega}.
\ee
Here $B_{\phi 0}$ is the toroidal field that is generated by linear winding of the imposed radial field over the 
 vertical diffusion time through the disk.  The numerical simulations 
of weakly magnetized disks ($\beta_0 \gg 1$) just described suggest $\delta = 1/2$.
In the calculations performed later in this paper, we encounter strong magnetization
($\beta_{\phi 0} \lesssim 1$) and use the alternative scaling $\delta = 1$ (Equation \ref{eq:alpha}).

Consider the simplest configuration with $B_{r0}$ uniform with height between the top
of the disk and a reconnection layer near the midplane.  Such a configuration
emerges in the vertical profiles to be presented in Section \ref{s:results}.  Taking 
\be\label{eq:bphieq}
B_{\phi0} = -(3/2)\Omega t_{\rm diff} B_{r0},
\ee
with diffusion time $t_{\rm diff} \sim \nu_{\rm MRI}^{-1}h_g^2$ across the thermal scale height $h_g = c_g/\Omega$,
one finds
\ba\label{eq:equil}
{B_{\phi0}^2\over 8\pi P} &=& \left({3\over 2\widetilde\alpha_{\rm MRI}}\right)^{2/(1+2\delta)} 
\left({B_{r0}^2\over 8\pi P}\right)^{1/(1+2\delta)};\nn
\nu_{\rm MRI} &=& \left[(3/2)^{2\delta}\widetilde\alpha_{\rm MRI}\right]^{1/(1+2\delta)}
\left({B_{r0}^2\over 8\pi P}\right)^{\delta/(1+2\delta)}{c_g^2\over \Omega}.
\ea
This demonstrates that a radial seed field can lead to a substantially enhanced viscous stress compared with 
an equivalent toroidal seed when $B_0^2\lesssim 8\pi P$. 

\subsection{Mass Transfer Induced by a Split-monopolar Wind Magnetic Field}

The magnetized wind that is emitted by a T-Tauri star may be sourced by the star's internal dynamo; 
or may be launched from the inner regions of the disk itself \citep{pudritz86}. Throughout most of this paper, we approximate the wind flow near the disk plane 
($\theta = \pi/2$ in spherical coordinates) as steady and spherical, with an embedded split-monopole magnetic field,
\be\label{eq:wind}
B_r^w = B(R_\star) \left({R_\star\over r}\right)^2\,{\rm sgn}(\pi/2-\theta).
\ee

Because the kinetic pressure of the wind is much less than the internal hydrostatic pressure of the disk,
the principal quantity of interest is the magnetic flux that threads it.  We consider radial fields
up to $\sim 1$ G at 0.1 AU from the protostar, corresponding to a surface dipole magnetic 
$\sim 10^2$ G at $R_\star \sim 2R_\odot$ \citep{valenti04,daou06}.  

The change of sign of $B_r$ across the disk plane is mirrored by the toroidal magnetic field that is wound up inside the disk.
Here a small imbalance in the amplitude of the applied radial field between the two hemispheres has a marked effect on the internal
structure of the disk.  The region near the midplane is very weakly ionized, leading to rapid reconnection of toroidal fields
of opposing signs.  The current sheet sits at the midplane in the case of perfect reflection symmetry, but otherwise 
is displaced to one side.

Now consider now a geometrically thin disk in which a viscosity given by the second half of Equation (\ref{eq:equil}) is excited.
We move to cylindrical coordinates $(R\simeq r,z\simeq \theta r)$.
The magnetically active gas column is assumed to be regulated to a characteristic value $\delta\Sigma_{\rm act}$
by external ionization, with a slow inward growth $\delta\Sigma_{\rm act}(R) \propto R^{-k_\Sigma}$ due to the increasing
ionization rate and angle of attack of X-rays on the disk near the protostar.  
In the inner disk, where the flaring of the dust absorption layer is smaller than 
$d(h_p/R)/d\ln R \lesssim  R_\star/2R$, the temperature of a passively irradiated disk scales as $T \sim R^{-3/4}$.  
Combining Equation (\ref{eq:equil}) and (\ref{eq:wind}) with these radial profiles of $\delta\Sigma_{\rm act}$ and $T$,
the local rate of mass transfer scales as
\be\label{eq:mdotscale}
\dot M_{\rm MRI} \sim 3\pi \nu_{\rm MRI} \delta\Sigma_{\rm act} \sim R^{(6-8k_\Sigma-(5+8k_\Sigma)\delta)/8(1 + 2\delta)}.
\ee

First let us consider how this result depends on the relation (\ref{eq:nuana}) between MRI-generated stress
and the linearly amplified toroidal magnetic field.  We take $k_\Sigma = 1/4$ from the detailed model developed 
later in the paper.  Equation (\ref{eq:mdotscale}) gives $\dot M_{\rm MRI} \propto R^{1/32}$ in a weakly magnetized disk 
with $\delta = 1/2$; and $\dot M_{\rm MRI} \propto R^{-1/8}$ in a strongly magnetized disk, for which we adopt $\delta = 1$.  
\begin{figure}[ht]
\epsscale{1.2}
\plotone{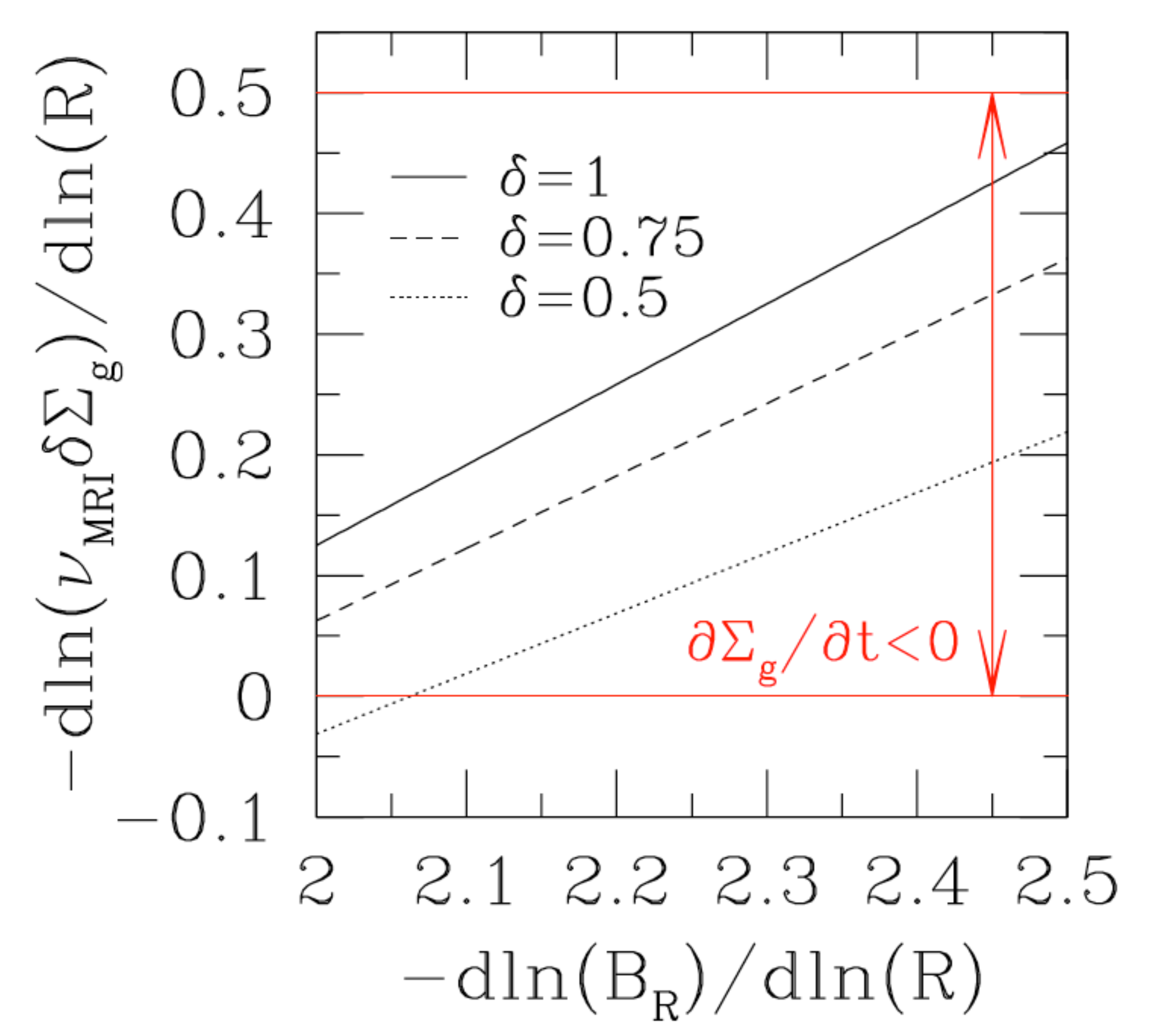}
\caption{Radial scaling of the mass transfer rate driven by MRI turbulence, given an imposed
radial magnetic field with a profile $-2 > d\ln(B_R)/d\ln R > -2.5$.  A profile slightly steeper than monopolar near
the midplane could result from collimation toward the rotation axis.  Black curves label different dependences
of $T_{r\phi}$ driven by the MRI on seed toroidal magnetic pressure, $\nu_{\rm MRI} \propto (B_{\phi 0}^2/8\pi P)^\delta$.
The vertical disk models constructed in this paper use $\delta = 1$.  The surface density of the inner disk
{\it decreases} with time between the two horizontal red lines.  Results plotted here depend mildly on the radial
profile of the X-ray irradiated active column, $\delta\Sigma_{\rm act} \propto R^{-1/4}$.}
\vskip .2in
\label{fig:coll}
\end{figure}

The laminar torque driven by the linearly amplified stress $B_{R0}B_{\phi0}/4\pi$ can also contribute significantly
to mass transfer, especially inside $R \sim 0.1$ AU.  The transfer rate is obtained from
\be
\dot M_{\rm lam} \Omega R^2 \sim 2\pi R^2 \int dz {B_{R0}B_{\phi0}\over 4\pi},
\ee
which gives $\dot M_{\rm lam} \sim B_{R0}B_{\phi0} c_g/2\Omega^2$.  Then
\ba
{\dot M_{\rm lam}\over \dot M_{\rm MRI}} &\sim& 2\left[(3/2)^{2\delta}\widetilde\alpha_{\rm MRI}\right]^{-2/(1+2\delta)}
\left({B_{R0}^2\over 8\pi P}\right)^{1/(1+2\delta)}\nn
  &\propto& R^{-(17-8k_\Sigma)/8(1+2\delta)}
\ea
in the inner disk.  This corresponds to $\dot M_{\rm lam} \sim R^{-3/4}$ for $\delta = 1$,
shifting to a slightly steeper profile $\dot M_{\rm MRI} \propto R^{-29/32}$ when $\delta = 1/2$ (both again assuming 
$k_\Sigma = 1/4$). Comparing with the preceding scalings for the MRI-generated stress, we see that the
laminar Maxwell stress increases more rapidly toward the protostar.

These results are displayed graphically in Figure \ref{fig:coll}.
Here we also relax the assumption of a monopolar wind magnetic field.
Collimation of the wind toward the stellar rotation axis would
steepen slightly the poloidal field profile near the midplane.  

The surface density in the inner disk decreases with time
when $-0.5 < d\ln(\nu_{\rm MRI}\delta\Sigma_{\rm act})/d\ln r < 0$.  We find that
is the case for radial indices of $B_R$ in the range $-2.5$ to $-2$ and 
most values of the MRI stress index $\delta$ between $0.5$ and 1.   The general effect of steepening
the profile of the imposed poloidal magnetic field  is to steepen the radial profile of $\dot M$,
thereby increasing the depletion rate in the inner PPD.

Although this analytic model is admittedly schematic, the scalings obtained from it correspond quite 
closely to those implied by the detailed vertical disk profiles described in Section \ref{s:results}.  
Combining the contributions from the turbulent and laminar torques gives a net mass transfer rate $\dot M \simeq \dot M_{\rm MRI} 
+ \dot M_{\rm lam}$ that decreases with radius.  The implications of this result are examined in 
Section \ref{s:summary}, and in more detail in Paper II.

\section{Disk Model:  Physical Ingredients}\label{s:model}

This section lays out our description of a PPD interacting with a T-Tauri wind and photoionizing X-rays
from the stellar corona (Table 1).  Our main goal is to calculate the steady-state vertical profile of a thin, 
Keplerian, and axially symmetric disk in response to an imposed radial magnetic field.   We consider 
how this profile changes over a range $10^{-1.25}-1$ AU of distances from the protostar.   

We construct vertical profiles assuming a single characteristic mass column,
\be\label{eq:sigtot}
\Sigma_{g,\rm tot} = 200~{\rm g~cm^{-2}},
\ee  
which is high enough for the disk midplane to be magnetically inactive.   
This allows us to apply our results to a wider variety 
of PPD states than if we had focused on the minimum mass solar nebula.  A low column is found to
be the endpoint of the MRI-driven evolution, and so we are able to study a PPD
before it emerges as a full-fledged `transition disk'.  In Paper II we also consider the forward 
evolution of a PPD that has emerged from an FU-Orionis like outburst, starting with a significant depletion of 
mass inside $\sim 1-2$ AU.  In any case, the structure of the MRI-active layer at the top of the 
disk depends slightly on the total column when it is larger than (\ref{eq:sigtot}), and so we expect
our results to apply qualitatively to the active layers of higher-column disks.

We consider PPDs in which the active layer of the inner disk is fed by a gas reservoir situated 
around $\sim 1$ AU.  Then the dust loading of the active layer is limited strongly by the sticking and 
settling of grains in this reservoir.  A characteristic grain abundance is obtained by balancing the 
time for grain-grain collisions in the reservoir against the radial flow time of gas.  One easily 
shows that the mass fraction of $\mu$m-sized grains is regulated to $\sim 10^{-6}$, with a stronger 
depletion for smaller grains.  This effect may be counterbalanced by the lofting and fragmentation 
of macroscopic particles from a settled layer in the disk midplane.   Further details of
the radial flow of solids are addressed in Paper II.

\begin{table}
\caption{Default Model Parameters} \label{tab:model} 
\begin{center}
\begin{tabular}{ |l|l| }
  \hline
  \multicolumn{2}{|c|}{Disk Parameters} \\
  \hline
  $\Sigma_{g,\rm tot}$ & $200$ g cm$^{-2}$ \\
  $T_{\rm bl}$ & $5000$ K \\
  $\mu_{g,\rm bl}$ & $1.27\,m_u$ \\
  $\mu_{g,c}$ & $2.32\,m_u$ \\
  $a_{\rm cool}$ & $5$ \\[1ex]
  \hline
  \multicolumn{2}{|c|}{Stellar Parameters} \\
  \hline 
  $M_*$ & $M_\odot$ \\
  $R_*$ & $2R_\odot$ \\
  $R_X$ & $10R_\odot$ \\
  $L_*$ & $L_\odot$ \\
  $L_X$ & $2\times 10^{30}$ erg s$^{-1}$ \\
  $\kB T_X$ & 1 keV \\[1ex]
  \hline
  \multicolumn{2}{|c|}{Wind Parameters} \\
  \hline
  $\dot{M}_w$ & $10^{-9}M_\odot$ yr$^{-1}$ \\
  $V_w$ & $400$ km s$^{-1}$ \\
  $B^w_R$ & $\epsilon_B\,(R/0.1~{\rm AU})^{-2}\;{\rm G}$ \\
  $\epsilon_B$ & 0.01-1 \\[1ex]
  \hline
  \multicolumn{2}{|c|}{Turbulence Parameters} \\
  \hline
  $\widetilde\alpha_{\rm MRI,0}$ & $0.1$ \\
  $\alpha_{\rm mix,0}$ & $1$ \\
  ${\rm Am}_{\rm crit}$ & $10$ \\
  $\Lambda_{\rm{O},\rm crit}$ & $100$ \\[1ex]
  \hline
\end{tabular}
\end{center}
\end{table}

\subsection{Wind-disk Boundary Layer}\label{s:wind}

A quasi-radial wind intersects a flared disk, creating a turbulent boundary layer 
(or `aeolosphere') within which the disk material is overturned.  
\cite{elmegreen1978}, \cite{hollenbach2000} and \cite{matsuyama2009} considered the dispersal of a 
PPD by a T-Tauri wind either by direct stripping from the disk surface or through 
a reduction in the angular momentum of disk material, thereby causing it to spiral inward.  

The work described here addresses for the first time the interaction of the embedded wind magnetic field
with the disk.  This embedding is driven by a Kelvin-Helmholtz instability between the wind and disk.
The disk mass that is overturned is limited by the stable stratification of the upper disk, and 
is further restricted by the loss of internal energy to radiation from the mixed layer.  
In the present context, we find that the stabilizing effect of radiation dominates.

Mass loss rates from young stars have been measured to be as high as $\dot M_w \sim 10^{-7}M_\odot$ 
yr$^{-1}$ \citep{gullbring1998}.  The net outflow from an accreting protostar is typically less 
than $10\%$ of the accretion rate \citep{koenigl1993,calvet1997}, although the measured spin rates 
of T-Tauri stars suggest that it cannot be much less than this \citep{matt05}.  We choose a
spherical flow profile,
\be\label{eq:wind2}
\rho_w = {\dot M_w\over 4\pi r^2 V_w},
\ee
and fix $\dot M_w$ to a nominal value $10^{-9}M_\odot$ yr$^{-1}$, which is appropriate for accretion 
at a rate 10 times higher.  Most of the wind acceleration is concentrated inside $0.1$ AU, and we 
fix the wind speed to $V_w \sim 400$ km s$^{-1}$.

We adopt a simple hydrostatic description of the mixing layer.   This is only a rough approximation;
indeed some shock heating is expected given that $V_w$ far exceeds the sound speed in the upper layer
of atomic gas ($c_g = (kT/\mu)^{1/2} \sim 6(T/5000 \,\rm{K})^{1/2}$ km s$^{-1}$).
 The boundary between wind and disk 
is taken to be the height $z_w$ where the component of the wind ram pressure normal to the disk surface 
balances the gas thermal pressure,\footnote{We neglect the contribution of the magnetic field to the pressure
balance, because the radial field is nearly continuous Parker instability limits
the toroidal field imbalance.}
\be \label{eq:normpressbal}
\rho_w V^2_w \sin^2\theta_w=\rho_g c^2_g = P.
\ee
Here $\theta_w$ is the angle between the wind velocity and the disk surface and $\rho_g$ is the gas density
at the top of the disk.  The solution to Equation (\ref{eq:normpressbal}) depends on the radial profile of $P$
and the scale height in the upper disk.  We expect the upper disk to be composed of atomic
gas with a nearly uniform temperature (Section \ref{s:temp}), so that the height at which
Equation (\ref{eq:normpressbal}) applies scales as $z_w \sim c_g/\Omega \propto R^{3/2}$.  We set\footnote{The 
coefficient is taken to be slightly less than $1/2$ to compensate for the greater magnetic support in the inner
disk.}
\be\label{eq:thetaw}
\theta_w = 0.3{z_w\over R}
\ee

Some previous authors have identified the base of the mixed layer with the surface where the gas pressure matches the 
full ram pressure $\rho_w V_w^2$ of the wind (e.g., \citealt{elmegreen1978}).  
This is not the best measure for two reasons.  
First, the wind carries a high enough flux of kinetic energy to overturn deeper layers of the disk.  But, second, 
the mixed layer cannot radiate more energy than is deposited by the 
wind and stellar X-rays.  \cite{glassgold2004} previously calculated the temperature profile of the mixed layer by 
imposing a local balance between a simplified heat source (proportional to density) and a detailed model of radiative cooling; 
but they did not set a cutoff on the column of heated material that reflects a balance between the
height-integrated rates of heating and cooling.

An upper bound on the column of the mixed layer is obtained by balancing the wind energy incident over a height 
$\Delta z \sim R \theta_w$  in a time $\sim \Omega^{-1}$ with the energy needed to overturn an annulus
of gas of radial extent $R$,
\be 
\frac{1}{2}\epsilon_{\rm kin}\rho_{w}v_{w}^{3}\Delta z(2\pi R) \sim \delta\Sigma_{g,\rm mix}c_g^2\Omega R(2\pi R).
\ee
Here $\epsilon_{\rm kin}$ is the fraction of wind energy converted to kinetic energy.  In hydrostatic equilibrium
\be 
P_{\rm mix} \sim {\delta\Sigma_{g,\rm mix}c_g^2\over c_g/\Omega},
\ee
so that
\be 
\frac{P_{\rm mix}}{\rho_w V_w^2} \sim
\frac{0.3\epsilon_{\rm kin}V_w}{R\Omega} \sim 4\epsilon_{\rm kin}\left(\frac{V_w}{400~{\rm km~s^{-1}}}\right)
\left(\frac{R}{\rm AU}\right)^{1/2}.
\ee

A more stringent bound on the mixed column is obtained by balancing heating with two-body cooling.
We use the atomic cooling function $\Lambda=10^{-25.9}$ erg cm$^3$ s$^{-1}$ for $T=5000$ K \citep{dalgarno1972}. 
The boundary layer maintains this temperature due to the steep increase in H electronic transitions at higher temperatures. 
A rapid transition at the base of the atomic layer to lower temperatures and higher densities is driven by molecular cooling,
notably CO roto-vibrational emission \citep{glassgold2004}.

We now balance the wind energy flux with the radiative emission from the corresponding volume of the mixed layer, thereby
defining a characteristic mass density 
\be \label{eq:rhocool}
\rho_{\rm cool}(R) \approx \mu_g\sqrt{\frac{\dot{M}_w V^2_w}{ 8\pi R^3 \Lambda}} \sim 10^3 \rho_w.
\ee
Here $\mu_g$ is the mean molecular weight.
The corresponding atomic H column is 
\be\label{eq:sigcool}
\delta\Sigma_{\rm cool} \sim \rho_{\rm cool} h_g = 1\times 10^{-3}\quad{\rm g~cm^{-2}}
\ee
for the adopted wind parameters, with a weak dependence on $R$.  The high value of the cooling density
(\ref{eq:rhocool}) compared with $\rho_w$ implies that the type of boundary outflow envisaged 
by \cite{matsuyama2009} has a small effect on the instantaneous disk structure.

\subsection{Temperature Profile}\label{s:temp}

We employ an analytic temperature profile in each hemisphere of the disk.  This consists of two isothermal 
layers that are separated by a sharp transition at a freely adjustable height $z_{\rm bl}$,
\be \label{eq:Tprofile}
T(z)=\left(\frac{T_{\rm bl}-T_{c}}{2}\right)\tanh\left(a_{\rm cool}\ln\frac{z}{z_{\rm bl}}\right)+\frac{T_{\rm bl}+T_{c}}{2}.
\ee
The transition is matched to the cooling density (\ref{eq:rhocool}), $\rho_g(z_t) = \rho_{\rm cool}(R)$.
\begin{figure}[ht]
\epsscale{1.2}
\plotone{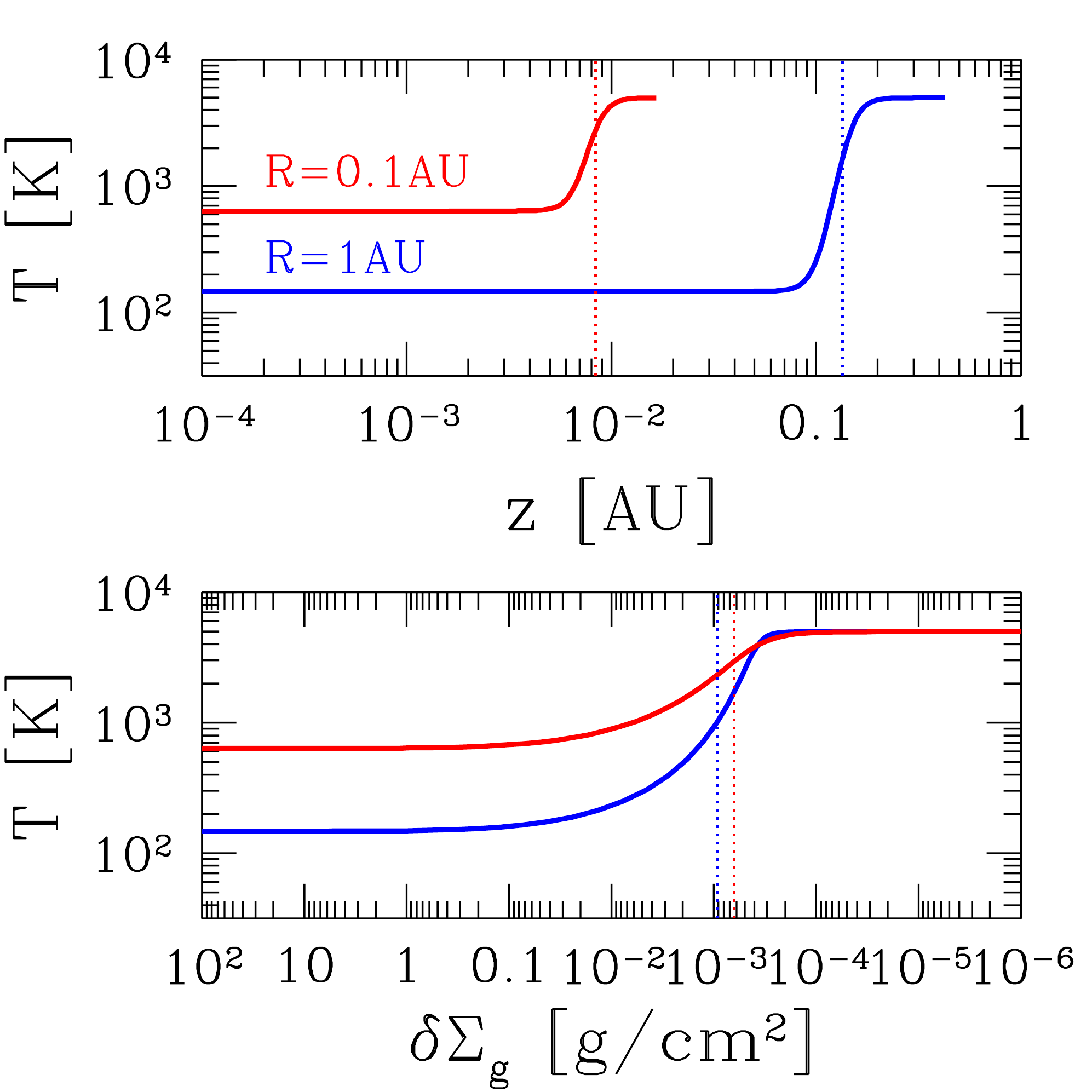}
\caption{Vertical temperature profile in an unmagnetized disk at $R=0.1,\,1$ AU. 
\textit{Top panel}: temperature versus height, Equation (\ref{eq:Tprofile}).   The upper atomic hydrogen layer is fixed to 
$T_{\rm bl}=5000$ K, and the lower temperature $T_c$ is set by radiative equilibrium of dust. 
Vertical dotted line:  transition height $z_{\rm bl}$ given by Equation (\ref{eq:rhocool}) with $a_{\rm cool}=5$. 
\textit{Bottom panel}:  temperature as a function
of column density below disk surface.}
\vskip .1in
\label{fig:Tprofile}
\end{figure}

Sample temperature profiles for an unmagnetized disk are shown in Figure \ref{fig:Tprofile}.
The parameter $a_{\rm cool}$ in Equation (\ref{eq:Tprofile}) sets the magnitude of the temperature gradient, 
with $a_{\rm cool}\sim 5$ giving profiles similar to those obtained by \cite{glassgold2004}. The 
transition between atomic and molecular gas is handled by varying the mean molecular weight $\mu_{g}= (1.27$--$2.32)m_u$ 
in parallel with Equation (\ref{eq:Tprofile}). The turbulent diffusivity in the mixing layer is cut off
sharply below height $z_{\rm bl}$ (Section \ref{s:activecri}).

The upper atomic layer has a temperature
\be\label{eq:tbl}
T_{\rm bl}(R) = 5000 \,\rm{K},
\ee
which is taken to be independent of $R$.  This represents the combined effect of X-ray and UV absorption, 
and the damping of wind-disk turbulence;  turbulent heating dominates in the present context.
The upper layer has a large enough scale height that the wind-disk boundary, as defined by Equation (\ref{eq:normpressbal}),
reaches a moderate fraction of the orbital radius.  

The temperature $T_c$ in the cooler layer below is regulated by the exchange of heat between irradiated dust grains
and gas particles \citep{chiang1997}.  Given an angle of incidence $\theta_{\rm rad}$ of the stellar optical light 
(luminosity $L_\star$) to the disk surface, the effective temperature of the radiation flowing through the disk surface is
\be\label{eq:teff}
T_{\rm eff} = \left(\frac{3\dot M\Omega^{2}}{8\pi\sigma_{\rm SB}} +
\sin\theta_{\rm rad} {L_\star\over 4\pi\sigma_{\rm SB} r^2}\right)^{1/4}.
\ee
Here $\sigma_{\rm SB}$ the Stefan-Boltzmann constant.  

Our focus here is on a later stage of accretion when (i) the accretion flow through the disk has dropped below
$\dot M \sim 10^{-8}\,M_\odot$ yr$^{-1}$; and (ii) the accreted material (as sourced by the disk outside
$\sim 1$ AU) is substantially depleted in dust.  We include the accretion term in Equation (\ref{eq:teff}) when
working out the vertical disk model, but find that it only contributes significantly inside $\sim 0.1$ AU.
For the purposes of calculating $T_{\rm eff}$, we simply set $\dot M = 10^{-8}~M_\odot$ yr$^{-1}$, which is typical
of the peak accretion rates that we obtain in the inner disk.

The second term in Equation (\ref{eq:teff}) depends on the disk thickness relative to the vertical extent of 
the stellar photosphere.  Close to the protostar, $\theta_{\rm rad} \sim (4/3\pi)R_\star/R$; but beyond a radius 
$R_{\rm flare} \sim 0.7(h_p\Omega/c_g)^{-8/9}$ AU the irradiation is dominated by disk flaring, $\theta_{\rm rad} \sim 
d(h_p/R)/d\ln R$ \citep{chiang1997}.  Then
\ba \label{eq:T1}
T_{\rm eff}(R) &=& 540 \,\rm{K}\,\left({L\over L_\odot}\right)^{5/16}\left(\frac{R}{0.1\,\rm AU}\right)^{-3/4}
               \quad R < R_{\rm flare};\nn
T_{\rm eff}(R) &=& 110 \,\rm{K}\,\left({h_p\Omega\over c_g}{L\over L_\odot}\right)^{2/7}
\left(\frac{R}{1\,\rm AU}\right)^{-3/7}\quad R > R_{\rm flare}.\nn
\ea
We take $h_p\Omega/c_g = 2$ to be consistent with the levels of dust depletion invoked in Section \ref{s:dust}. 

We find that the vertical profile has neither very large nor small optical depth to IR radiation:  this is 
suppressed compared with the radial optical depth $\tau_{\rm opt,r}$ of stellar photons by a factor
$\sim (\kappa_{\rm IR}/\kappa_{\rm opt})(c_g/\Omega R) \sim 10^{-2}$.   Therefore we simply take
\be\label{eq:tc}
T_c = T_{\rm eff}
\ee
in the denser molecular layer.

\subsection{Ionization Model}\label{s:ionrate}

Several sources of ionization operate in protostellar disks.  We focus on X-rays from the stellar corona,
whose absorption is the dominant source of free electrons in the inner disk at columns $\delta\Sigma_g \lesssim 50$
g cm$^{-2}$ below the surface (\citealt{glassgold1997}, and references therein).

Radioactive decay could supply an ionization rate $\Gamma_i\sim 10^{-19}$ s$^{-1}$ that
would dominate X-ray photo-absorption at $\delta\Sigma_g \gtrsim 50$ g cm$^{-2}$ \citep{umebayashi2009}.  It is 
suppressed if the abundance of refractory elements is reduced by particle settling in the outer gas reservoir, 
and is neglected here.

Cosmic rays dominate X-rays as a source of ionization only at larger radii ($R \gtrsim 5$ AU), and
would in any case be heavily shielded from the disk by the impacting solar wind considered here \citep{cleeves2013}. 
Finally, we neglect thermal ionization, which is only relevant at temperatures $\gtrsim 1000$ K
\citep{fromang02}.  We find that such temperatures are reached only inside $\sim 0.1$ AU, due to a combination
of modest accretion rate and (partial) dust depletion.

Details of the X-ray ionization model are given in Appendix \ref{s:ionrate2} and are summarized here.
The rate $\Gamma_i$ of photo-absorption per nucleon of disk gas is calculated as a function of $\delta\Sigma_g$
at each radius using a simplified numerical integration of the radiation transfer equation.
We assume that solid refractory grains are strongly depleted, and focus on absorption by gas phase volatiles
such as C, N, O, Ne, and S.  These species are assumed to be depleted by a factor of 0.01 compared with solar abundances. The results are consistent with the Monte Carlo simulations of \cite{glassgold1997}.

The stellar X-ray source has an optically thin bremsstrahlung spectrum with default temperature $\kB T_X = 
1$ keV and luminosity $2 \times 10^{30}$ erg s$^{-1}$, typical of T-Tauri stars \citep{feigelson03,preibisch05}.  This source is
displaced outward from the stellar photosphere, so that the ionization rate at a given column varies 
significantly with $R$.  Photons travelling directly from this corona impact the disk surface at relatively 
large angles close to the protostar, but small angles ($\theta_X \approx 0.04$) at $R=1$ AU.  

\begin{figure}[!]
\epsscale{1.2}
\plotone{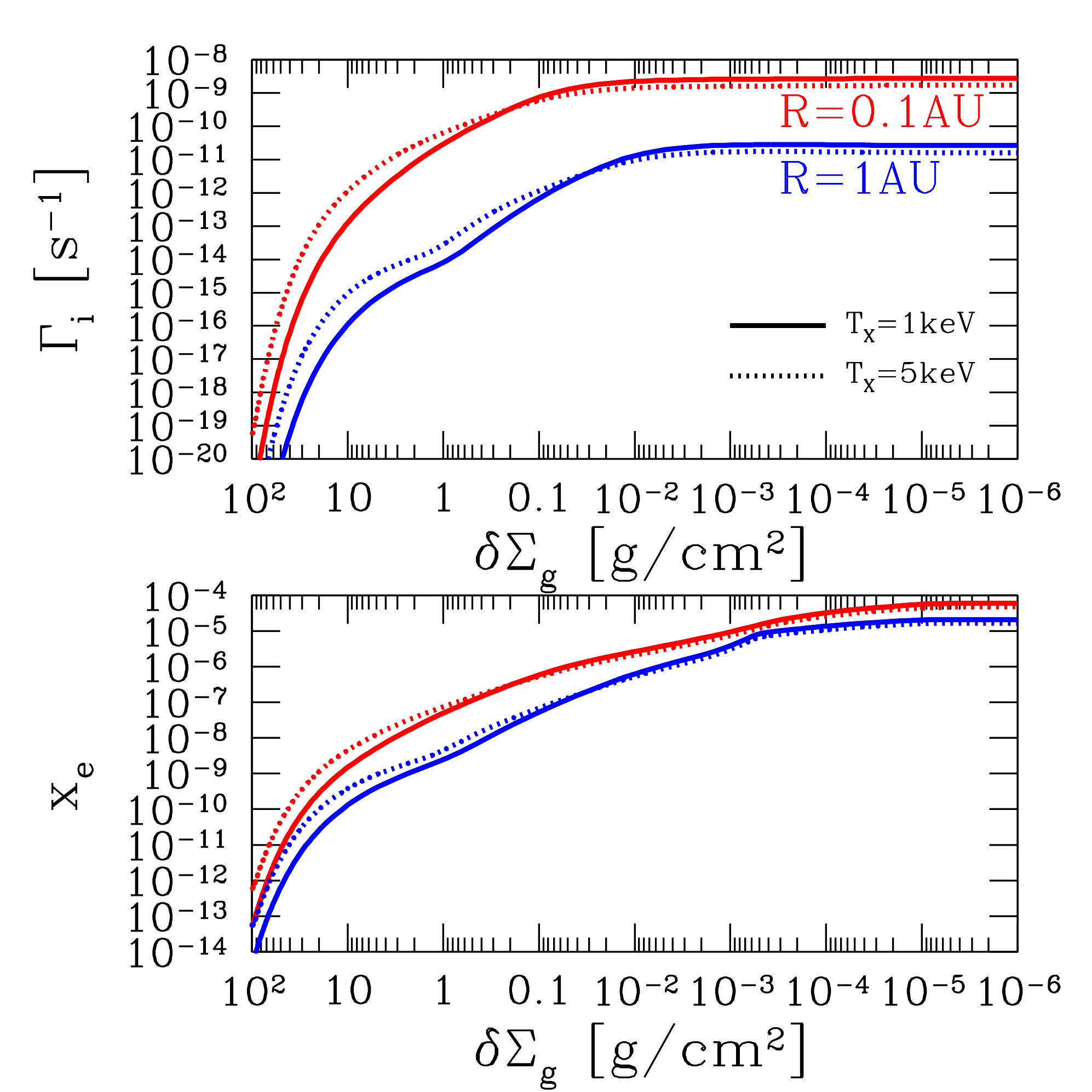}
\caption{\textit{Top panel}: ionization rate per nucleon as a function of column density below disk surface,
shown at $R = 0.1$, 1 AU.
Solid (dotted) lines correspond to stellar X-ray bremsstrahlung temperature $\kB T_X = 1$ (5) keV with fixed luminosity
$2\times 10^{30}$ erg s$^{-1}$.
\textit{Bottom panel}: corresponding ionized fraction, as obtained from Equation (\ref{eq:Xi}).   Vertical disk
profile is here assumed to be unmagnetized.  Ionization calculation is repeated for magnetized disks following
the method described in Appendix \ref{s:ionrate2}.}
\vskip .1in
\label{fig:Giprofile}
\end{figure}

The profile of $\Gamma_i$ is shown in Figure \ref{fig:Giprofile} for both $\kB T_X=1$ and 5 keV.
The column that experiences a given ionization rate increases with $T_X$.  
Direct photo-absorption occurs at a vertical column $\delta\Sigma_g \gtrsim \sin\theta_X \mu_g/\sigma_X \sim 
4\times 10^{-3}(\theta_X/0.04)(E_X/{\rm keV})^3$ g cm$^{-2}$.  The secondary bump in the ionization profile seen in
Figure \ref{fig:Giprofile} results from Compton scattering of a fraction of the photons into the normal direction,
where the ionized column is increased by a factor $\sim \theta_X^{-1}$.   The amplitude of this bump scales
as $\sim R^{-3}$ with distance $R$ from the protostar because it is proportional to $\theta_X/R^2$.  This 
represents a faster drop-off in the ionization rate than in the calculations of \cite{baigood2009}, where
an $\sim R^{-2.2}$ envelope is applied to both the direct and scattered components of the ionizing flux.  

These details influence the scaling of mass transfer 
rate with distance from the protostar.  We find that the active column is larger in the inner disk, 
whereas \cite{baigood2009} find the opposite trend (see their Figure 2).  This difference also reflects the
choice of threshold parameter for MRI activity:  here we use the Ohmic and ambipolar Elsasser numbers, which
are proportional to $v_{\rm A}^2$; whereas, \cite{baigood2009} use the magnetic Reynolds number, which is 
proportional to $c_g^2$ (see Equations (\ref{eq:Lambda}), (\ref{eq:Lambda2})  below).  The stronger magnetization
that we find in the inner disk helps to steepen the radial dependence of the critical parameter by a factor $\sim R^{-1}$.

The abundance of free metal atoms is a major uncertainty in determining the ionization level in PPDs
\citep{fromang02,ilgner2006}.  Atomic Mg is not effectively adsorbed on grain surfaces below
a temperature $\sim 150$ K \citep{baigood2009}, but no simple mechanism regulating its abundance
has been proposed.

We choose a fixed metal abundance 
$x_M = 10^{-8}$ (about $10^{-3}$ the solar abundance of Mg by number), and use the numerical fit to the (grain-free)
equilibrium ionized fraction obtained by \cite{baigood2009} to define an effective recombination 
coefficient, $\alpha_{\rm eff}=1.65\times 10^{-10}(\rho_g/10^{-10}~{\rm g~cm^{-3}})^{-0.2}$ cm$^3$ s$^{-1}$.
This gives
\ba\label{eq:Xi} 
x_e &=& \sqrt{\frac{\Gamma_i}{\alpha_{\rm eff}n_H}}  \nn
     &\approx& 10^{-10}\left({\rho_g\over 10^{-10}\,{\rm g~cm^{-3}}}\right)^{-0.4} 
    \left({\Gamma_i\over 10^{-16}\,{\rm s^{-1}}}\right)^{0.5}.\nn
\ea
This metal depletion is consistent with the level of dust depletion that must be reached for dust
to have a modest effect on the equilibrium ionization level (Section \ref{s:dust}).

\newpage
\subsection{Dust Depletion in Active Layer}\label{s:dust}

Dust grains constitute a very small fraction $X_d$ of the material in a PPD by mass, but have great influence on 
the temperature profile of the disk, its interaction with the embedded magnetic field, and the brightness of 
re-emitted infrared and submillimeter radiation.  

Here we compare the constraints on $X_d$ that result from two considerations:  (i) free electrons are not significantly 
depleted by recombination with ions on the surfaces of grains; and (ii) the PPD is optically thick to stellar 
light at a certain column $\delta\Sigma_g$ below its surface.  We show that a PPD can retain enough dust to 
efficiently reprocess stellar light into the infrared band, while remaining strongly magnetically active.  

In the first instance, grains extract free electrons from the gas through adsorption on their surfaces.  This 
direct channel is, however, suppressed by the charging up of grains.  Even at solar abundance, the space density of
grains is smaller than the free electron density when the gas is sufficiently ionized to be MRI active.  Grains quickly
build up multiple charges, and the probably of electron adsorption is significantly suppressed for a negative charge 
larger than 2--3 \citep{ilgner2006}.  From this point on, charge depletion is mediated by the transfer of positive
charge from molecular and metal ions to negatively charged grains.

An upper bound on $X_d$ is obtained by balancing the ion collision rate $\Gamma_{\rm ads}$ with grain surfaces with 
the recombination rate $\alpha_{\rm eff} n_e n_H$ between free electrons and free ions (which in the absence of
grains sets the equilibrium value of $x_e$).  We consider, for simplicity, a single population of spherical grains
of radius $a_d$, number density $n_d$, and material density $\rho_s$.  Then
\be
\Gamma_{\rm ads} = \pi a_d^2 n_d \left({8\mu_g\over \pi m_i}\right)^{1/2} c_g
                 \sim 0.3 {X_d \delta\Sigma_g\over \rho_s a_d} \Omega.
\ee
Here we take $m_i = 29\,m_u$, representing a mixture of HCO$^+$, Mg$^+$, and heavier ions, and
estimate $\delta\Sigma_g \sim \rho_g h_g$.   This gives
\be\label{eq:maxloading}
\frac{X_{d}}{a_{d}}\lesssim 4\times 10^{-4}\left(\frac{x_{e}}{10^{-9}}\right)
\left(\frac{\delta\Sigma_{g}}{10\,{\rm g~cm^{-2}}}\right)^{-0.2}\left(\frac{R}{\rm AU}\right)^{1/2}~{\rm \mu m^{-1}}.
\ee
Such a dust depletion is consistent with measurements of the outer, more transparent parts of PPDs, which yield
$X_d \sim 0.1-10^{-3}$ times the solar abundance $\sim 10^{-2}$ \citep{furlan2006,dalessio2006}.

Note that $\Gamma_{\rm ads}$
is closely related to the radial optical depth, which we estimate by setting the Mie factor of the optical light
interacting with a $\mu$m-sized grain to unity:
\be
\tau_{\rm opt,r} \sim {3X_d\rho_g r\over 4\rho_sa_d} \sim {3\Omega r\over 4c_g} {X_d\delta\Sigma_g\over\rho_sa_d}.
\ee
The disk is optically thick only if
\be\label{eq:xmin}
{X_d\over a_d} > 1\times 10^{-6} { (T/200 \,\rm{K})^{1/2} (R/{\rm AU})^{1/2}\over 
(\delta\Sigma_g/10~{\rm g~cm^{-2}})}\; {\rm \mu m}^{-1},
\ee
which is significantly smaller than the limit implied by Equation (\ref{eq:maxloading}).

The vertical disk profiles obtained in this paper imply that most of the accretion (driven both by the MRI
stress and an ordered Maxwell stress $B_RB_\phi/4\pi$) is concentrated at a column $\delta\Sigma_g \sim 10$--30 g cm$^{-2}$
at all radii.  Although Equation (\ref{eq:xmin}) suggests that higher optical depths can be sustained at lower
gas columns, in practice we expect the dust to be well-mixed throughout the active layer at the top of the disk.  
For the low $X_d$ considered here, collisions between particles are slow compared with vertical diffusion.
   
Calculations of the spectrum
of reprocessed IR radiation from our model disks are presented in Paper II.

\section{Disk Model: Equilibrium Vertical Profile}\label{s:model2}

To calculate the vertical profile of the disk, we divide the magnetic field into two components:
(i) a background field  ${\bf B} = (B_R,B_\phi)$ with both radial and toroidal components (and therefore 
nearly horizontal); and (ii) a turbulent field which is excited by two independent mechanisms:  the 
MRI and wind-disk shear.

The background field evolves in response to linear winding of the radial field in combination with
vertical diffusion.  It is predominantly toroidal, with a pressure that approaches (or even exceeds) 
the thermal gas pressure, $\beta_\phi \sim 0.1-10$.  The Parker instability and the MRI will certainly generate
a secondary vertical field but -- in contrast with most disk simulations -- a vertical field is not being
imposed ab initio.  

Wind-driven turbulence is concentrated in a thin column near the top of the disk (Equation (\ref{eq:sigcool})), 
and its main role is to provide a diffusive `pump' that pushes the background field into the denser
molecular gas below.   There we include the additional effects of (i) Ohmic drift; (ii) ambipolar drift; and
(iii) MHD turbulence that is self-consistently generated by the MRI.  Vertical Hall drift is neglected for the
reasons described in the following.

\subsection{Magneto-hydrostatic Equilibrium: Stability}\label{s:mhe}

A thin PPD attains an approximate state of magneto-hydrostatic equilibrium against the vertical 
component of the stellar gravity.  The self-gravity of the disk can be neglected in the later stages 
of evolution considered here, but the magnetic field makes a significant contribution to the vertical pressure gradient.

In the one-dimensional approximation, where radial gradients are neglected, the equation of 
magneto-hydrostatic equilibrium reads
\be \label{eq:HE1}
\partial_{z}\left(\rho_g c_{g}^{2}+\frac{B_R^{2}}{8\pi} + 
{B_\phi^2\over 8\pi}\right)=-\rho_g\Omega^{2}z,
\ee
which can be written as
\be \label{eq:weqn}
\partial_{z}\rho_g =-\rho_g\left(\frac{\Omega^{2}}{c_{g}^{2}}z+\frac{\partial_{z}B^{2}}
{8\pi P}+\partial_{z}\ln c_{g}^{2}\right).
\ee
We make no distinction between gas density and total density. 

A vertical profile that satisfies Equation (\ref{eq:HE1}) may be subject to 
the (Newcomb-)Parker instability.  A perfectly conducting fluid, sitting in a gravitational
field $g$ and threaded by a horizontal magnetic field, is unstable to a perturbation with 
wavevector ${\bf k} \parallel {\bf B}$ if \citep{newcomb61}
\be \label{eq:Parker}
\frac{\partial \rho_g}{\partial z} > -\frac{\rho_g^2 g}{\gamma P}.
\ee
Here $\gamma$ is the adiabatic index, which we take to be $7/5$.  The stability threshold
of this `undular' mode is equivalent to the Schwarzschild criterion for convective 
instability.  There is no explicit dependence on the magnetic field when long-wavelength
modes are permitted -- which is the case for a geometrically thin disk.

We find that a Parker-unstable layer may be localized in the vertical direction.  In this case,
the density structure may adjust through an internal redistribution of magnetic flux.   We 
account for the Parker instability by replacing Equation (\ref{eq:weqn}) with the adiabatic density 
profile (Equation (\ref{eq:Parker}) with the inequality saturated).   This is reasonable as long 
as the unstable region has a vertical extent that is of order the disk scale height (or larger), 
and the instability growth rate is larger than the growth rate of the toroidal field by shearing.

The growth rate of the undular mode, as measured by this criterion, turns out to be relatively high 
in the deeper parts of the disk. 
On the other hand, the shearing growth of the toroidal field is relatively fast in the highly magnetized 
wind-disk mixing layer, implying that an adiabatic density profile would tend to {\it underestimate} 
the magnetization there.  We make the practical choice of limiting the density gradient 
throughout the disk whenever Equation (\ref{eq:weqn}) implies inequality (\ref{eq:Parker}).
Layers with high and low growth rate cannot easily be separated from each other.

Relatively few three-dimensional simulations have studied the onset and saturation of the Parker instability in rotating 
disks; and none to our knowledge have included resistive effects in a shearing disk.  
\cite{kim02} studied a shearing disk in the ideal MHD approximation, finding that the instability saturates in about 
two orbital periods.  This is still slow enough to allow for significant growth of $B_\phi$ by shearing in the wind-disk boundary layer.  \cite{kowal03} considered a uniformly rotating galactic disk and found that the Parker instability is suppressed 
for large resistivities, $\eta/\Omega R^2\gtrsim 10^{-5}$, which is comparable to the values obtained here for the active layer
of a PPD. 

Concrete examples of Parker instability are discussed in Section \ref{s:Parker}.

\subsection{Vertical Transport of Magnetic Flux}\label{s:transport}

We adopt a single-fluid description of non-ideal MHD effects, suitably modified
to include the effects of turbulent transport.
Three different conductive regimes are defined by the magnetization parameters 
$\tau_\alpha|Z_\alpha|eB/m_\alpha c$ of the electrons and ions ($\alpha = e,i$).
Here $\tau_\alpha = (\gamma_\alpha \rho)^{-1}$ is the mean free time for collisions with neutrals 
(mainly H or H$_2$, and He).   The collisional drag coefficient 
\be\label{eq:gamma}
\gamma_\alpha \equiv {\langle\sigma v\rangle_\alpha \over m_\alpha +\mu_g}
\ee
depends on the particle mass $m_\alpha$.  One has \citep{bai11},
\ba\label{eq:ccoeff}
\langle\sigma v\rangle_i &=& 2.0\times10^{-9}~{\rm cm^3~s^{-1}}; \nn
\langle\sigma v\rangle_e &=& 8.3\times10^{-9}\left(\frac{T}{100\, \rm{K}}\right)^{1/2}~{\rm cm^{3}~s^{-1}}.
\ea

The induction equation reads
\be \label{eq:induction0}
\frac{\partial{\bf B}}{\partial t}={\bf \nabla}\times\left[{\bf v}\times{\bf B}
-{4\pi \eta\over c}{\bf J} - {{\bf J}\times{\bf B}\over en_e}
+ {({\bf J}\times{\bf B})\times{\bf B}\over \gamma_i\rho_i \rho_g c}\right].
\ee
Here ${\bf v}$ is the velocity of the neutrals, and ${\bf J} = c\bnabla\times{\bf B}/4\pi$ 
is the current density.  The charged species are represented by electrons and a single 
type of positively charged ion of mass $m_i$ and mass density
$\rho_i = (m_i/m_u) x_e\rho$.
The diagonal resistivity, which we write as
\be\label{eq:eta}
\eta \equiv \eta_{\rm O} + \nu_{\rm MRI} + \nu_{\rm mix},
\ee
has contributions from Ohmic diffusion, MRI-driven turbulence, and shear-driven mixing 
between the stellar wind and the disk.  

When calculating turbulent transport of the magnetic field, we make the approximation of 
unit magnetic Prandtl number, ${\rm Pm}=\nu/\eta = 1$.  This choice is supported by 
shearing box simulations of MRI turbulence, which find that the turbulent kinematic viscosity 
is less than or comparable to the turbulent magnetic diffusivity, so that
${\rm Pm}$ approaches unity \citep{fromang09,guan09,lesur09}.  We also make no
distinction between the vertical and radial diffusion of momentum, so that
vertical magnetic field redistribution and radial angular momentum transport are described
by the same diffusivity.

The relative importance of Ohmic, Hall, and ambipolar drift -- the three non-ideal terms in 
Equation (\ref{eq:induction0}) -- can be readily gleaned by defining an anisotropic conductivity 
tensor (e.g., \citealt{wardle12}),
\be \label{eq:induction_ch5}
\frac{\partial{\bf B}}{\partial t}={\bf \nabla}\times\left[{\bf v}\times{\bf B}
-\frac{4\pi}{c}\left(\eta{\bf J}+\eta_{\rm H}{\bf J}\times\hat B + \eta_a{\bf J}_{\perp}\right)\right]
\ee
with coefficients\footnote{It should be kept in mind that only Ohmic and ambipolar drift here are strictly 
diffusive processes.}
\ba \label{eq:etamatrix}
\eta_{\rm{O}}=\frac{c^{2}m_{e}\gamma_{e}\mu_{g}}{4\pi e^{2}}x_e^{-1}; \quad\quad 
\eta_{\rm H}=\frac{c\mu_{g}}{4\pi e}x_e^{-1}\left(\frac{B}{\rho}\right); 
\nn\eta_a=\frac{m_u}{4\pi\gamma_{i}m_{i}}x_e^{-1}\left(\frac{B}{\rho}\right)^{2}.\quad\quad\quad\quad\quad
\ea
Non-diagonal terms in the resistivity tensor associated with hydromagnetic turbulence are neglected, and
dust grains are assumed to have a negligible effect on charge transport.  A critical discussion of
this second assumption is given in Section {\ref{s:dust}.

The effects of Ohmic diffusion are measured by the magnetic Reynolds number $R_M$, 
or by the closely related Elsasser number $\Lambda_{\rm O}$,
\be\label{eq:Lambda}
R_M \equiv {c^2_g\over \eta_{\rm{O}}\Omega};\quad\quad \Lambda_{\rm O} \equiv \frac{v^2_{\rm A}}{\eta_{\rm O} \Omega}.
\ee 
Here $v_{\rm A}=B/\sqrt{4\pi\rho}$ is the Alfv\'en speed.  The Elsasser number
provides a more direct description of the saturation of the MRI.\footnote{The
MRI growth rate in a vertical field reaches a maximum 
$\sim 3\Omega/4$ at a vertical wavenumber $k_z \sim \Omega/v_{\rm{A},z}$
\citep{balbus91}, meaning that $\Lambda_{\rm{O},z} = {v^2_{\rm{A},z}/\eta_{\rm O}\Omega} \sim 1$ 
corresponds to a balance between growth and Ohmic drift over a scale 
$\sim k_z^{-1}$.}  

The effect of ambipolar drift on the MRI is encapsulated by 
\be\label{eq:Lambda2} 
{\rm Am}\equiv \Lambda_a \equiv \frac{v^2_{\rm A}}{\eta_a\Omega}.
\ee
When electrons and positive ions are the only charged species, this reduces to 
\be
{\rm Am} = \frac{m_i\gamma_i x_e \rho}{m_u\Omega}.
\ee
The Elsasser number for Hall drift, $\Lambda_{\rm H} = v^2_{\rm A}/\eta_{\rm H} \Omega$, enters into our
treatment of linear magnetorotational instability in Appendix \ref{s:nonsym}.

Small charged dust grains can significantly reduce ${\rm Am}$ by raising the
ion molecular weight, but the effect is small in the dust-depleted PPD considered here.
\cite{wardle99} obtained a large reduction in the conductivity $\sigma \propto {\rm Am}$ 
for a solar abundance of $0.1\,\mu$m grains.  The corresponding dust loading $X_d/a_d$ is much ($10^2$--$10^4$ times)
larger than the one posited here.   Small grains with a higher abundance will rapidly adhere to form
larger conglomerates because their drift speed $v_{d-g}$ through the gas is smaller than the critical
sticking speed ($\sim 1$ m s$^{-1}$).  In an active layer of column $\delta\Sigma_g$ below the disk surface, one has
\ba
v_{d-g} &\sim & (\epsilon_{\rm dr} \alpha)^{1/2} \left({\rho_s a_d\over \delta\Sigma_g}\right)^{1/2}c_g  \nn
   &\sim&   0.3\alpha_{-1}^{1/2}\,\left({a_d\over 0.1~\mu{\rm m}}\right)^{1/2} \left({\delta\Sigma_g\over 10~{\rm g~cm^{-2}}}\right)^{-1/2}
\quad {\rm m~s^{-1}}.\nn
\ea
Here $\epsilon_{\rm dr} \sim 0.4$ is the Epstein drag coefficient.

\subsection{Relative Importance of Ohmic, Hall and Ambipolar Drift}\label{s:vhall}

We find that ambipolar drift and Hall drift are moderately fast in the active layer:
the respective Elsasser numbers are ${\rm Am} \sim 10^2-10^3$ and $\Lambda_{\rm H} \sim 10-10^2$.
Ohmic transport dominates in the midplane region of the PPD, where the density is highest and $B/\rho_g$ 
smallest; see Equation (\ref{eq:etamatrix}).  Figure \ref{fig:elsass} shows examples obtained
from our vertical disk model, along with the boundary of the active layer as defined below.

\begin{figure}[!]

\epsscale{1.2}
\plotone{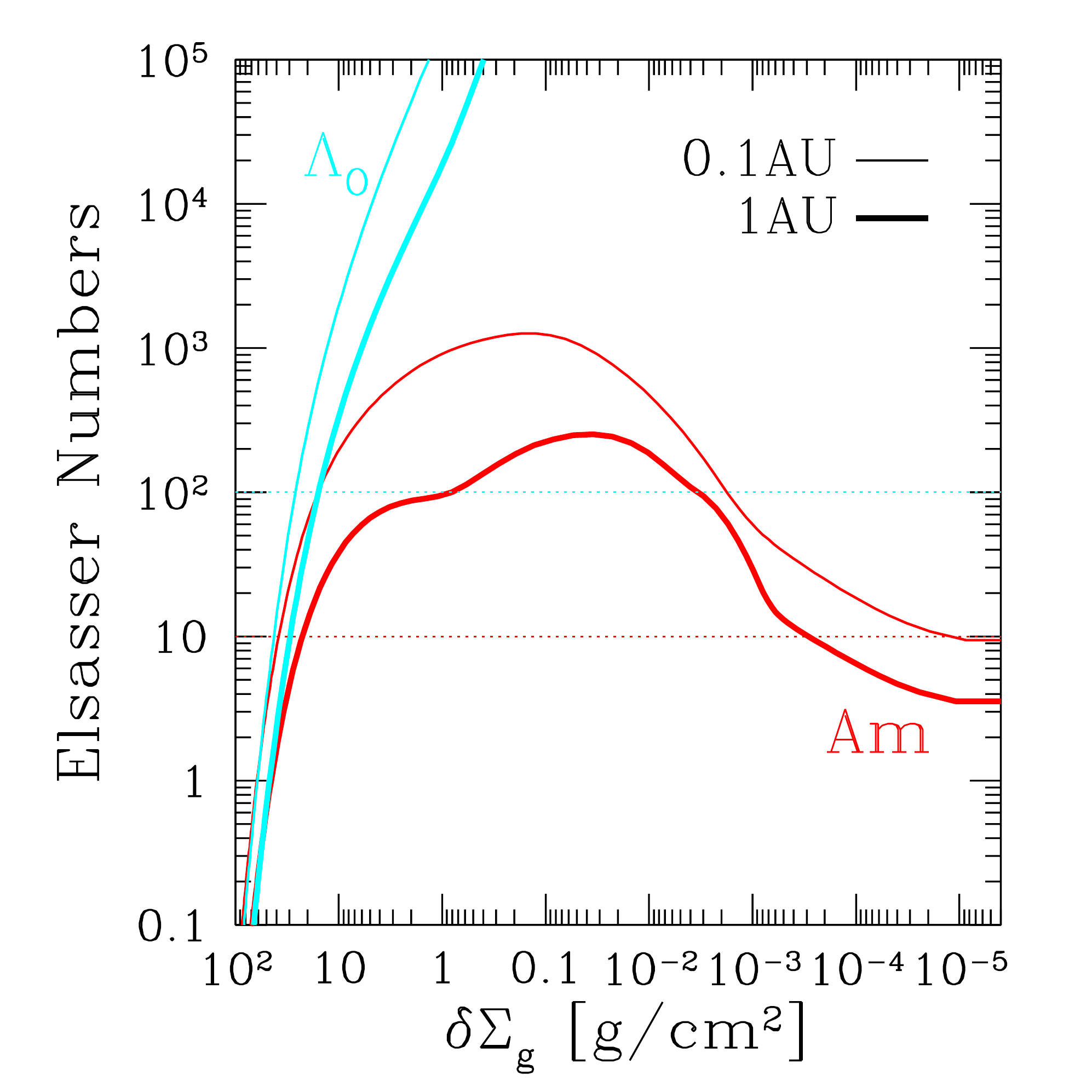}
\caption{Ohmic (cyan) and ambipolar (red) Elsasser numbers shown at $R = 0.1,1$ AU in the more strongly magnetized
hemisphere.  MRI diffusivity is linearly reduced in our parametric model when $\Lambda_{\rm O}$ or ${\rm Am}$ falls below a
critical value $\Lambda_{\rm{O},\rm crit}=100$ (dotted cyan), $\rm{Am}_{\rm crit}=10$ (dotted red), according to Equation(\ref{eq:alphaMRI}).}
\vskip .1in
\label{fig:elsass}
\end{figure}

The linear winding of a radial magnetic field can be sustained even in
some places where MRI activity is effectively quenched.   Balancing the winding 
term in equation (\ref{eq:inductor}) with the vertical Ohmic drift term gives 
\be
B_\phi \lesssim R_M B_R.
\ee
We find $B_\phi \sim 10^2B_R$ in some parts of the disk, implying the modest constraint $R_M \gtrsim 10^2$.  Some
have argued that global modes may be present in stratified disks even if the local MRI is quenched by diffusion 
\citep{gammie94,salmeron05}; but we do not consider such effects here.

Both Hall and ambipolar drift are typically faster than Ohmic drift in the active layer,
but Hall drift is fastest, with a somewhat lower Elsasser number:
\ba
{\Lambda_{\rm H}\over {\rm Am}} &=& {eB\over \rho\langle \sigma v\rangle_i c}\left({\mu_g\over m_u}\right)^{-1}\nn
   &=& 0.3\,{(B^2/8\pi P)^{1/2}\;(T/200 \,\rm{K})^{3/4}\;(R/{\rm AU})^{3/4}\over (\delta\Sigma_g/10~{\rm g~cm^{-2}})^{1/2}}.\nn
\ea
We find $\Lambda_{\rm H} \sim 0.1\,{\rm Am}$ in the layer where radial mass transport is concentrated.

Hall drift has a direct influence on the vertical transport of magnetic flux, as well as an indirect effect
by facilitating MRI turbulence.  Here geometrical factors play an important role, and distinguish our
disk model from those which invoke a vertical magnetic field \citep{lesur14}.

One notices in Equation (\ref{eq:induction_ch5}) that the relative size of the Hall and ambipolar terms depends on
the relative orientation of ${\bf J}$ and ${\bf B}$, not just on the ratio $\eta_{\rm H}/\eta_a$.  The direction
of ambipolar drift is perpendicular to the background magnetic field, which in this case is horizontal.
Therefore, ambipolar drift involves the transport of magnetic flux at the speed
\be
v_{{\rm amb},z} = -\eta_a\frac{\partial_{z}{B}}{{B}}.
\ee

Hall drift involves the advection of the magnetic field with a speed ${\bf v}_{\rm H} = -{\bf J}/en_e$.  
Within the interior of the disk, the current is mainly vertical, and 
\be 
v_{{\rm H},z}=-\frac{c}{4\pi en_{e}R}\partial_R \left(RB_{\phi}\right).
\ee
The direction of Hall drift is approximately anti-symmetric about the disk midplane (as measured in a 
cylindrical coordinate system).  The imposed radial magnetic field has opposing signs in the two hemispheres.
For example, a positive radial field at $z > 0$ is sheared by the disk into a negative toroidal field, which
typically decreases more rapidly than $\sim R^{-1}$.  As a result $v_{\rm{H},z} < 0$, and both the
radial and toroidal fields are advected toward the disk midplane.  Hall drift has the opposing sense at $z < 0$, 
and so is also toward the midplane.  

Because $v_{\rm{H},z}$ involves the radial gradient of the field, its magnitude is suppressed compared
with ambipolar drift, even where $\eta_{\rm H} > \eta_a$:
\ba 
\frac{v_{\rm{H},z}}{v_{{\rm amb},z}} &\approx &
\frac{h_g}R\frac{\eta_{\rm H}}{\eta_a} \nn
&=& 0.01\left({\delta\Sigma_g\over{\rm g~cm^{-2}}}\right)
\left(\frac{B}{1~\rm G}\right)^{-1}\left(\frac{R}{1~\rm AU}\right)^{-1}.
\ea

This only approaches unity for $\delta\Sigma_g \gtrsim 100$ g cm$^{-2}$,
at which point Ohmic drift is the dominant non-ideal effect.   
Vertical transport of large-scale fields can, therefore, be first
approached by neglecting Hall drift.

A radial Hall drift, at a speed $v_{\rm{H},R} \sim (c/4\pi n_e) \partial_z B_\phi$,
is a source of a radial magnetic field in the presence of a mean
vertical field, due to differential bending of the vertical field \citep{kunz08,lesur14}.  
Even when the seed poloidal field is predominantly radial, a small vertical field is present, 
$B_z \sim (h_g/R) B_R$.  We can nonetheless show that this additional source term for $B_R$ is subdominant compared
with vertical ambipolar drift of the radial field.  The respective terms in the induction equation are
\be
{\partial_t B_R|_{\rm Hall}\over \partial_t B_R|_{\rm amb}}
 = {\partial_z(B_z \eta_{\rm H} \partial_z\ln B_\phi)\over\partial_z(B_R \eta_a\partial_z\ln B)} \sim  {h_g \over R}{\eta_{\rm H}\over \eta_a},
\ee
with the same scaling as for vertical Hall drift. 

\subsection{Turbulence in the Wind-disk Boundary Layer}

Hydrostatic support is only a rough approximation in the wind-disk
boundary layer.  The turbulent viscosity is taken to be
\be \label{eq:num}
\nu_{\rm mix}  = \alpha_{\rm mix} {c_g (c_g^2 + v_{\rm A}^2)^{1/2}\over \Omega}.
\ee
Details of the boundary layer structure are relatively insensitive to the normalization of $\nu_{\rm mix}$.
Although the boundary layer is very strongly magnetized in the inner disk, $B^2/8\pi \gg P$, 
we find that the magnetic field profile is fairly flat.  The scale height 
remains close to $c_g/\Omega$ as a result of our imposition of marginal Parker stability.

The diffusivity (\ref{eq:num}) dominates in the boundary layer, but must decrease
significantly below the transition height $z_{\rm bl}$ between atomic and molecular gas, where the heat that is deposited by mixing is rapidly radiated by molecular lines.  We take
\be \label{eq:alpham}
\alpha_{\rm mix} = \alpha_{{\rm mix},0}\frac{T}{T_{\rm bl}}e^{-\frac{1}{10}\left(\frac{z_{\rm bl}}{z}\right)^{7}};
\quad\quad \alpha_{{\rm mix},0} = 1.
\ee

\subsection{Prescription for MRI Turbulence}

The combination of a (nearly) toroidal background magnetic field, strong magnetization, and 
moderate ${\rm Am}$ leads us to treat the MRI-driven magnetic field as a perturbation to the 
background field.  Numerically derived scalings between the $\widetilde\alpha$ stress coefficient and the 
imposed magnetic field are not available in this regime, and so we adopt a simple proportionality
\be \label{eq:alpha}
\nu_{\rm MRI} = \widetilde\alpha_{\rm MRI} {B_\phi^2\over 8\pi \rho_g\Omega}.
\ee
The coefficient is taken to be
\be
\widetilde\alpha_{\rm MRI} = \widetilde\alpha_{{\rm MRI},0} = 0.1
\ee
in the ideal MHD regime, with a cutoff in ${\rm \Lambda_{\rm O}}$ and ${\rm Am}$ 
as prescribed by Equation (\ref{eq:alphaMRI}). 

The rate of vertical ambipolar drift is taken to be proportional 
to the gradient of the background magnetic pressure, with the contribution of the MRI-generated field 
neglected.

Independent mechanisms for exciting MHD turbulence can be considered.  The toroidal magnetic field
obtained in solving for the vertical disk profile is generally unstable to the undular Newcomb-Parker mode
(Section \ref{s:mhe}).   The radial magnetic field carried out by the T-Tauri wind could
also have an azimuthal structure, which after embedding in the disk shear would facilitate internal
magnetic reconnection.  These additional effects are effectively subsumed into the parameter $\widetilde\alpha_{\rm MRI}$.

\begin{figure}[ht]
\epsscale{1.2}
\plotone{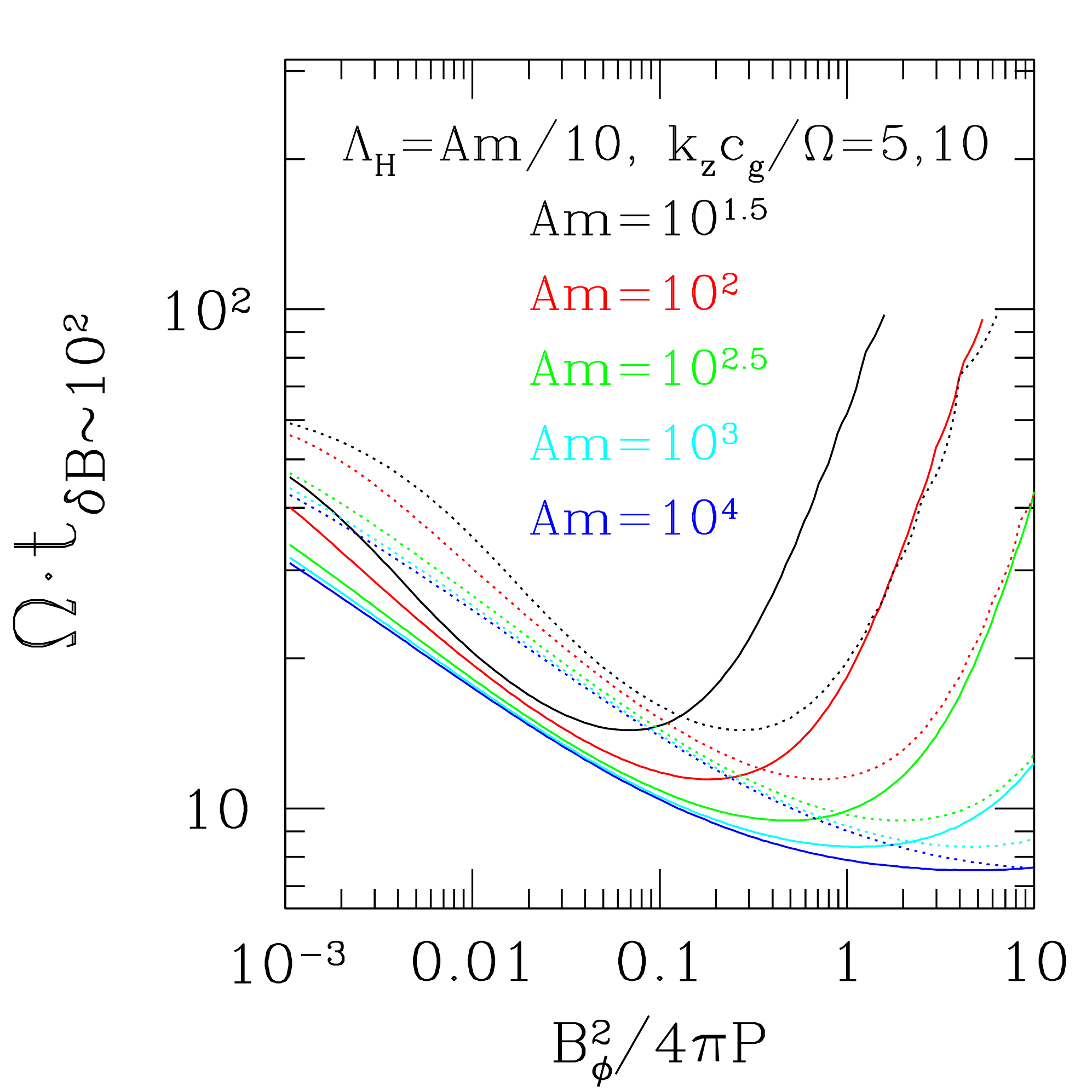}
\caption{Dependence of minimum growth time (for any $m$) on background
magnetization, for a range of ambipolar and Hall Elsasser numbers.  Here
$\Lambda_{\rm H} = {\rm Am}/10$ for each value of ${\rm Am}$.  Solid lines:  $k_zc_g/\Omega=10$;
dotted lines: $k_zc_g/\Omega=5$.}
\vskip .1in
\label{fig:growthmag}
\end{figure}

\begin{figure*}[ht]

\epsscale{.95}
\plottwo{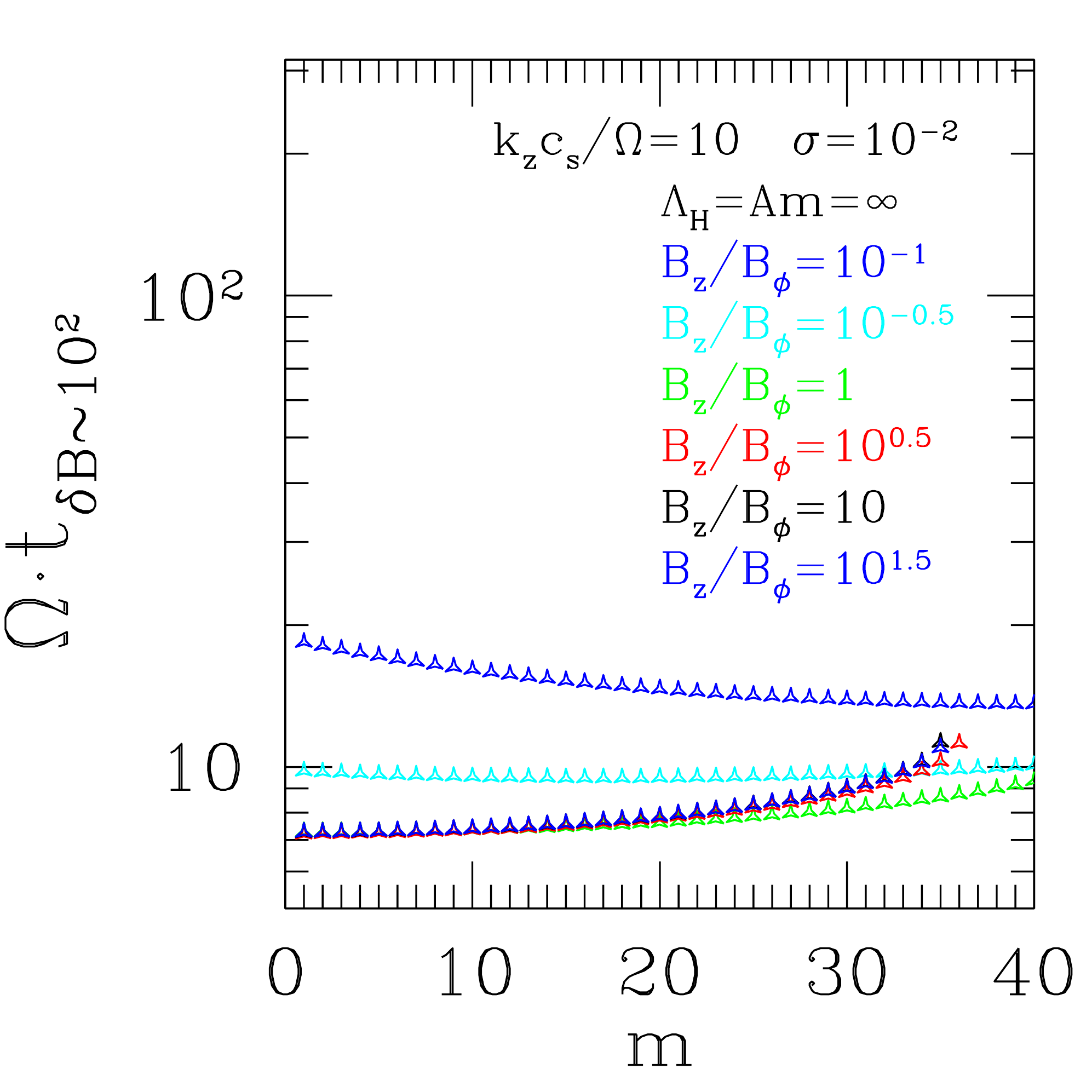}{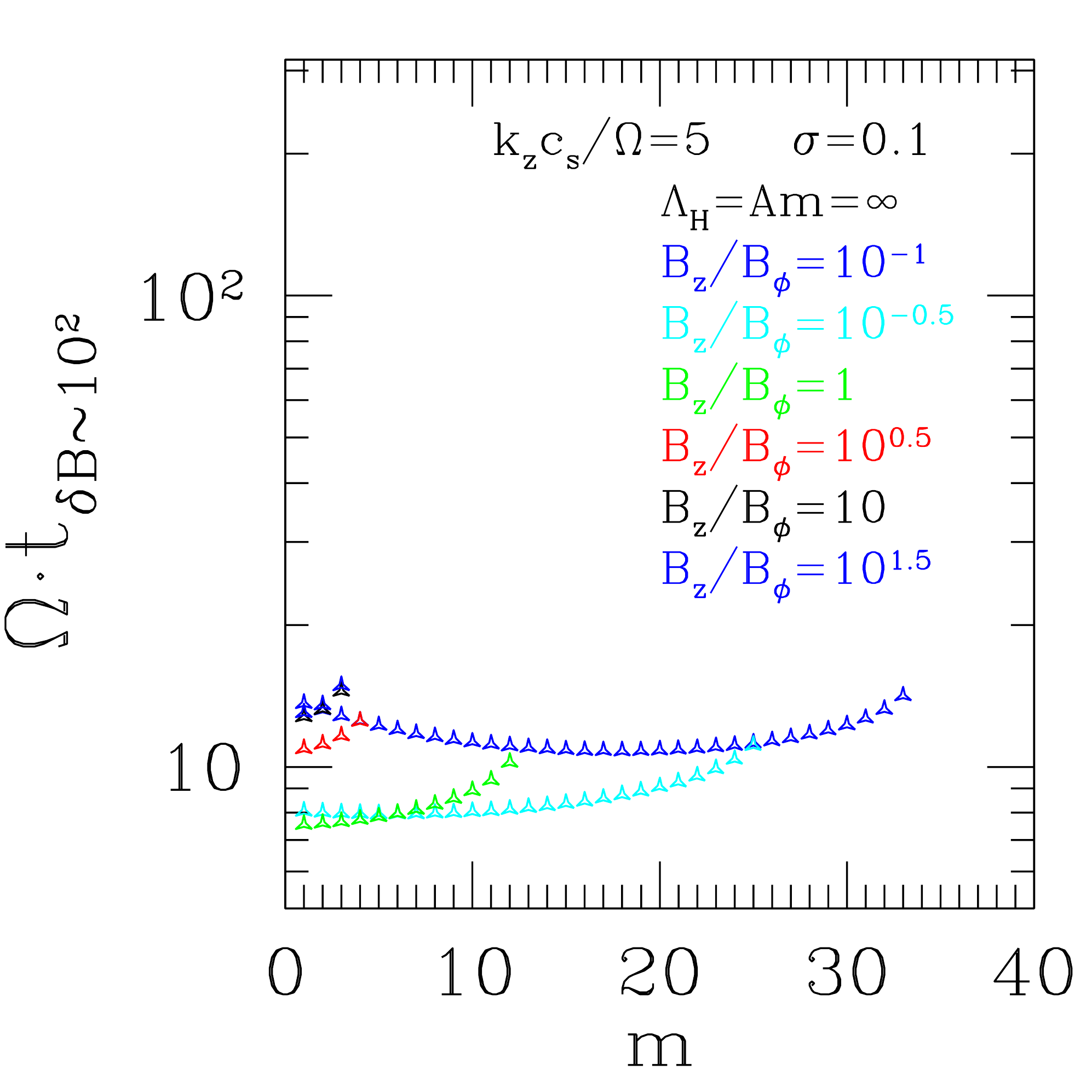}
\plottwo{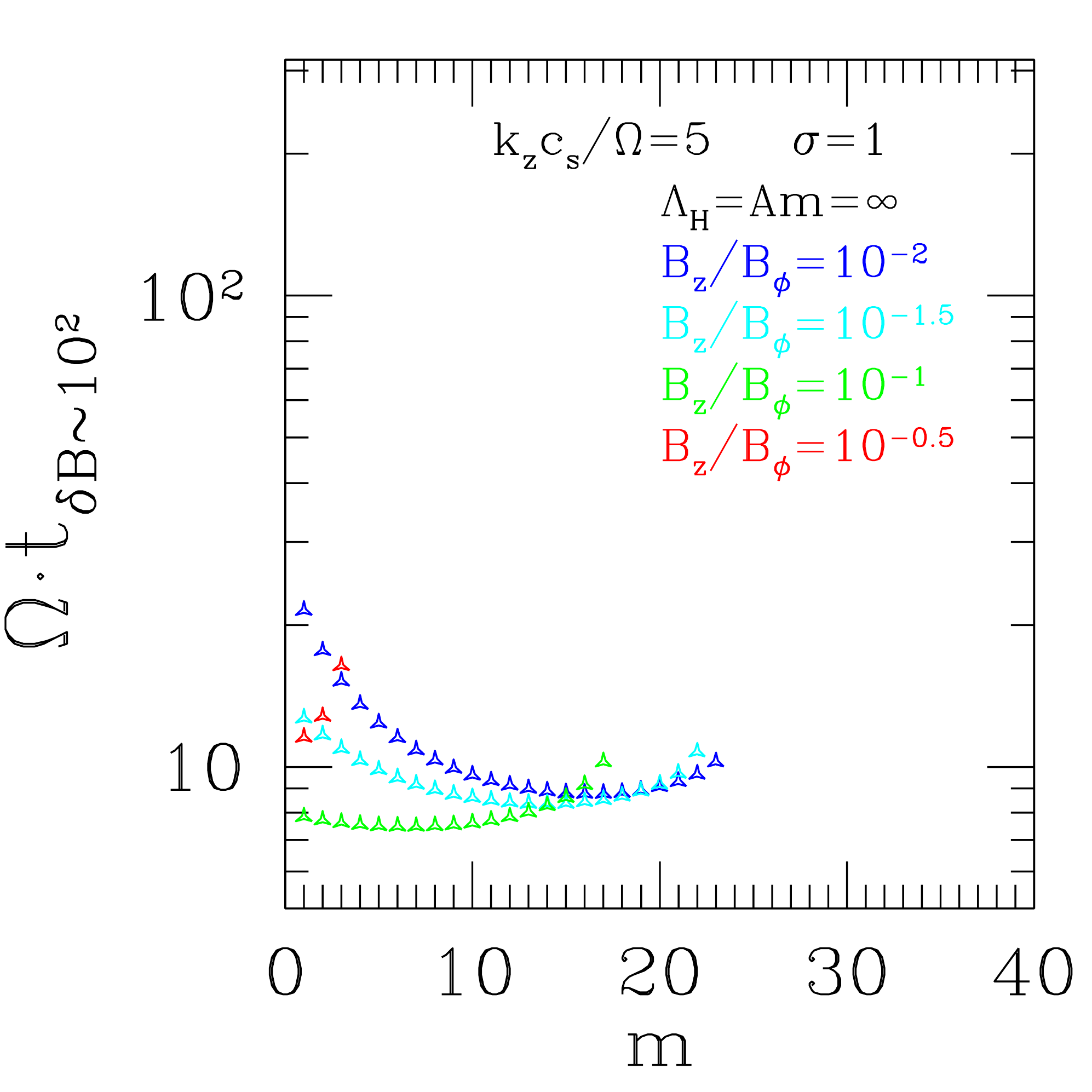}{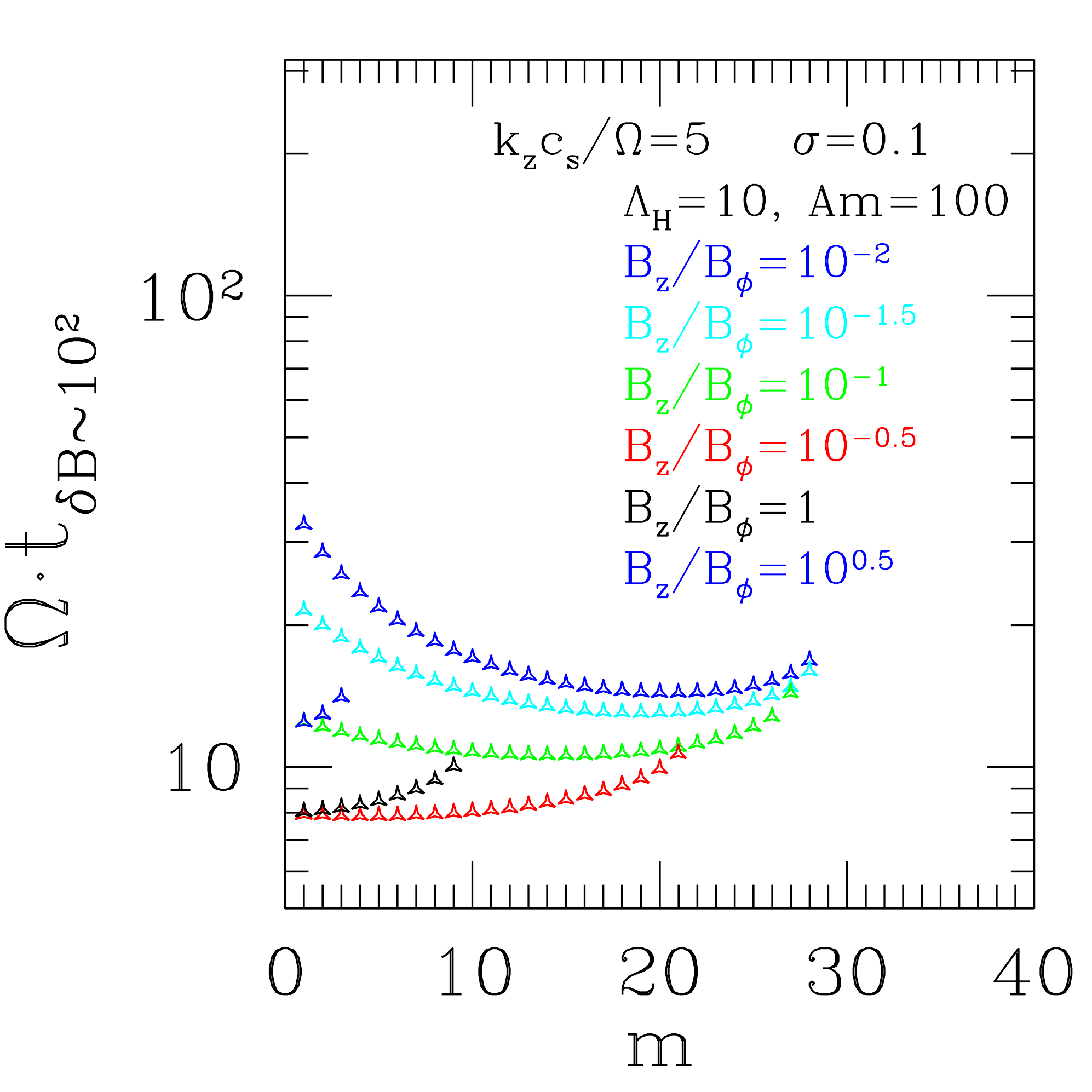}

\caption{Effect of background field orientation on MRI growth.  Plotted is time for a magnetic perturbation 
$\delta B_R = \delta B_0$ with azimuthal wavenumber $m/R$ to grow to an amplitude 
$|\delta{\bf B}| = 10^2\delta B_0$.  Vertically unstratified Keplerian disk with background field
$B_\phi\hat\phi + B_z\hat z$ composed of ideal MHD fluid (${\rm Am} = \Lambda_{\rm H} = \infty$).  {\it Top left panel:} 
magnetization $\sigma = (B_\phi^2 + B_z^2)/4\pi P = 10^{-2}$. {\it Top right, bottom left panels:} $\sigma = 0.1$, 1.
MRI growth is shut off in predominantly vertical background magnetic fields
at larger $\sigma$.  {\it Bottom right panel:} effect of turning on ambipolar and Hall drift.}
\vskip .1in
\label{fig:modegrowth}
\end{figure*}

\begin{figure}[ht] 
\epsscale{1.23}
\plotone{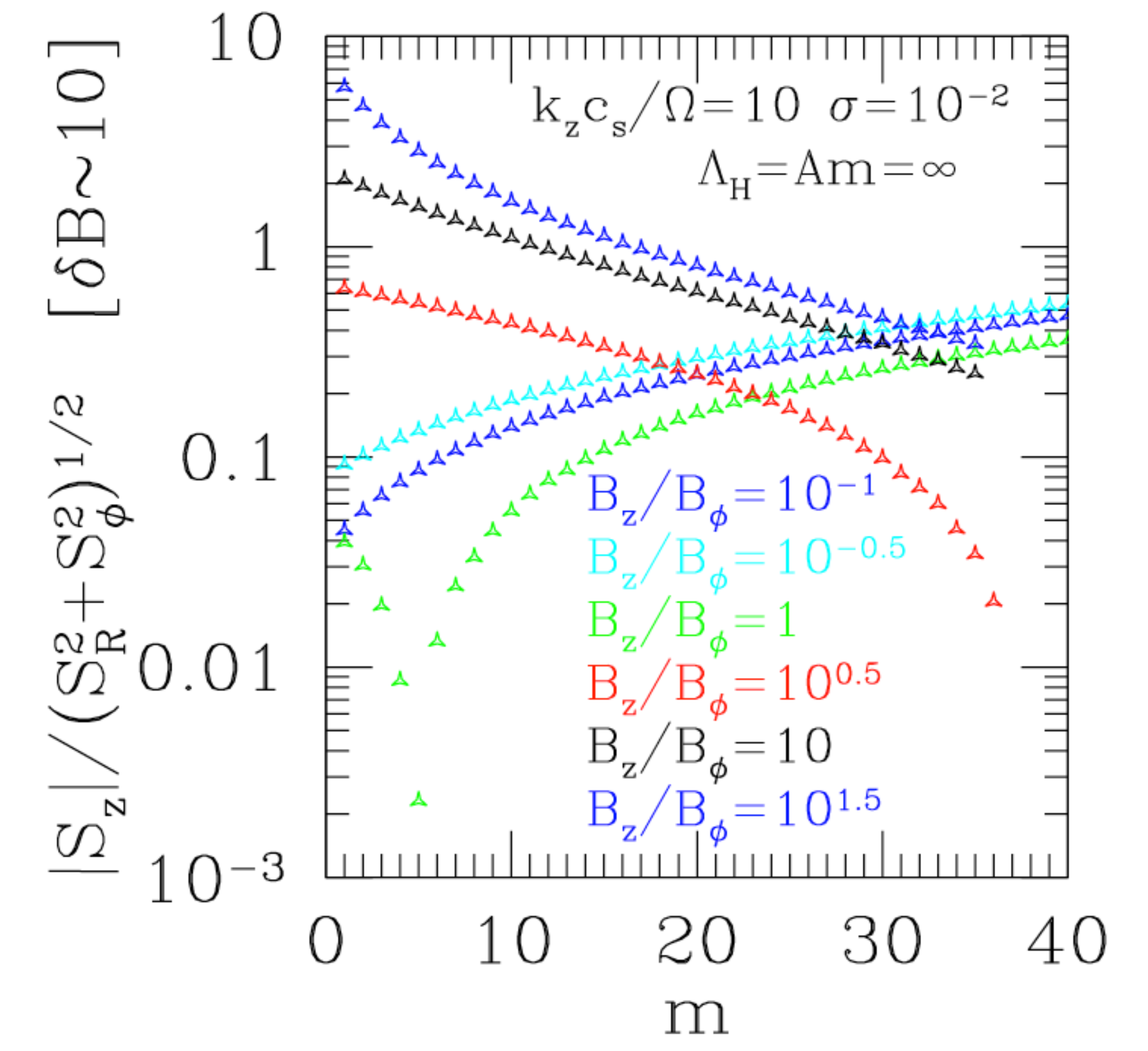}
\caption{Direction of Poynting flux, $|S_z|/(S_R^2+S_\phi^2)$, corresponding to the top left panel of Figure
\ref{fig:modegrowth}.  The MRI mode transports energy horizontally in a toroidally dominated
magnetic field, suggesting confinement of the mode within a vertically stratified disk.}
\vskip .1in
\label{fig:poynting}
\end{figure}

\newpage
\subsubsection{Insights from Linear Stability Analysis}\label{s:linsum}

The vigor of the MRI in a toroidally magnetized disk needs careful consideration.  
Direct numerical simulations imposing a vertical seed field find that strong magnetization
($\beta \ll 10^2$) suppresses the MRI \citep{bs13,lesur13}.  In a first approximation, this is because
the dominant growth wavelength $\lambda \sim 2\pi v_{\rm{A},z}/\Omega \sim 2\pi \beta^{-1/2} h_g$ 
cannot fit inside a scale height $h_g \sim c_g/\Omega$.  

The MRI of a toroidal magnetic field is fundamentally non-axisymmetric \citep{balbus92}.
The range of azimuthal wavenumber $m$ that experiences strong growth is limited by winding up of the 
mode by the disk shear. The analysis of \cite{papaloizou97} suggests that the fastest growth is concentrated at 
$m \sim \Omega R/v_{\rm{A},\phi}$;  we give a refined estimate below.

The introduction of ambipolar drift imposes additional constraints on growth.  Shearing box simulations 
with a toroidal magnetic field and moderate magnetization ($\beta \gtrsim 10^2$) find a vigorous MRI 
instability for ${\rm Am} \gtrsim 10$ \citep{baistone11}.
Recent simulations by \cite{lesur14} and \cite{bai14} that adopt somewhat faster ambipolar and Hall drift
(${\rm Am} \lesssim 10$) and assume a vertical background field, find a mostly laminar behavior.
But as yet there are no numerical explorations of the MRI in a strong toroidal magnetic field,
reaching higher values of ${\rm Am}$, and including the effect of Hall drift as well as ambipolar drift.

Linear theory therefore offers some guidance here.  Our stability analysis, which includes both
ambipolar and Hall drift, is laid out in Appendix \ref{s:nonsym}.  We adapt the approach of \cite{balbus92} 
(who considered a disk composed of ideal MHD fluid) and \cite{balbus01} (who added in the combined effects of 
Hall drift and ohmic drift).

This analysis implies rapid MRI growth at a high toroidal magnetization ($\beta_\phi \sim 1$--$10$) when
the Elsasser numbers ${\rm Am} \sim 10\Lambda_{\rm H}$ are as low as 30--100.   MRI growth is 
more restricted when the Hall term is neglected, requiring ${\rm Am} \gtrsim 10^2$ for modes with a
vertical wavenumber $k_z \sim 5$--$10\,h_g^{-1}$.  The growth rate peaks at an azimuthal wavenumber
\be
m \lesssim {k_z R\over 2\ln(\delta B/\delta B_0)}
   \sim 30\,\left({k_z\over 10\, h_g^{-1}}\right) \left({h_g\over 0.03 R}\right)^{-1}.
\ee
Here $\delta B_0$ is the seed perturbation; see Appendix \ref{s:analytic} for a derivation.

Figure \ref{fig:growthmag} shows the
minimum time for the magnetic perturbation to reach amplitude $|\delta {\bf B}| \sim 10^2 \delta B_0$
for any azimuthal wavenumber $m$, as a function of magnetization and for a range of Elsasser numbers.
Rapid growth shuts off at $\beta \lesssim 1$ once ${\rm Am}$ drops below $\sim 10^2$.   
On this basis, our model disk, which achieves peak ${\rm Am} \sim 300$--$10^3$ (Figure \ref{fig:elsass}),
will sustain a vigorous MRI.

The orientation of the background magnetic field has an important influence on MRI growth.  
Figure \ref{fig:modegrowth} shows that growth slows as the field
develops a significant vertical component.  At high magnetization, growth is shut off 
as $B_z$ approaches $B_\phi$.   

This effect is confirmed analytically in Appendix \ref{s:analytic}
by examining the time evolution equations for the magnetic perturbations $\delta B_{z,R}$.  When
the magnetic tension term $({\bf k}\cdot{\bf v}_{\rm A})^2\delta B_{z,R}$ exceeds the dominant
term driving the MRI in magnitude, then the instability shuts off.  The corresponding upper bound on $B_z$
is $\sim (\beta/k_z h_g)^{1/2} B_\phi$.  This effect is enhanced when ambipolar and Hall drift are 
turned on.  Locally stronger $B_z$ could be generated by a Parker instability of the toroidal field, but the
resulting feedback on the MRI cannot easily be handled by the method adopted in this paper. 

A vertical magnetic field in a radially sheared disk experiences a direct instability when
the effects of vertical stratification are neglected \citep{balbus91}.  But stratification is present,
which raises the possibility that mode growth is inhibited by vertical propagation.  It is therefore
useful to consider the orientation of the Poynting flux associated with a growing MRI mode.
The Poynting flux vector is
\be
{\bf S} \equiv {{\bf E}\times{\bf B}\over 4\pi}c = {B^2{\bf v} - ({\bf B}\cdot{\bf v}){\bf B}\over 4\pi},
\ee
where ${\bf E} = -{\bf v}\times{\bf B}/c$ is the electric field.  We work in the frame co-rotating with the
local disk angular velocity $\Omega(R)$, so that ${\bf v}$ is the velocity perturbation 
$\delta{\bf v}$ that is induced by the MRI.  Then the piece of ${\bf S}$ that is quadratic in the perturbation is
\ba
S_R &=& B_\phi(2\delta B_\phi \delta v_R - \delta B_R\delta v_\phi) + B_z(2\delta B_z\delta v_R - 
   \delta B_R\delta v_z);\nn
S_\phi &=& -B_\phi(\delta B_R \delta v_R + \delta B_z\delta v_z) + B_z(2\delta B_z\delta v_\phi - 
   \delta B_\phi\delta v_z);\nn
S_z &=& B_\phi(2\delta B_\phi \delta v_z - \delta B_z\delta v_\phi) - B_z(2\delta B_R\delta v_R + 
   \delta B_\phi\delta v_\phi).\nn
\ea

The relative magnitudes of the vertical and horizontal compoents of ${\bf S}$, obtained from the stability
analysis of Appendix \ref{s:nonsym}, are plotted in Figure \ref{fig:poynting}.  One finds that ${\bf S}$ is
nearly horizontal for modes with $m \lesssim 10$ and $B_z \lesssim B_\phi$,
but becomes vertical as the background magnetic field is rotated into
the vertical direction.  This behavior
parallels that of an Alfv\'en wave, whose Poynting flux is also
directed along the background field.  It suggests that a MRI mode 
in a toroidal magnetic field will remain confined to the disk, and its
growth will not be shut off by propagation across the vertical density
gradient.

\newpage
\subsubsection{Cutoff in MRI Diffusivity due to Non-ideal Effects}\label{s:activecri}

We apply cutoffs in the Ohmic and ambipolar Elsasser numbers to represent 
the suppression of the MRI:
\renewcommand*{\arraystretch}{1.5}
\ba\label{eq:alphaMRI}
&\widetilde\alpha_{\rm MRI} = \widetilde\alpha_{{\rm MRI},0}\left(\frac{\Lambda_{\rm O}}{\Lambda_{\rm O}+
   \Lambda_{\rm{O},\rm crit}}\right)
\left(\frac{\rm Am}{\rm Am+Am_{\rm crit}}\right);\\[5mm]
&\Lambda_{\rm{O},\rm crit} = 100; \quad {\rm Am}_{\rm crit} = 10.
\ea
The scaling in $\Lambda_{\rm O}$ is motivated by Figure 20 of \cite{sano02}.

A threshold value ${\rm Am}_{\rm crit} \sim 10$ 
(in combination with Hall drift of a strength $\Lambda_{\rm H} \sim {\rm Am}/10$)
implies a minimal ionized fraction
\ba\label{eq:xmin2}
x_e &>& {\rm Am}_{\rm crit} {m_u c_g\over \delta\Sigma_g \langle \sigma v\rangle_i} \nn 
    &=& 7\times 10^{-11}\,\left({{\rm Am}_{\rm crit}\over 10}\right)
    \left({\delta\Sigma_g\over 10~{\rm g~cm^{-2}}}\right)\left({T\over 200\, \rm{K}}\right)^{1/2}.\nn
\ea
Equation (\ref{eq:Xi}) gives the  corresponding lower bound on the ionization rate 
\be
\Gamma_i > 1.6\times 10^{-17}\,{ ({\rm Am}_{\rm crit}/10)^2\; (T/200 \,\rm{K})^{0.4} \over
(\delta\Sigma_g/10~{\rm g~cm^{-2}})^{1.2}\; (r/{\rm AU})^{1.2} } \quad{\rm s^{-1}}.
\ee

As regards Ohmic drift, which dominates in the midplane regions of the disk,
shearing box simulations indicate a threshold $\Lambda_{\rm O} \sim 10$ 
for MRI activity in a vertical magnetic field \citep{sano02}, increasing 
to $\Lambda_{\rm O} \sim 10^2$ in a toroidal field \citep{simon09,flock12}.  
The peak growth rates are a few times smaller in a toroidal field.
We choose $\Lambda_{\rm{O},\rm crit} = 100$ to mark 
the MRI-active layer of the disk.   

Figure \ref{fig:elsass} shows that the depth of the active layer is insensitive to the choice 
of $\Lambda_{\rm{O},{\rm crit}}$ given the rapid drop in $x_e$ and $\Lambda_{\rm O}$ above 
$\delta\Sigma_g \sim 10$--30 g cm$^{-2}$.

\newpage
\subsection{Analytic Estimate of Sheared Magnetic Field}\label{s:equil}

The upper disk is strongly magnetized as a result of the imposed magnetic field in the T-Tauri wind.  
The Alfv\'en Mach number of the wind itself is
\be
{V_w\over B_R/(4\pi\rho_w)^{1/2}} =  11\,\epsilon_B^{-1}\left({R\over {\rm AU}}\right)
\left({\dot M_w\over 10^{-9}\,M_\odot~{\rm yr}^{-1}}\right)^{1/2},
\ee
where $V_w \sim 400$ km s$^{-1}$.  Estimating the height of the wind-disk interface as 
$z_w \sim 2c_s(T_{\rm bl})/\Omega$, and taking Equation (\ref{eq:pb}) for the boundary pressure, we obtain
\be\label{eq:magnetize}
{B_R^2\over 8\pi P}\biggr|_{z_w} \sim 0.5\,\epsilon_B^2\left({R\over {\rm AU}}\right)^{-3}\,
                                    \left({T_{\rm bl}\over 5000\,\rm{K}}\right)^{-1}.
\ee
Even without including the contribution from the toroidal magnetic field, we see that the magnetic pressure approaches
the gas pressure at $R \sim 1$ AU, and greatly exceeds it at $R \sim 0.1$ AU.  

The radial magnetic field is pumped downward from the wind-disk boundary layer, and maintains
a fairly flat profile in the active region.  The equilibrium toroidal field is fixed by a competition 
between growth by shearing of the radial field, and vertical diffusion.

Where the disk is thermally supported, an analytic approximation is easily obtained from Equation 
(\ref{eq:bphieq}).  The diffusion time is $t_{\rm diff} \sim \eta^{-1} h_g^2$,
where $\eta$ is the maximum of $\nu_{\rm MRI}$ and $\eta_{\rm amb}$.  In either case of MRI or ambipolar dominated
transport, the diffusivity itself depends on the shear-amplified magnetic field.  

First considering turbulent diffusion, the equilibrium toroidal field is
\ba \label{eq:Beq1}
B_{\phi,\rm eq} &\;\approx\;& 
-\left(\frac{12\pi\delta\Sigma_g\Omega c_{g}B_R}{\widetilde\alpha_{\rm MRI}}\right)^{1/3}\quad\quad\quad ({\rm MRI})\nn
&\approx& -0.9\,\epsilon_B^{1/3}{(\delta\Sigma_g/10\,{\rm g~cm^{-2}})^{2/3}\,(T_c/200 \,\rm{K})^{1/6} \over
  (R/{\rm AU})^{7/6}\,(\widetilde\alpha_{\rm MRI}/0.1)^{1/3}}\;{\rm G},\nn
\ea
within the mass-transferring layer of column $\delta\Sigma_g \sim 10$--30 g cm$^{-2}$.
The magnetization reaches a local maximum at a much shallower depth in the disk, at $\delta\Sigma_g \sim 10^{-2}$ g cm$^{-2}$, 
\ba \label{eq:betaeq1}
\frac{B_{\phi,\rm eq}^{2}}{8\pi P} &\;\approx\;& 1.7\,\epsilon_B^{2/3}
\left({R\over {\rm AU}}\right)^{-5/6}\left({\delta\Sigma_g\over 10^{-2}~{\rm g~cm^{-2}}}\right)^{-1/3}\nn
&& \quad\quad \times \left({T_c\over 200\, \rm{K}}\right)^{-1/6}\left({\widetilde\alpha_{\rm MRI}\over 0.1}\right)^{-2/3}.
\ea

Similar estimates are obtained if ambipolar diffusion balances growth by winding, 
\ba
B_{\phi,\rm eq} &\;\approx\;& \left({6\pi \delta\Sigma_g^2 \Omega x_e \langle \sigma v\rangle_i B_R \over m_u}\right)^{1/3}
\quad\quad ({\rm ambipolar})\nn
&\;=\;& -0.8\,\epsilon_B^{1/3}{ (\delta\Sigma_g/10~{\rm g~cm^{-2}})^{2/3}(x_e/10^{-10})^{1/3}\over (R/{\rm AU})^{7/6}}
\;{\rm G},\nn
\ea
and 
\ba \label{eq:betaeq2}
\frac{B_{\phi,\rm eq}^{2}}{8\pi P}  &\;\approx\;&  6.4\,\epsilon_B^{2/3}
\left({R\over {\rm AU}}\right)^{-5/6}\left({\delta\Sigma_g\over 10^{-2}~{\rm g~cm^{-2}}}\right)^{1/3}\nn
&&\quad\quad\times \left({x_e\over 10^{-6}}\right)^{2/3}\left({T_c\over 200\, \rm{K}}\right)^{-1/2}.\nn
\ea

Comparing Equations (\ref{eq:betaeq1}) and (\ref{eq:betaeq2}) shows that 
transport of the background toroidal field in the active layer is dominated by magnetorotational turbulence:
$\nu_{\rm MRI} > \eta_a$.

\subsection{Broadening of the Active Layer by Downward Mixing of Free Charges}

Here we examine if downward advection of free charges that are liberated by X-ray absorption 
at $\delta\Sigma_g \lesssim 10$ g cm$^{-2}$ can broaden the active column.  Some numerical evidence for this
effect has been found by \cite{turner07}, but starting with a different magnetic field geometry. 

This effect grows in importance with increasing depth in the disk.  For example, stellar FUV photons raise 
the ionized fraction to $x_e\sim10^{-5}$--$10^{-4}$ in the upper $\sim 10^{-2}$ g~cm$^{-2}$ of the disk 
\citep{perez11};  but recombination in situ is fast compared with downward advection at such high $x_e$.

We require that the time to reach ionization equilibrium exceed the mixing time, $x_e/\Gamma_i > 
(\alpha_{\rm MRI}\Omega)^{-1}$.   Setting $x_e$ to the equilibrium value, one obtains
\be\label{eq:xmax}
x_e < {\alpha_{\rm MRI}\Omega\over\alpha_{\rm eff} n_H}.
\ee
This condition turns out not to be satisfied for our baseline metal abundance.  However, it
can be satisfied if we adjust $x_M$ to a high value, so that $\alpha_{\rm eff} 
\rightarrow \alpha_{\rm rad} = 3\times 10^{-12} (T/100 \,\rm{K})^{-1/2}$ cm$^3$ s$^{-1}$ \citep{fromang02}.
Then Equation (\ref{eq:xmax}) becomes
\be
x_e < 1.5\times 10^{-9}\alpha_{\rm MRI,-1}
{(T/200 \,\rm{K})^{1/2}\over \delta\Sigma_g/10~{\rm g~cm^{-2}}}.
\ee
Balancing this with the minimal ionization level that will sustain MRI turbulence 
(according to the condition (\ref{eq:xmin2}) that is derived in Section \ref{s:activecri}), we obtain
\be
\delta\Sigma_g < 50\,\alpha_{\rm MRI,-1}^{1/2}\left({{\rm Am}_{\rm crit}\over 10}\right)^{-1/2}
       \left({T\over 200\, \rm{K}}\right)^{1/4}\;{\rm g~cm^{-2}}.
\ee
This is modestly higher (by a factor $\sim 1.5-3$) than the active column that emerges from
our baseline ionization model, under the assumption of local ionization equilibrium.


\section{Calculational Method}\label{s:steady}

A steady-state approximation to the vertical profile of the disk is found by solving for the background
magnetic field $(B_R,B_\phi)$ and gas density in combination with the temperature profile (\ref{eq:Tprofile}).
We idealize the PPD as being geometrically thin, and neglect the $R$-derivatives of all quantities in 
comparison with $z$-derivatives.

Further ignoring the effects of Hall drift, for the reason described in Section \ref{s:vhall}, the induction
equation becomes
\be \label{eq:inducrad}
\partial_t B_R
=\partial_{z}\left(\eta\partial_{z}B_R\right)+\partial_{z}\left(\eta_a\frac{B_R}{B}\partial_{z}B\right)
\ee
and
\be \label{eq:inductor}
\partial_t B_\phi =\partial_{z}\left(\eta\partial_{z}B_{\phi}\right)+\partial_{z}
\left(\eta_a\frac{B_{\phi}}{B}\partial_{z}B\right)-\frac{3}{2}B_R\Omega.
\ee
We set $\partial_t B_R = \partial_t B_\phi = 0$ and then combine Equations
(\ref{eq:inducrad}) and (\ref{eq:inductor}) with the equation of magnetostatic equilibrium (\ref{eq:weqn}).  

This set of equations is highly nonlinear:  the temperature profile depends on the density
profile through the cooling condition (\ref{eq:rhocool}), and the rates of ambipolar drift and 
turbulent diffusion -- as determined by Equations (\ref{eq:ccoeff})-(\ref{eq:eta}), (\ref{eq:num}) 
and (\ref{eq:alphaMRI}) -- depend on $(B_R,B_\phi)$.  The solution to the disk
structure involves a determination of $\nu_{\rm MRI}$ as well as the laminar Maxwell stress
$B_RB_\phi/4\pi$.  

Our first step is to integrate Equations (\ref{eq:inducrad}) and (\ref{eq:inductor}) across the disk.  
The radial magnetic flux
\be \label{eq:phiR}
\Phi_R(z)\equiv\int_{z_w^-}^{z}B_R(z^{\prime})dz^\prime 
\ee
becomes an auxiliary variable.  Here $z_w^-$ is wind-disk interface in the $z<0$ hemisphere.  Then
\be 
\eta\partial_{z}B_R+\eta_a\frac{B_R}{B}\partial_{z}B=C_-
\ee
\be
\eta\partial_{z}B_{\phi}+\eta_a\frac{B_{\phi}}{B}\partial_{z}B-\frac{3}{2}\Omega\Phi_R= D_-
\ee
where $C_-$ and $D_-$ are the left-hand side evaluated at $z_w^-$.  We re-express these equations in
terms of the vertical gradients of the field components,
\be  \label{eq:dBr}
\partial_{z}B_R=\frac{\left(\eta+\frac{B_{\phi}^{2}}{B^{2}}\eta_a\right)C_--\frac{B_RB_{\phi}}{B^{2}}\eta_a
\left(D_-+\frac{3}{2}\Omega\Phi_R\right)}{\eta\left(\eta+\eta_a\right)}
\ee
\be \label{eq:dBphi}
\partial_{z}B_{\phi}=\frac{\left(\eta+\frac{B_R^{2}}{B^{2}}\eta_a\right)
\left(D_-+\frac{3}{2}\Omega\Phi_R\right)-\frac{B_RB_{\phi}}{B^{2}}\eta_aC_-}{\eta\left(\eta+\eta_a\right)},
\ee
which can be integrated along with $\partial_{z}\Phi_R=B_R$.

In each iteration we re-solve for the the vertical ionization profile $x_e(z)$, which feeds into $\eta_{\rm O}$ and $\eta_a$
and which, for a fixed external source of X-rays, depends on the scale height of the disk.  The ionization is most
directly a function of the column density $\delta\Sigma_g$ as measured from the upper or lower surface.

We also impose marginal stability for the undular Newcomb-Parker mode, meaning that $\partial_z\rho_g$ is limited
to the adiabatic density gradient (see Equation (\ref{eq:Parker})). 

\subsection{Upper and Lower Boundary Conditions}\label{s:boundaryvalue}

The disk profile is obtained over the range $z_w^- < z < z_w^+$, where $z_w^+$ is the height of the upper
wind-disk boundary.  The locations of the upper and lower boundaries can only be obtained in 
an iterative manner, because they depend, for example, on the contribution of magnetic pressure gradients to the vertical
disk support.  

Here we specify the boundary conditions that are imposed at $z_w^\pm$.
The key parameter controlling the profile -- in particular, how it departs from a simple,
unmagnetized disk -- is the radial magnetic field imposed at the wind-disk boundary.  We consider
a range of field strengths\footnote{We define $B_R$ as a function of $R$ so as to 
side-step details of wind collimation near the disk surface.}
\be\label{eq:boundary1}
B_R^w=0.01\,\epsilon_B\left({R\over \rm AU}\right)^{-2}\;{\rm G};\quad z = z_w^+ > 0.
\ee
by varying 
\be
\epsilon_B = 0.01-1.
\ee

The radial field has the opposing sign in the lower hemisphere,
but requiring it to have equal magnitude would impose an unrealistic and unstable symmetry. 
The Ohmic diffusion time through the weakly ionized midplane region is even shorter than the orbital period,
meaning that a slightly stronger field in one hemisphere will quickly diffuse into
the opposing hemisphere and modify the flux distribution there.  The zero of $B_R$ is then displaced from
the midplane into the hemisphere with the weaker field.  

To illustrate this asymmetry, we take $B_R^w$
in the subdominant (lower) hemisphere to be $50\%$ in magnitude of the strength in the dominant (upper) hemisphere,
\be\label{eq:boundary2}
B_R^w(z_w^-) = -{1\over 2}B_R^w(z_w^+).
\ee

The boundary value of the toroidal magnetic field is determined by balancing the linear winding and vertical
diffusion terms in the induction equation (\ref{eq:inductor}), $\partial_z\left(\nu \partial_z B_\phi\right)\approx 3B_R\Omega/2$.
In the boundary layer $\nu~h_g^2\Omega$ and after approximating vertical gradients as $\partial_z\sim h_g^{-1}$ 
we have
\be\label{eq:boundary3}
B_\phi(z_w) = B_{\phi,\rm eq} \approx -\frac{3}{2}B_R^w.
\ee 
\begin{figure*}[ht]
\epsscale{1.}

\plottwo{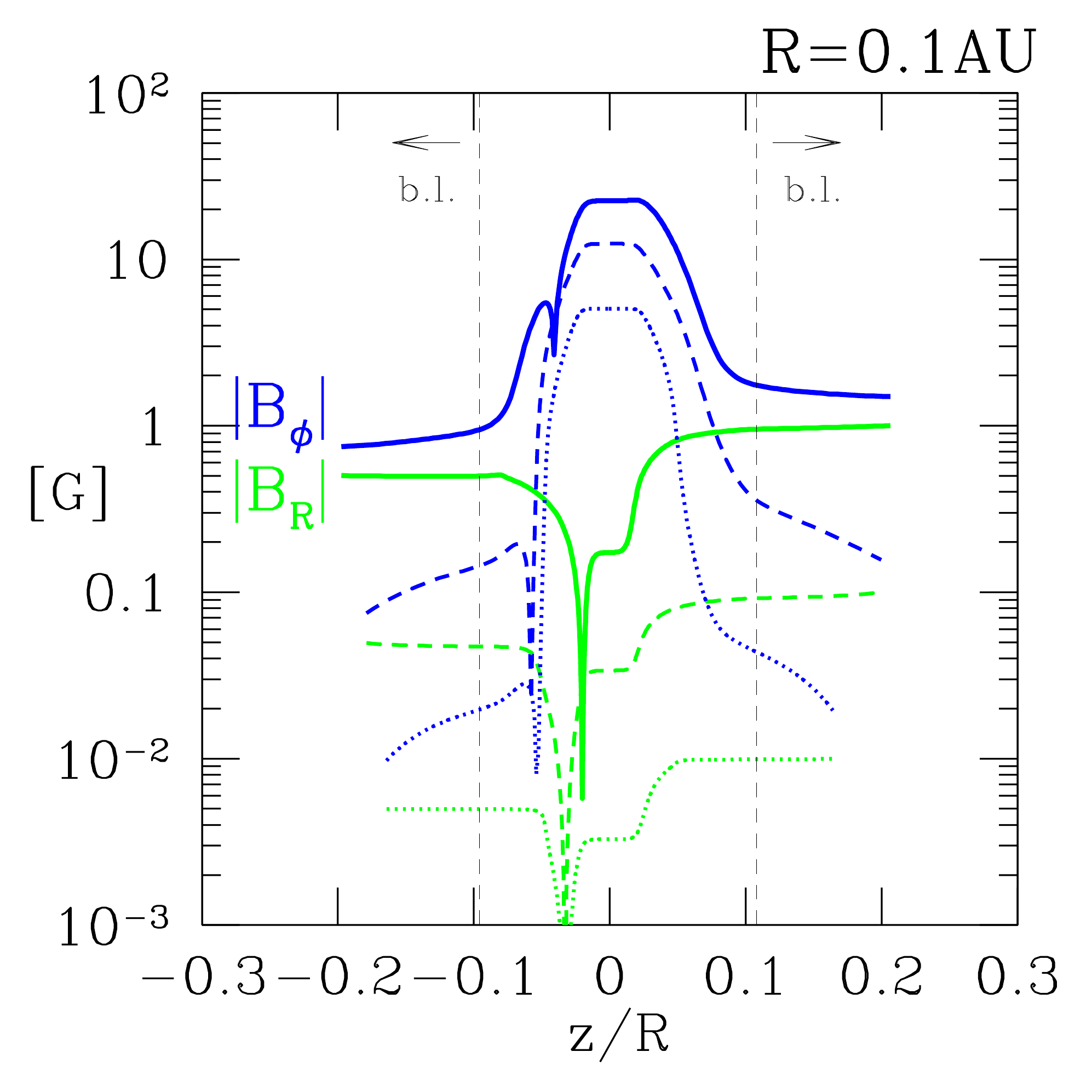}{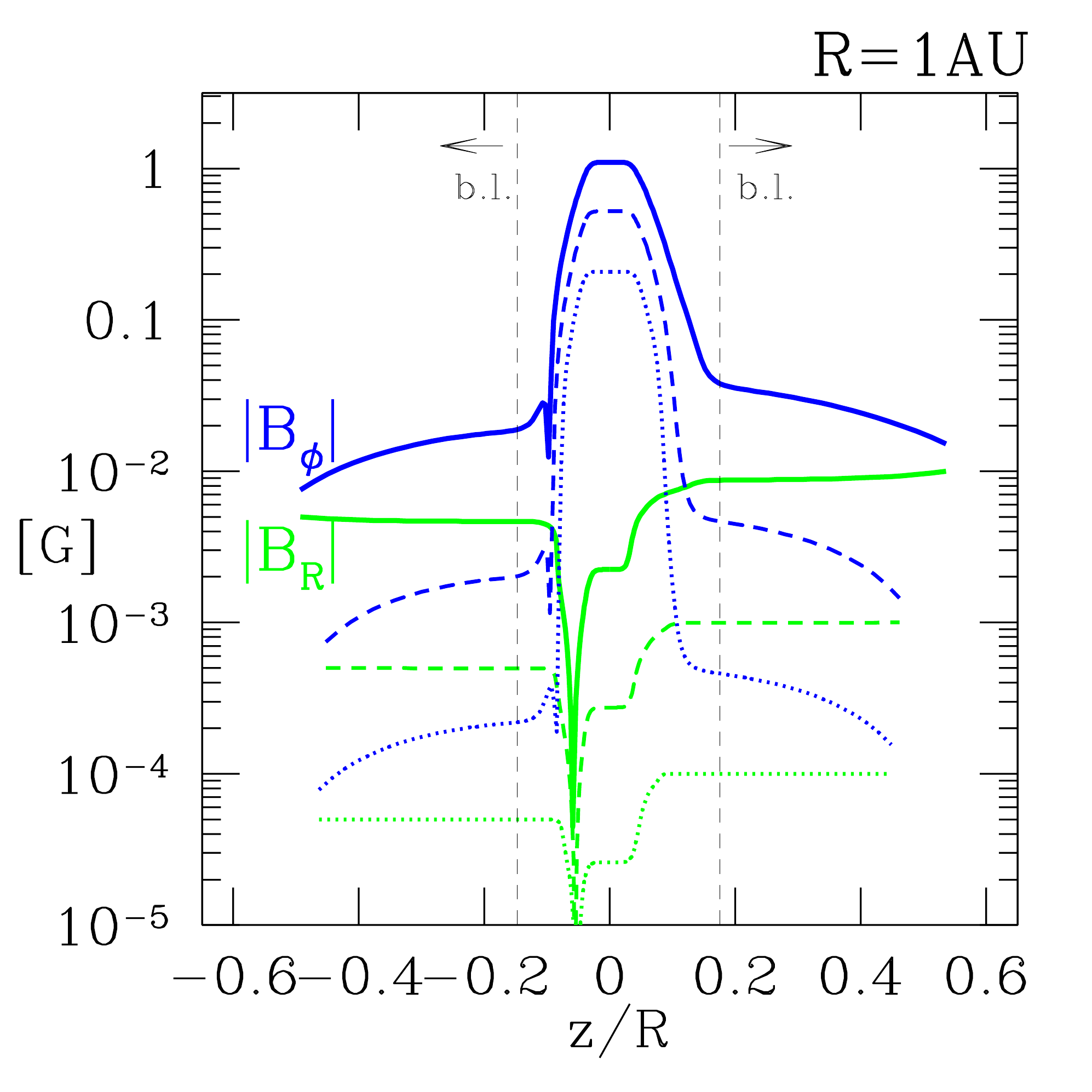}
\caption{Vertical profiles of background toroidal and radial magnetic field at $R=0.1, 1$ AU.  
Solid, dashed, and dotted lines: boundary field strength $\epsilon_B=1,0.1,0.01$ in Equations 
(\ref{eq:boundary1})-(\ref{eq:boundary3}), respectively. Dashed vertical lines mark base $z_{\rm bl}$ of wind-disk boundary layer.}
\vskip .1in
\label{fig:fields}
\end{figure*}

We emphasize that the peak toroidal field deeper in the disk is insensitive to this boundary condition:  the
radial field maintains a fairly flat profile within the disk, so that the toroidal field at each depth is
obtained from a quasi-local balance between linear winding and diffusion.

The radial flux (\ref{eq:phiR}) satisfies $\Phi_R(z_w^-) = 0$, but the value at the top of the disk
depends on the full vertical disk profile.  Finally, we note that the boundary values of the gas pressure are obtained from Equations
(\ref{eq:normpressbal}) and (\ref{eq:thetaw}),
\be\label{eq:pb}
P(z_w^\pm) = \left(0.3{z_w^\pm\over R}\right)^2 \rho_w V_w^2,
\ee
and the temperature from Equation (\ref{eq:tbl}).

\subsection{Hybrid Numerical Method}\label{eq:numerical}

We obtain vertical profiles of the gas variables, ionization rate, and horizontal magnetic field
in three separate stages, which are iterated until convergence.  

The first step involves solving the equations of hydrostatic equilibrium (Equation \ref{eq:weqn}) and of continuity.
To obtain the boundaries $z_w^-$/$z_w^+$ self-consistently, we employ the relaxation method described in
\cite{london82}, which treats $z$ as a dependent variable at the cost of introducing two new ODEs. 
The expanded system is written as a finite difference equation and solved on a fixed grid by iteratively improving 
on an initial trial solution.

When solving for hydrostatic equilibrium, the dependence of $(B_R, B_\phi)$ on $\delta\Sigma_g$ is held fixed.
We first guess the midplane mass density $\rho_g(0)$, and then integrate both upward and downward until the 
pressure boundary condition (\ref{eq:normpressbal}) is reached.  
We then improve the consistency of the temperature profile by re-calculating the base of the boundary layer $z_{\rm bl}^\pm$ 
in each hemisphere by matching the density to the transition value (\ref{eq:rhocool}) between atomic and molecular gas.  
Given the density structure, we then obtain the ionization rate at each column density, as described in Appendix 
\ref{s:ionrate2}, using a simple Riemann sum.  These last two steps involve strong nonlinear changes in the disk
structure if first performed for a finite value of the imposed radial magnetic field, but convergence is obtained
by starting with a purely hydrostatic disk, and then slowly increasing $B_R(z_w^\pm)$ from a small value.  

Having obtained a self-consistent hydrostatic profile, the total column through the disk generally 
differs from the fixed column $\Sigma_{g,\rm tot}$ (Equation (\ref{eq:sigtot})), and so we repeat this series of steps with
different values of $\rho_g(0)$ until $\Sigma_g = \Sigma_{g,\rm tot}$.

The next step is to recalculate the profiles of $B_R$ and $B_\phi$ using a shooting method based on a fifth order 
Runge-Kutta algorithm.  Starting at the top of the disk, where the radial and toroidal fields are given by Equations
(\ref{eq:boundary1})-(\ref{eq:boundary3}), we integrate $\partial_z B_r$ downward across the entire disk.
This step is repeated while adjusting $\partial_z B_r|_{-}$ (the constant $C_{-}$ in Equation (\ref{eq:dBr})) until 
the desired radial field $B_R(z_w^-)$ is achieved at the lower boundary.   The equations for $\partial_z B_\phi$ and
$\partial_z \Phi_R$ are then integrated upward from the lower boundary, where $\Phi_R=0$.  

We start with the a small value of the imposed magnetic field, typically $\epsilon_B = 10^{-2}$
in Equation (\ref{eq:boundary1}).
The entire process is repeated several times at each radius until well converged solutions to both the hydrostatic 
and induction equations are obtained.   Finally $\epsilon_B$ is gradually raised to unity, resulting in a sequence
of disk profiles for a range of imposed radial magnetic fields.


\section{Results}\label{s:results}
\subsection{Magnetic Field and Density Structure}

We obtained vertical profiles at six radii, uniformly spaced
in logarithmic radius between $R=10^{-1.25}$ AU and 1 AU, using the numerical method described in
Section \ref{s:steady}.  Results are shown in Figures \ref{fig:fields}-\ref{fig:magz}.
The magnetic field profiles are qualitatively similar at all radii, but the magnetization is much
higher in the inner disk.  

\begin{figure*}[!]
\epsscale{1.05}
\plottwo{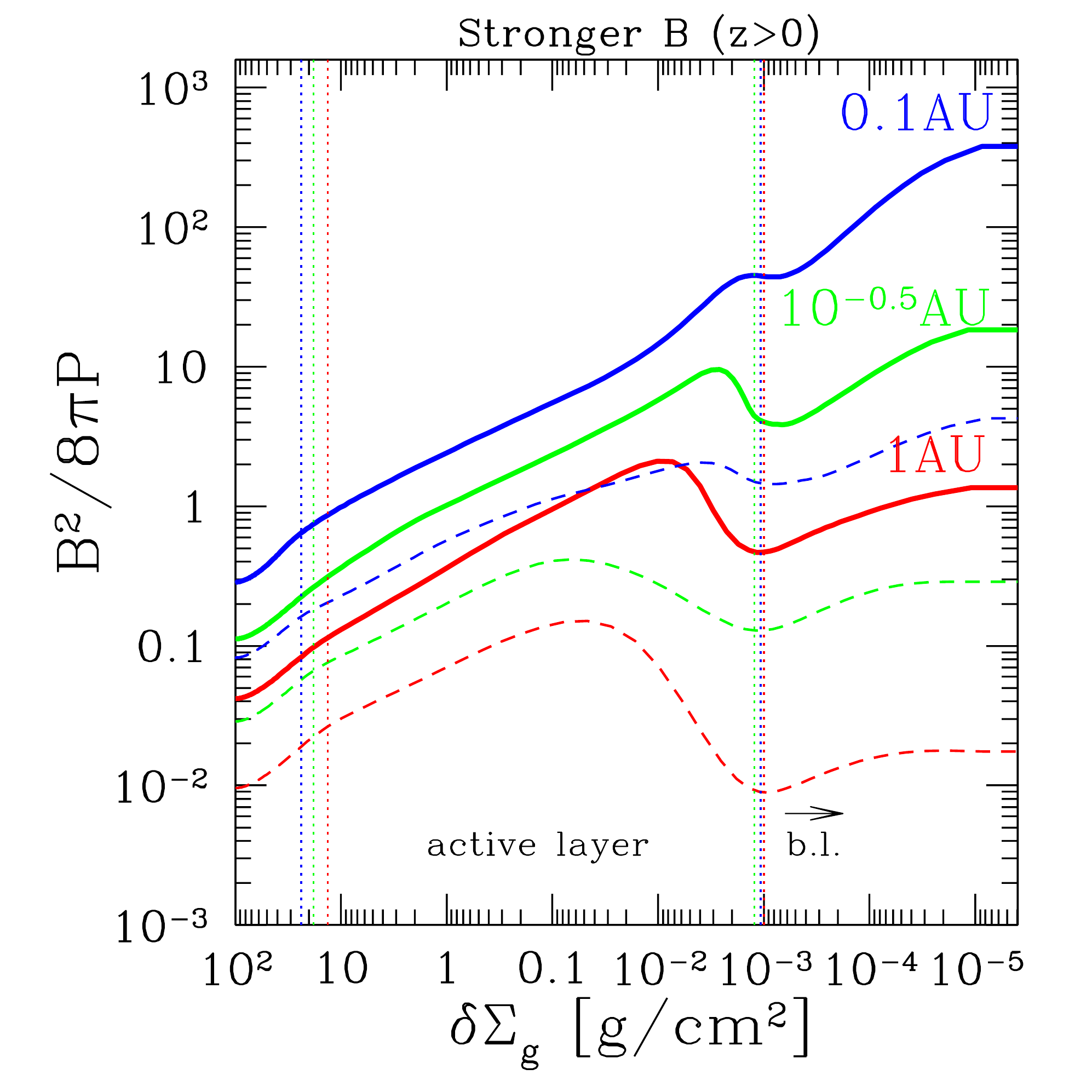}{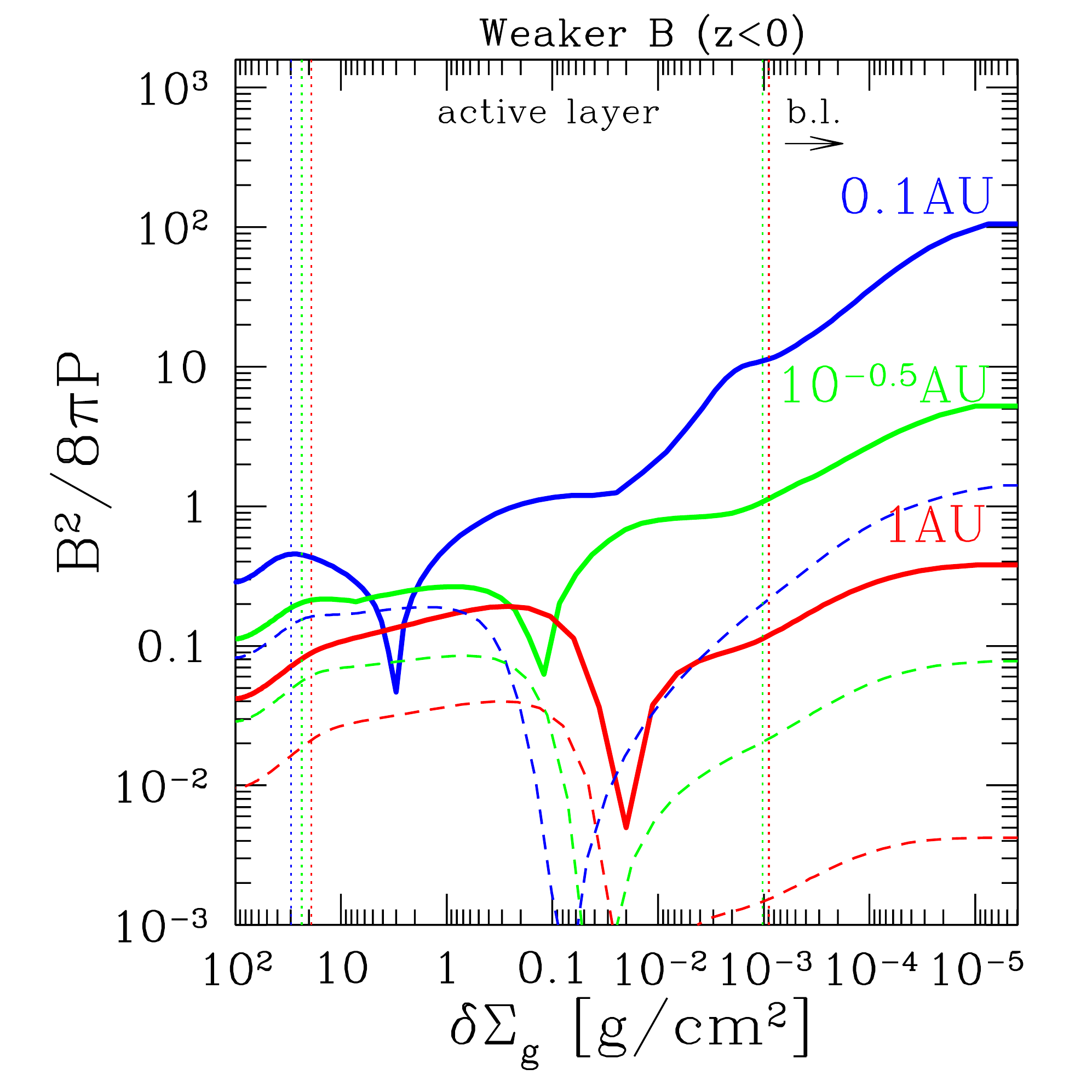}
\caption{\textit{Left panel:}  vertical magnetization profile in the more strongly magnetized disk hemisphere 
(with $\epsilon_B=1$) at radii $R=0.1,\,10^{-0.5}$ and $1$ AU. Column density increases toward the midplane. 
Dashed lines:  imposed boundary magnetic field reduced by factor 0.1.   Left vertical dotted lines:  base of the active
layer ($\Lambda_{\rm O} = 10^2$).  Right vertical dotted lines:  base $z_{\rm bl}$ of wind-disk boundary layer.
\textit{Right panel:}  same as top panel, but in the more weakly magnetized hemisphere ($z<0$).}
\vskip .1in
\label{fig:magSig}
\end{figure*}
\begin{figure*}[!]

\epsscale{1.12}

\plottwo{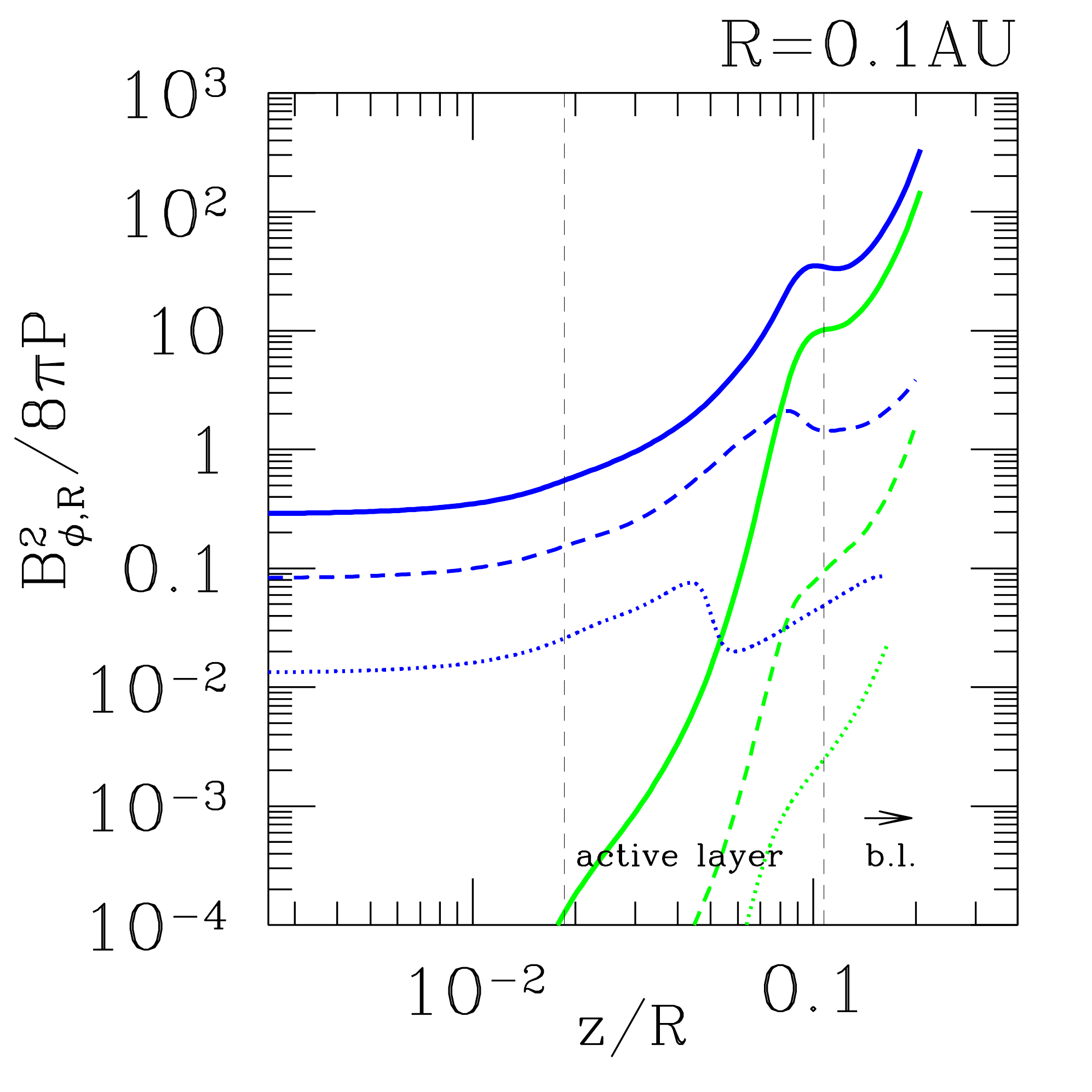}{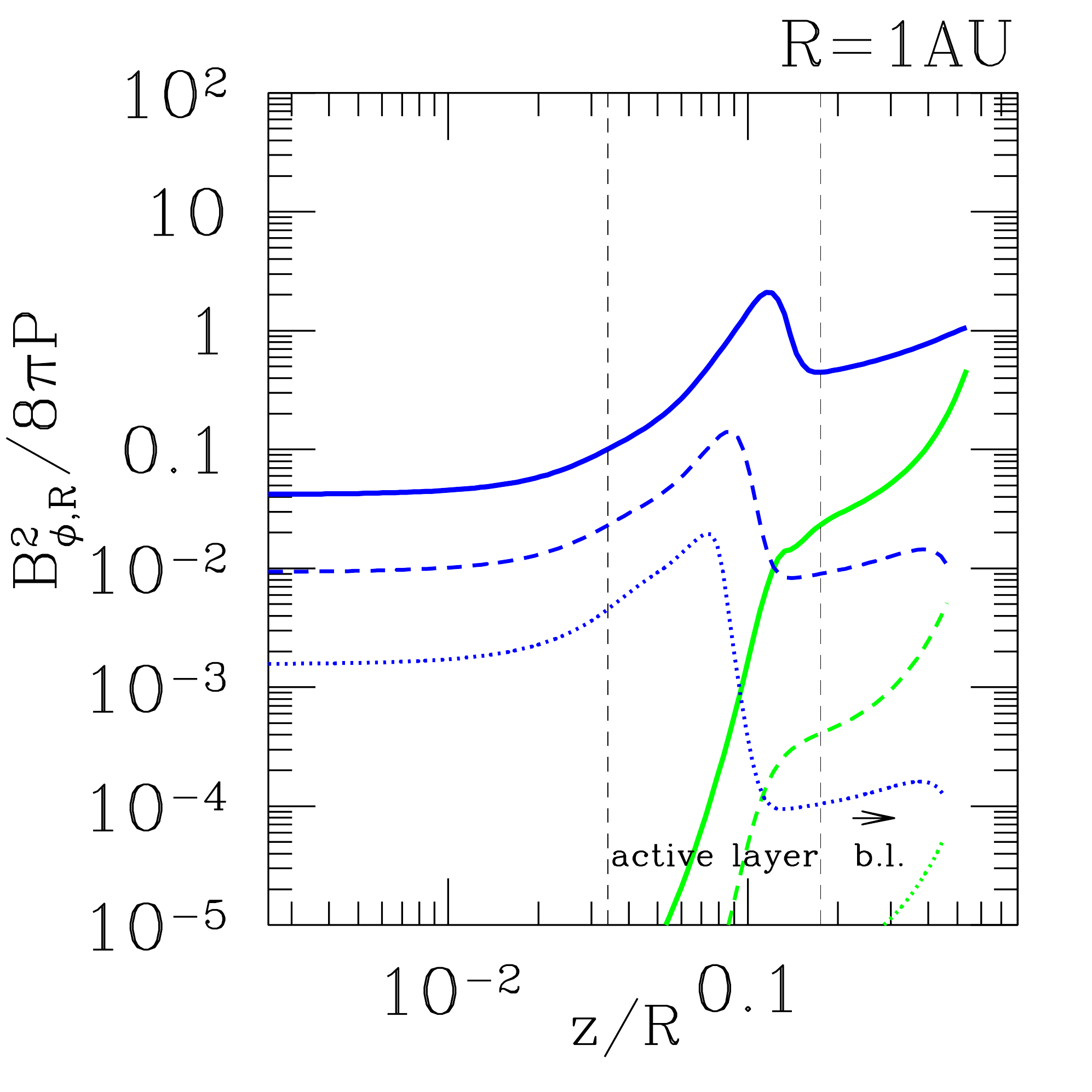}

\caption{\textit{Left panel:} toroidal (blue) and radial (green) magnetization vs. 
height at $R=0.1$ AU in the more strongly magnetized hemisphere. 
Solid, dashed and dotted lines: boundary field strength $\epsilon_B=1,0.1,0.01$ in
Equations (\ref{eq:boundary1})-(\ref{eq:boundary3}).  
\textit{Right panel:} same as top panel but at $R=1$ AU.}
\vskip .1in
\label{fig:magz}
\end{figure*}

The fields remain fairly constant throughout the wind-disk boundary layer, 
where the diffusion time is comparable to the orbital period.
The shear-driven diffusivity $\nu_{\rm mix}$ drops dramatically at the base of the boundary layer, 
resulting in strong downward pumping of the toroidal magnetic field.  The equilibrium $B_\phi$
also increases to higher densities, where MRI turbulence dominates, which is in agreement
with the analytic arguments of Section \ref{s:equil}.  

The disk is fully turbulent from the wind-disk interface to the base of the active layer,
which we define as the surface $\Lambda_{\rm O} = 10^2$.  The range of the active layer is
marked by vertical dotted lines in Figures \ref{fig:magSig} and \ref{fig:magz}.

Because the imposed magnetic field has a bipolar symmetry, the background field $(B_R,B_\phi)$ must switch sign
at some depth in the disk.  The examples shown here start with a relatively strong asymmetry in the
applied field, with $B_R^w(z_w^-)=-0.5B^w_R(z_w^+)$.  We see that the stronger magnetic field in the upper
hemisphere diffuses through the weakly conducting midplane layer into the lower hemisphere.  $B_R$ and
$B_\phi$ switch sign near the base of the active layer in the lower hemisphere.

As the imposed field is made more symmetric, the zero of $(B_R,B_\phi)$ moves toward $z=0$, but only in the
case of precise symmetry does it remain in the midplane layer where the Ohmic timescale is shorter than 
the orbital period.  The height-integrated torque is less dependent on this detail, and more rapidly 
approaches symmetry between the two hemisphere as $|B_R^w(z_w^-)|$  approaches $|B_R^w(z_w^+)|$.

\begin{figure*}[!]

\epsscale{1.07}
\plottwo{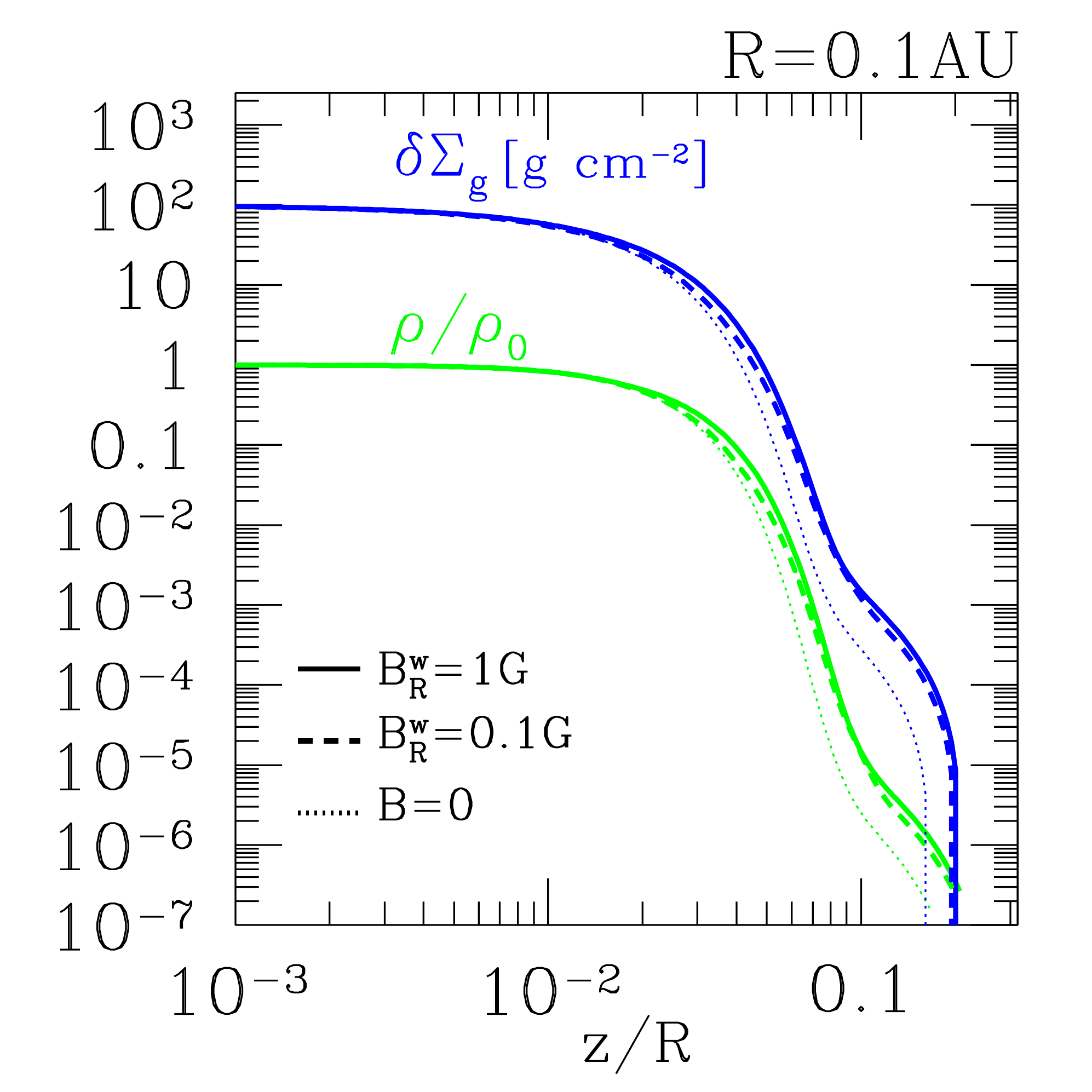}{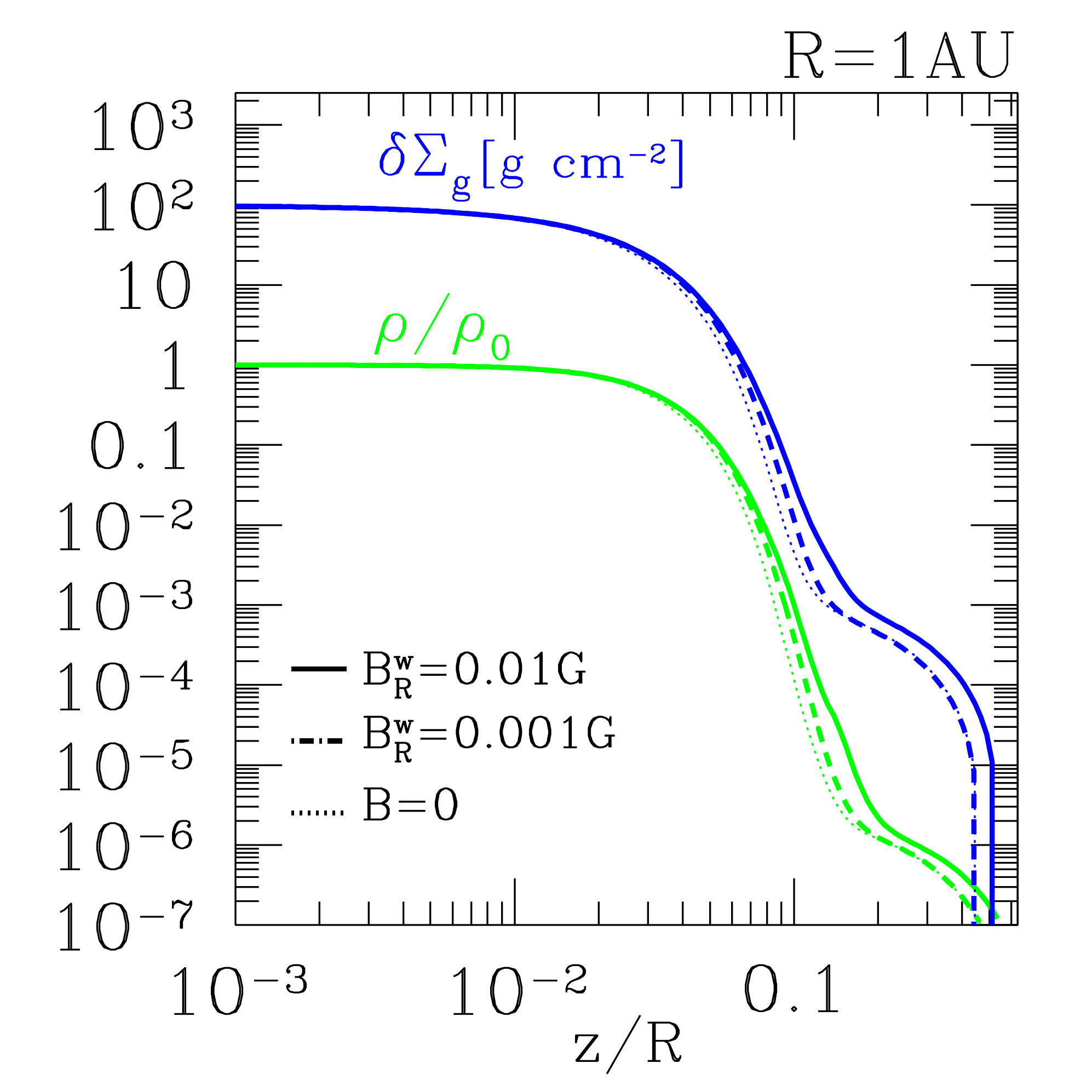}
\caption{Vertical profiles of density and column density (below disk surface), showing the effect of magnetic inflation
at $R=0.1$ AU (left panel) and 1 AU (right panel). Density is normalized to the midplane value 
$\rho_0\approx 3\times 10^{-9}$ g cm$^{-3}$ at $0.1$ AU and $\rho_0\simeq 2.3\times 10^{-10}$ g cm$^{-3}$ at $1$ AU.
Solid and dashed lines correspond to the imposed magnetic field $\epsilon_B=1$ and 0.1.  Dotted line: unmagnetized 
hydrostatic structure.}
\vskip .1in
\label{fig:density}
\end{figure*}

\begin{figure*}[!]

\epsscale{0.98}
\plottwo{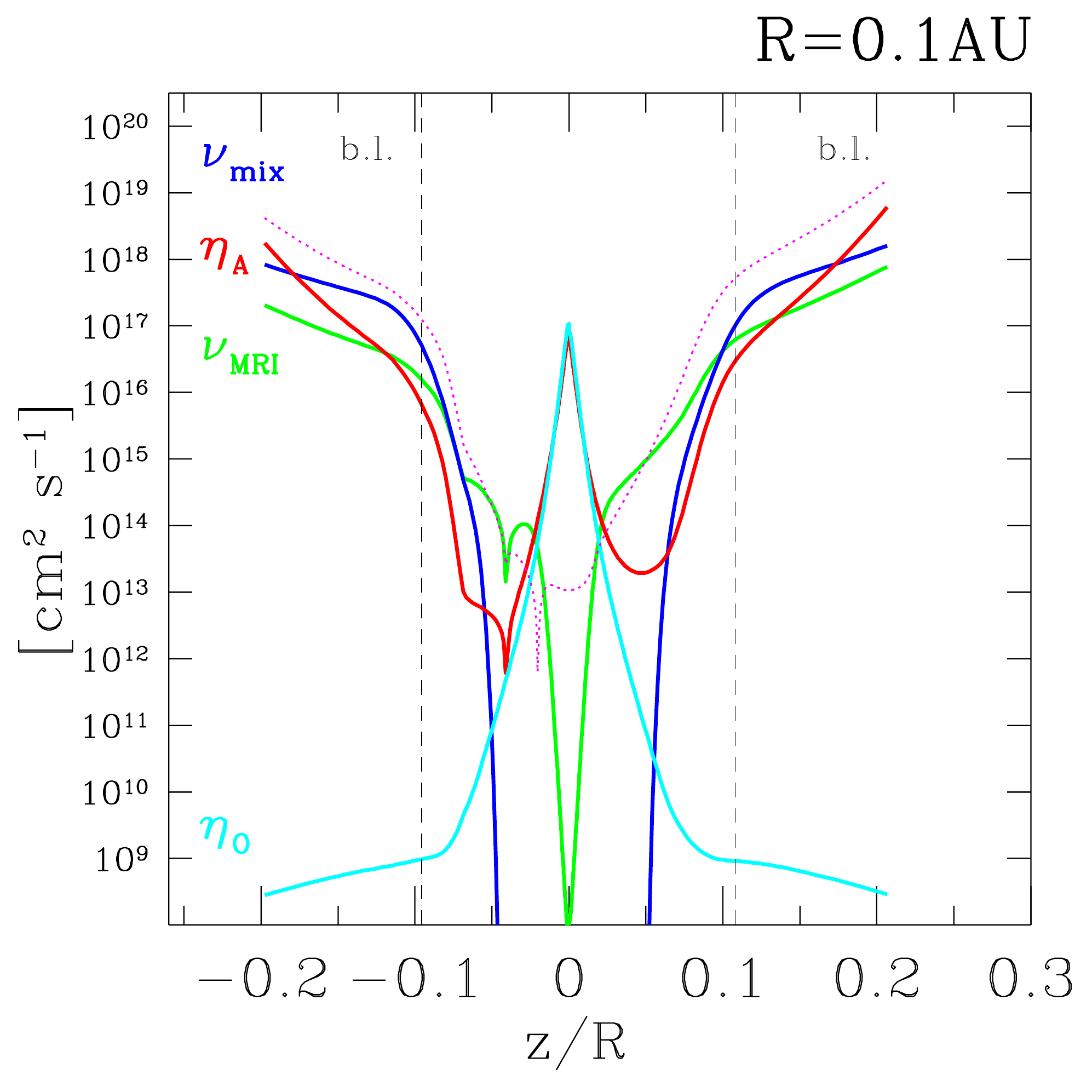}{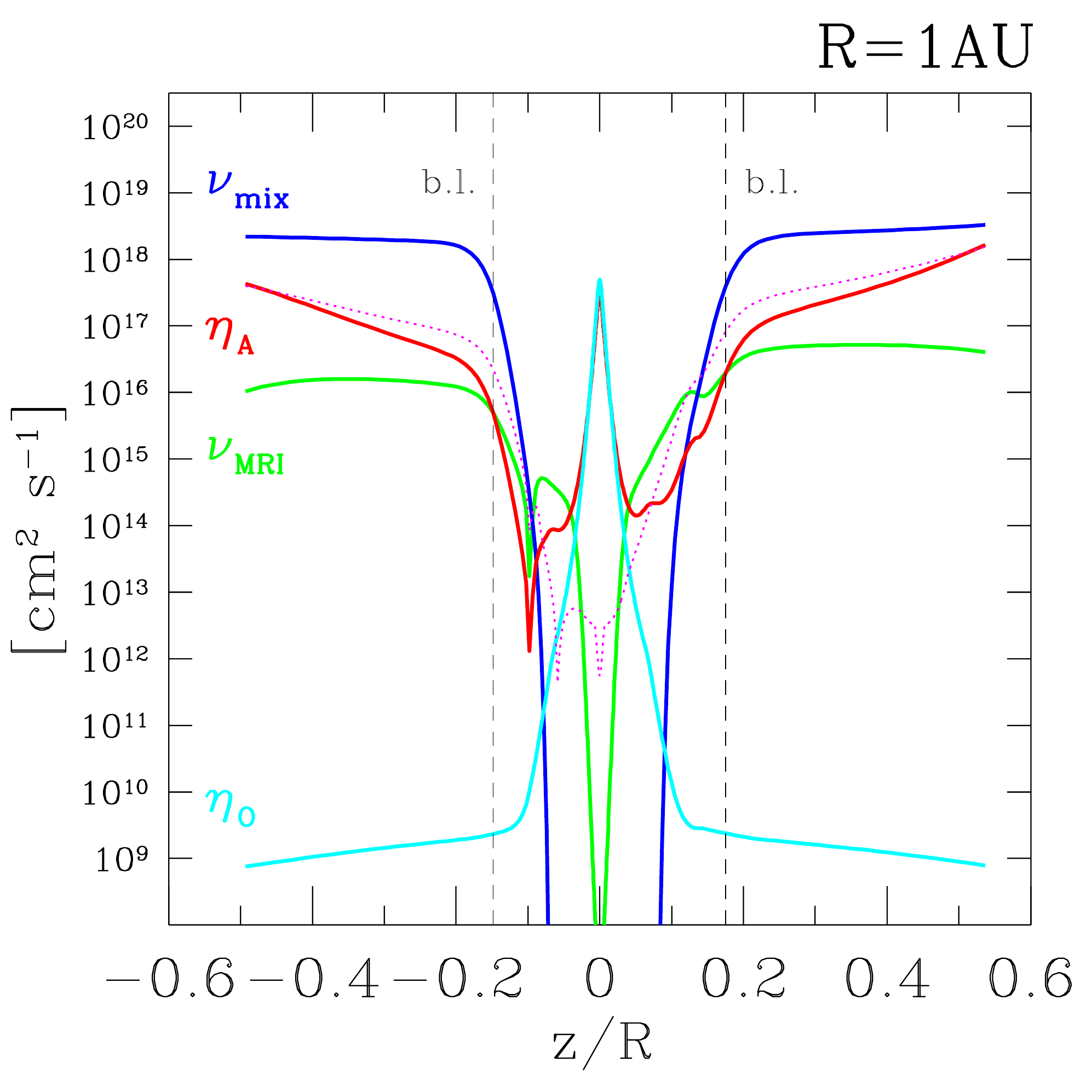}
\caption{Vertical profiles of various transport coefficients in the disk model obtained in Section \ref{s:results},
shown at $R = 0.1,1$ AU.  Ohmic diffusivity $\eta_{\rm 0}$ (cyan),  MRI turbulent diffusivity $\nu_{\rm MRI}$
(green), ambipolar diffusivity $\eta_a$ (red), and turbulent diffusivity $\nu_{\rm mix}$ in the wind-disk boundary layer
(blue) are shown.  Magenta dotted line shows viscosity that would drive the same steady radial mass transfer as the laminar 
Maxwell stress, corresponding to $T_{R\phi} = B_RB_\phi/4\pi = {3\over 2}\nu \rho\Omega$.  
Applied radial magnetic field has full strength ($\epsilon_B = 1$) in the upper hemisphere ($z > 0$); $-0.5$ this value
in lower hemisphere.}
\vskip .1in
\label{fig:diff}
\end{figure*}

The magnetization (shown as a function of $\delta\Sigma_g$ in Figure \ref{fig:magSig}, and of coordinate height
in Figure \ref{fig:magz}) directly influences the MRI diffusivity, as well as the strength of the density
gradient.   The wind-disk boundary layer
is very strongly magnetized, especially in the innermost disk, which is consistent with the estimate in 
Equation (\ref{eq:magnetize}).   (It is worth re-emphasizing that this strong magnetization in the upper disk results
from the interaction with the magnetized T-Tauri wind, not from the buoyant expulsion of 
MRI-amplified magnetic fields that is seen in vertically stratified shearing box calculations; e.g. \citealt{miller00,flaig10}.)

\begin{figure}[ht]
\epsscale{1.2}
\plotone{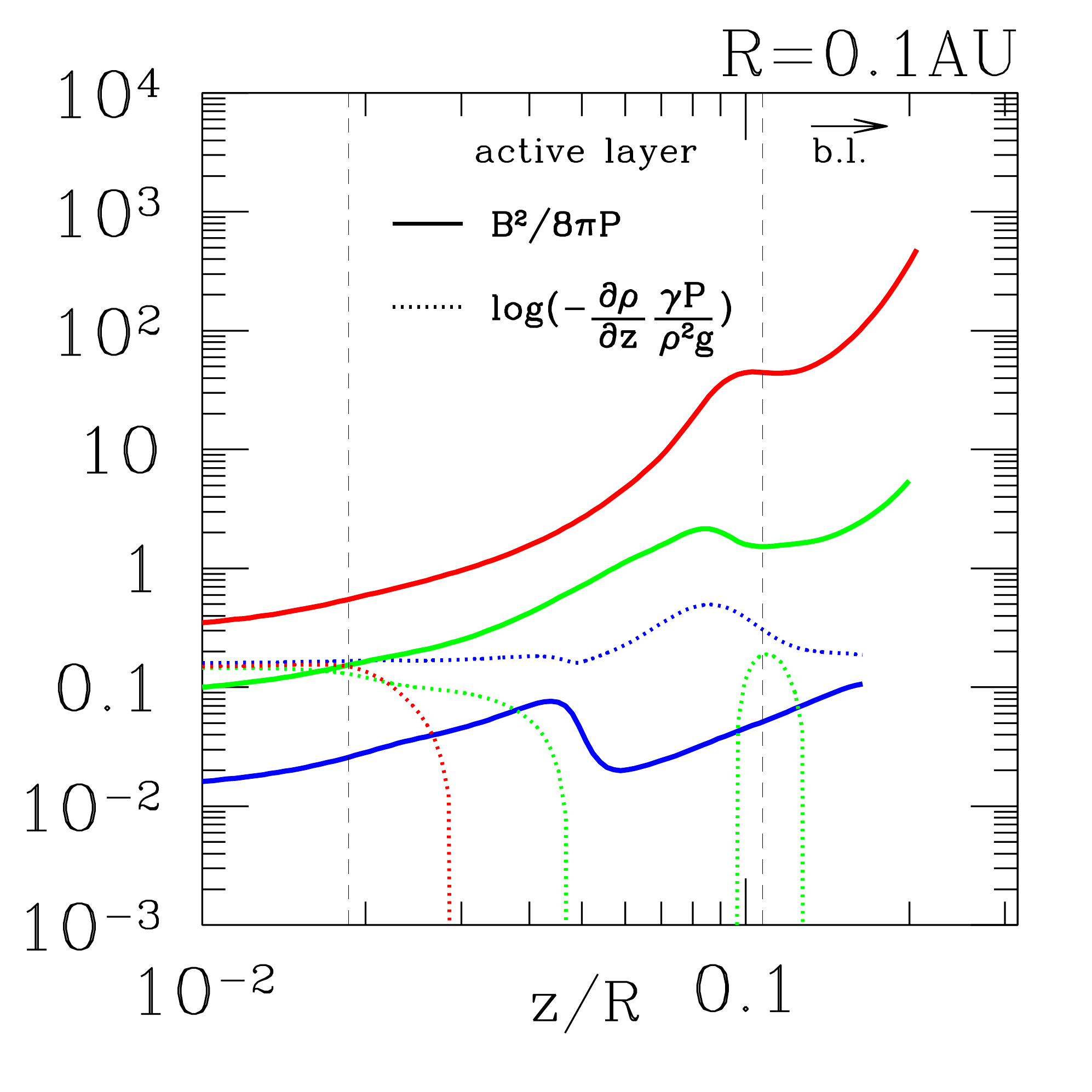}
\caption{Expansion of Parker-unstable layers in the disk with increasing imposed magnetic field, shown at $R =0.1$ AU.
Solid blue, green, and red lines correspond to $B_R^w(z_w^+) = 0.01, 0.1, 1$ G, respectively.  Dotted curves
mark out zones where $|\partial\rho/\partial z| > \rho g/\gamma c_g^2$ and the disk material
is stable to an undular Newcomb-Parker mode.  The entire disk is stable at the lowest imposed field strength,
with a localized unstable layer appearing at the intermediate field strength and then expanding to fill the entire
upper disk when $B_R^w(z_w^+) = 1$ G.}
\vskip .1in
\label{fig:parker}
\end{figure}

\begin{figure*}[ht]

\epsscale{1.05}
\plottwo{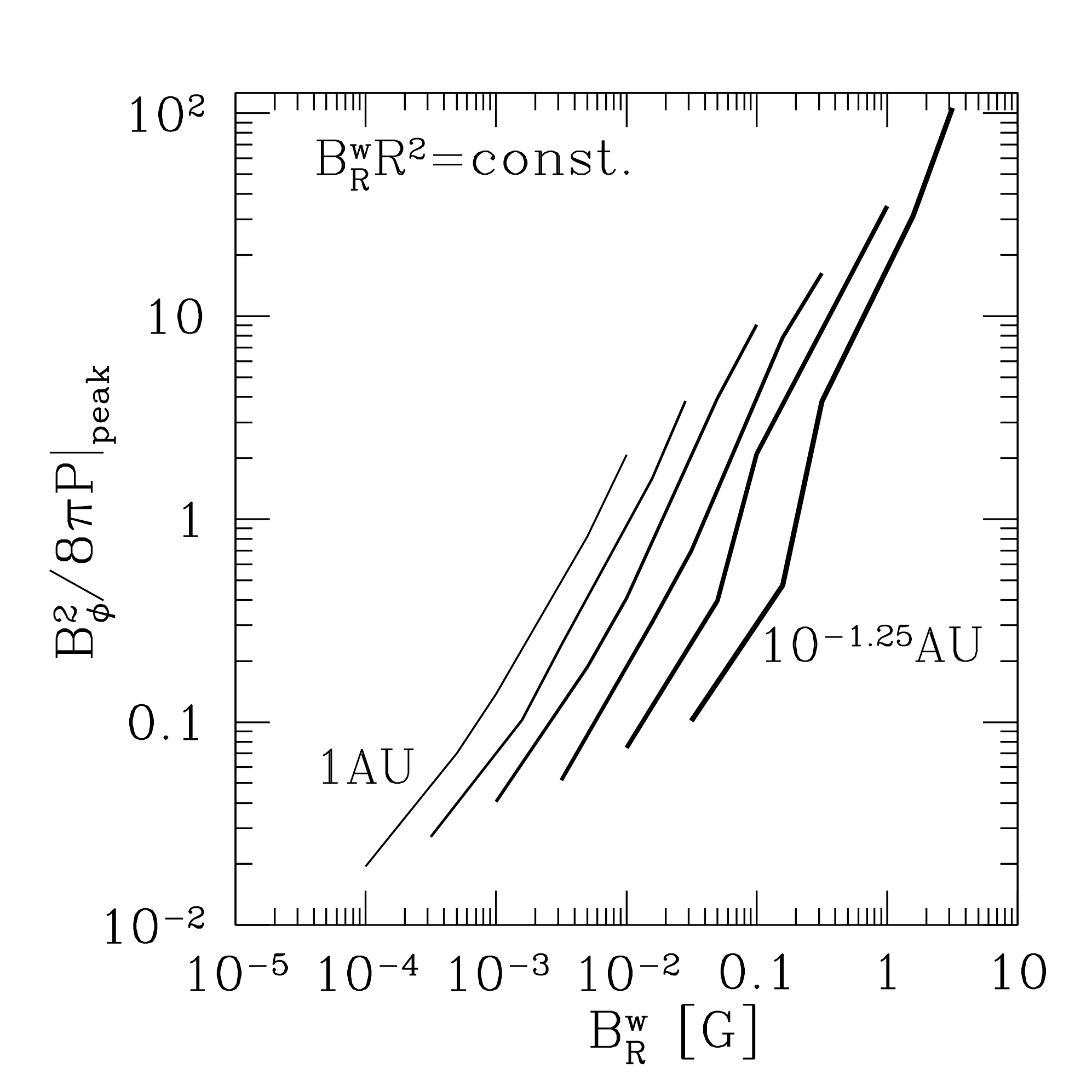}{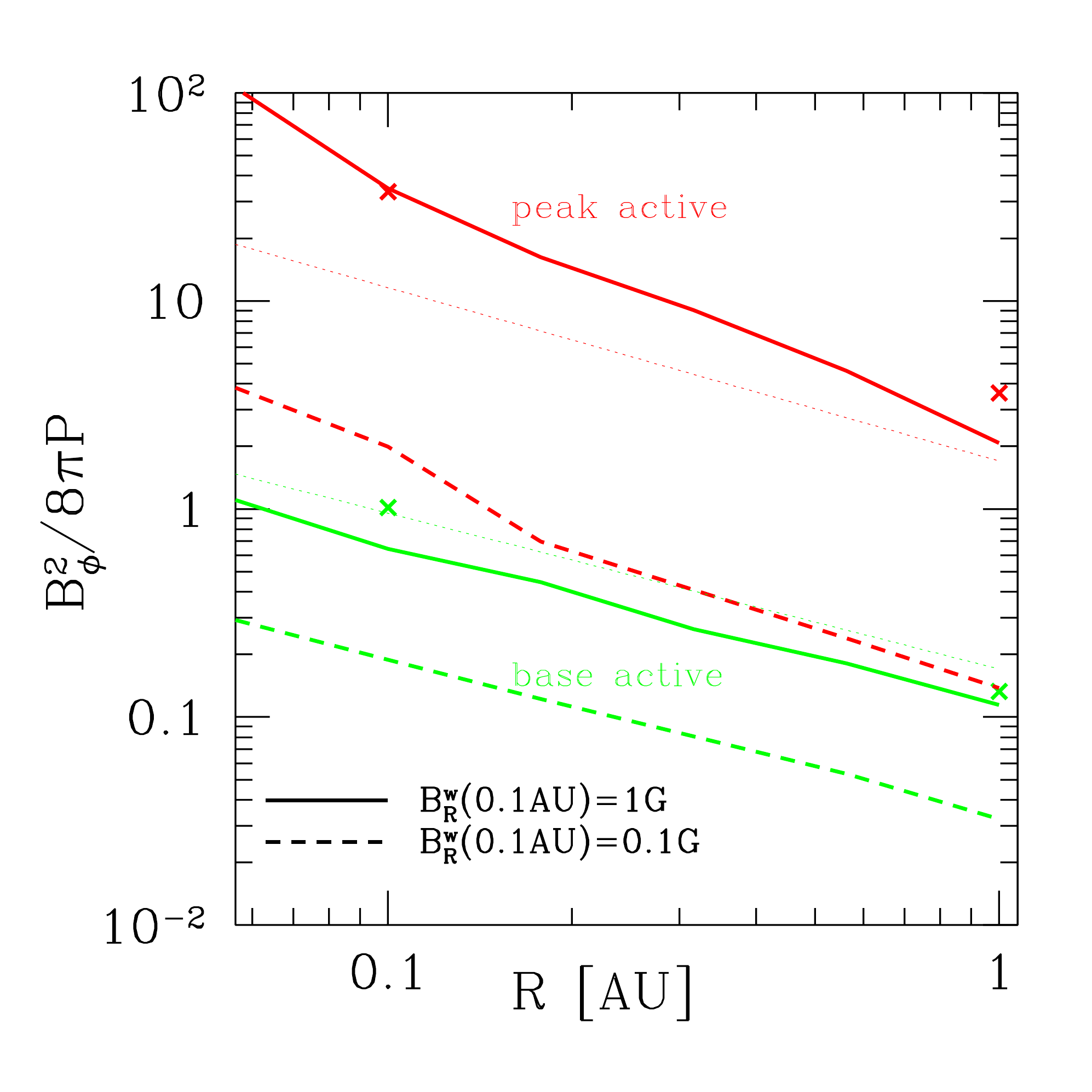}
\caption{{\it Left panel:} relation between peak toroidal magnetization in the disk active region,
and strength of the imposed wind magnetic field.   $B_R^w(z_w^+)$
is sampled at $\epsilon_B = 0.01, 0.05, 0.1, 0.5, 1$ in Equations (\ref{eq:boundary1})-(\ref{eq:boundary3}).
Radii uniformly spaced in $\log(R)$ between $R = 10^{-1.25}$ AU and 1 AU.
{\it Right panel:}  peak magnetization (red) and magnetization at base of the active layer (green)
for an imposed field $\epsilon_B=0.1$, 1 (dashed, solid). 
Dotted lines: estimate of equilibrium toroidal field in Equation (\ref{eq:betaeq1}) 
with $\delta\Sigma_{\rm act}=10^{-2}(R/{\rm AU})$  g cm$^{-2}$ (at peak magnetization)
and $10(R/{\rm AU})^{-1/4}$  g cm$^{-2}$ (at base of active layer). `X' marks the peak magnetization that results
from raising $\kB T_X$ to 5 keV from 1 keV.}
\vskip .1in
\label{fig:betapeak}
\end{figure*}

The magnetization shows a local maximum at a column $\lesssim 10^{-2}$ g cm$^{-2}$.
This results in some inflation of the disk, although it is limited to a $20-30\%$ increase 
in scale height given our imposition of marginal Parker stability.  Moving inward toward the star, the 
peak magnetization moves to lower $\delta\Sigma_g$ and then degenerates into a plateau 
(Figure \ref{fig:magSig}).  

At larger columns, the density increases dramatically as the temperature decreases, 
and the magnetization drops. Most of the cumulative torque, as measured by the 
product $\nu_{\rm MRI} \delta\Sigma_g$, is concentrated at 
$\delta\Sigma_g \sim 10$ g cm$^{-2}$ (see also Paper II).  

The density structure is compared in Figure \ref{fig:density} with the profile of an umagnetized disk.  
The wind-disk boundary layer with $5000$ K temperature appears as an extended, low-density shoulder.  
Larger scale heights are obtained in the upper disk if marginal Parker stability
is not imposed on the density profile.

The vertical dependence of the various transport coefficients is shown in Figure \ref{fig:diff}. Transport by MRI-driven turbulence dominates the active region, but is quenched near the midplane by steeply 
increasing Ohmic and ambipolar drift.   Mixing in the boundary layer is dominated by wind-driven turbulence,
except in the innermost disk where ambipolar drift is comparable.

\begin{figure*}[ht]

\epsscale{1.05}
\plottwo{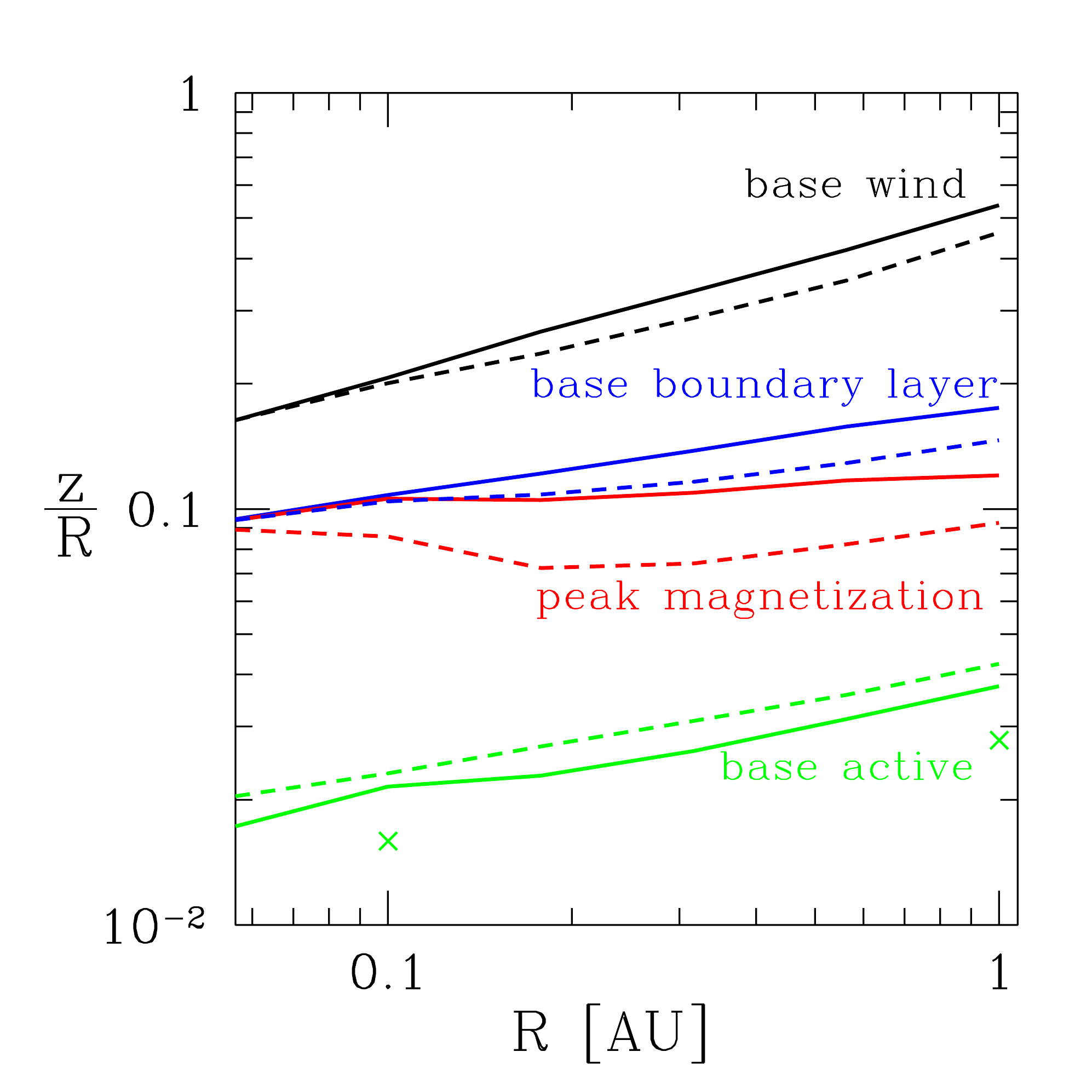}{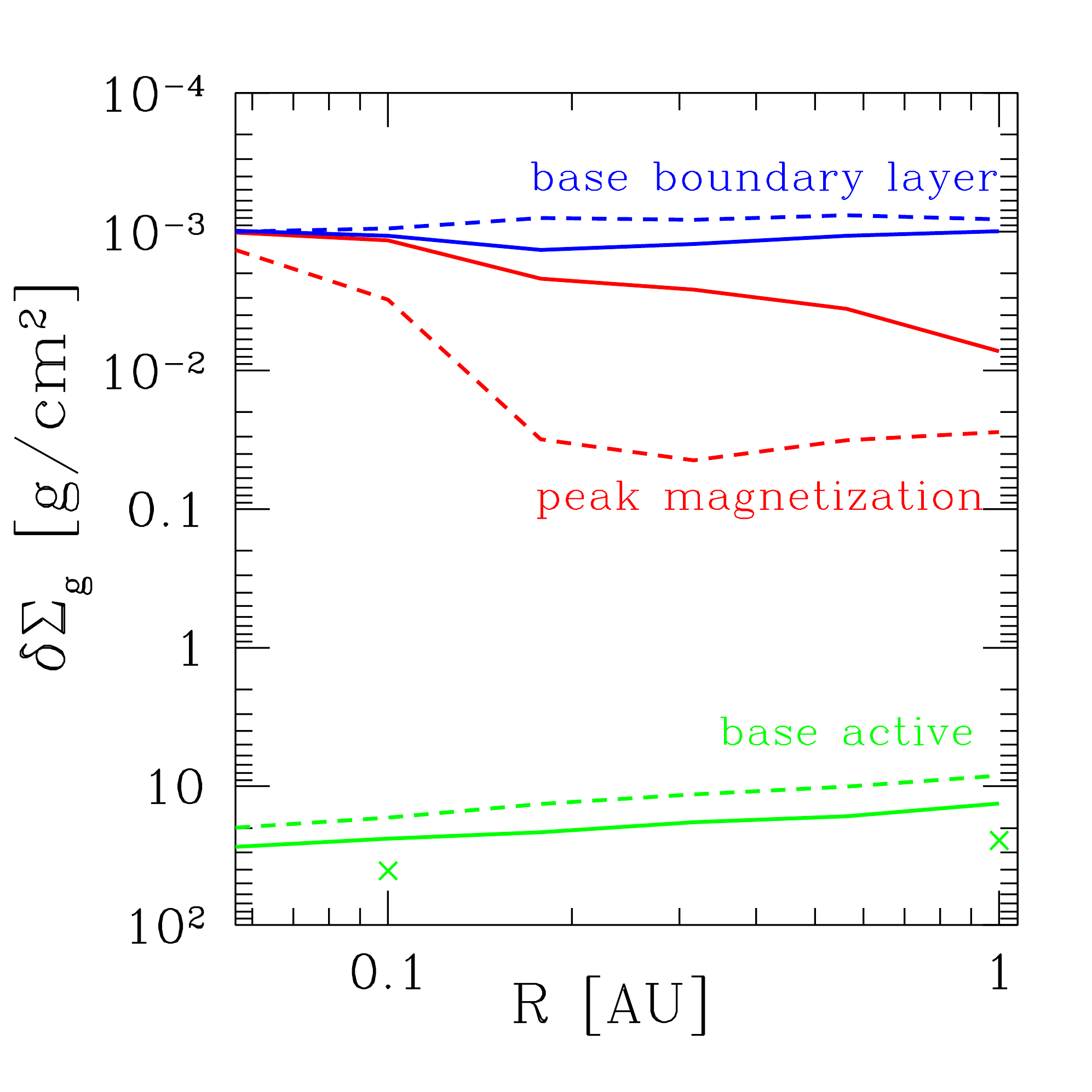}
\caption{
{\it Left panel:}  characteristic heights marking the disk profile, as a function of $R$.
{\it Right panel:} corresponding mass column measured from the top of the disk.
Solid (dashed) curves:  imposed magnetic field $\epsilon_B=1$ (0.1) in Equations 
(\ref{eq:boundary1})-(\ref{eq:boundary3}).  Black curves: height of wind-disk normal pressure balance, 
Equation (\ref{eq:normpressbal}); blue curves: base of wind-disk boundary layer, Equation (\ref{eq:rhocool}); 
red curves: height of peak magnetization; green curves: base of active layer ($\Lambda_{\rm O} = 10^2$).
`X' marks the base of the active region corresponding to an incident X-ray temperature $\kB T_X=5$ keV.}
\vskip .1in
\label{fig:heights}
\end{figure*}

\subsection{Parker Stability}\label{s:Parker}

Large magnetic field gradients cause the disk material to become unstable first to the undular 
Newcomb-Parker mode, and then to the interchange mode.   We find that the onset of the undular mode 
occurs in the active region for an imposed field of about $2\%$ of the maximum value at $0.1$ AU;
this threshold field strength increases roughly linearly with radius.  

As the unstable region grows 
with increasing applied $B_R$, the peak magnetization moves to lower $\delta\Sigma_g$ until the entire 
active region becomes (marginally) Parker unstable (Figure \ref{fig:parker}).  
The density profile then becomes effectively adiabatic.  This happens when the imposed field exceeds 
$B_R \sim 0.1$ G at $0.1$ AU; higher field strengths have a minimal effect on the density structure.

Note that the shearing rate of the imposed radial magnetic field is higher
than the growth rate of the undular mode in the wind-disk boundary layer as obtained from \citep{newcomb61}.  For this
reason, the magnetization that is obtained by imposing marginal stability of the undular mode 
may represent a lower bound to the true magnetization in this layer.

\begin{figure}[!]
\epsscale{1.2}
\plotone{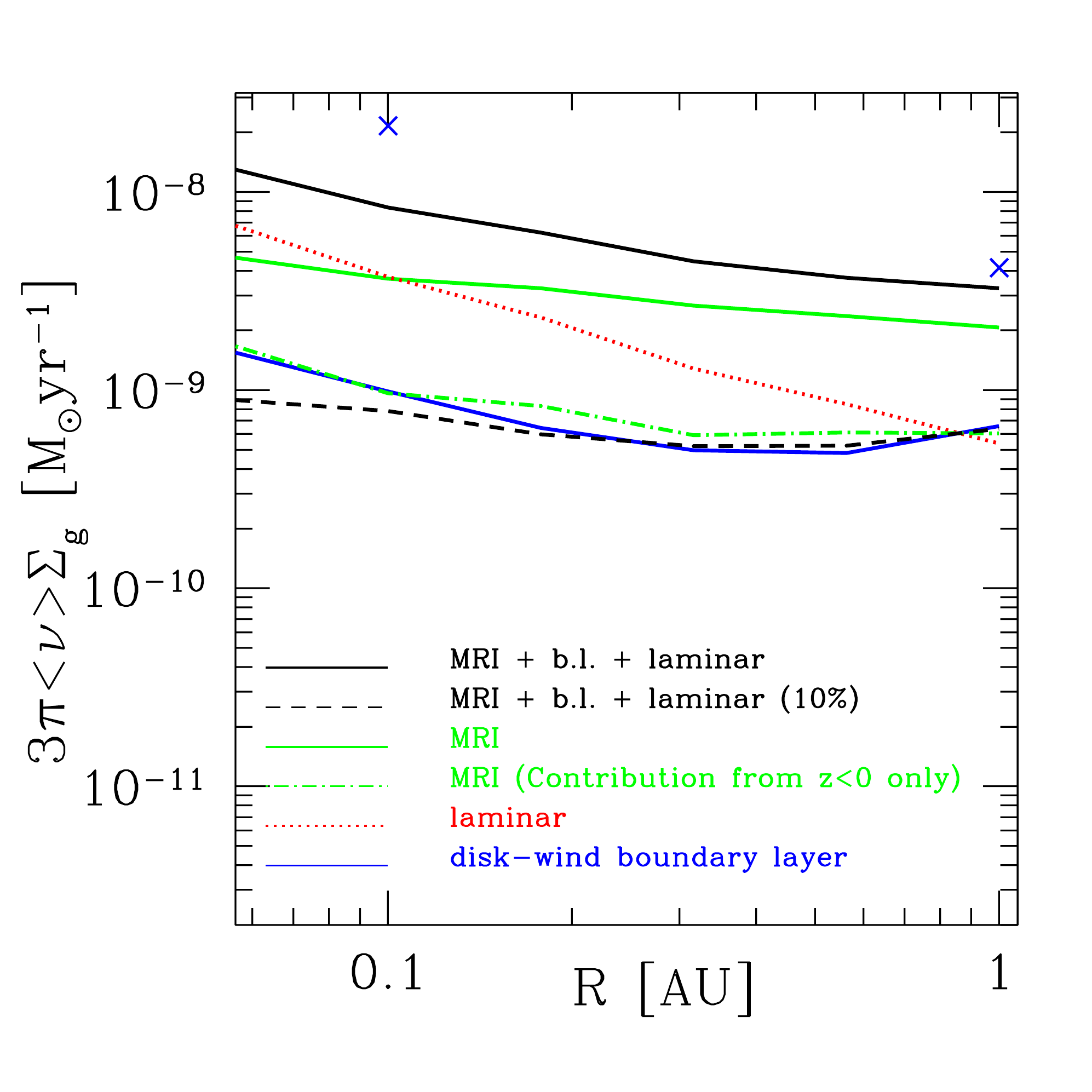}
\caption{Estimate of the local rate of radial mass transfer as a function of orbital radius for the imposed 
magnetic field $\epsilon_B = 1$, obtained by doubling the contribution from the more strongly magnetized
hemisphere ($z>0$).  Top curve:  combined contribution from MRI stress, laminar stress $B_RB_\phi/4\pi$, and
wind-disk boundary turbulence.  Separate contributions shown as green solid, cyan dotted, and blue solid curves.
Dashed black curve:  imposed field reduced by a factor of 0.1.  Dotted-dashed curve: separate contribution from 
MRI-driven torque in the lower (more weakly magnetized) hemisphere with $\epsilon_B = 1$.}
\vskip .1in
\label{fig:MdotR}
\end{figure}

\section{Summary}\label{s:summary}

We have presented a realistic model of a thin accretion disk 
with an imposed
radial magnetic field.  This degree of freedom has long been neglected in shearing box simulations in part for
technical reasons:  including it in such calculations leads to explosive and unlimited growth of the toroidal magnetic 
field.  Nonetheless, any accreting magnetized star that emits a strong wind (as does a T-Tauri star) provides an natural --
indeed inevitable -- external source for $B_R$ to a surrounding, centrifugally supported disk.   Some of the 
results obtained here may therefore have close analogs in the accretion flows feeding neutron stars or white dwarfs.

A key motivation for our approach is that the radial dependence of the imposed magnetic field is
much better defined than it is in the vertical-field geometry.  A quasi-spherical divergence of the wind magnetic field
implies a relatively strong magnetization in the inner disk.  The evolution of PPD with this type of radial structure is
examined in Paper II.  Here we summarize the radial structure that emerges.  

The peak magnetization that is reached in the wind-disk boundary layer, and at the base of the 
active layer, is shown in Figure \ref{fig:betapeak} as a function of distance from the star.  The peak 
magnetization increases approximately linearly with the applied radial field $B_R^w$ at 1 AU, sharpening to 
$\sim (B_R^w)^{3/2}$ at 0.1 AU.  The base of the MRI-active layer is mainly controlled by the rapid onset of
strong Ohmic drift, even though ambipolar drift remains faster throughout the bulk of the active layer.  

Diffusion of magnetic flux is very rapid through the weakly ionized
layer straddling the midplane.  A slight imbalance in the magnitude
of the imposed radial magnetic field causes toroidal field from the more
strongly magnetized hemisphere to push through to the base of the active layer
in the opposing hemisphere.

We also show in Figure \ref{fig:heights} the radial variation of several characteristic surfaces in the disk:
the height $z_w$ at which the momentum flux of
the wind normal to the disk balances the thermal pressure;  the base $z_{\rm bl}$ of the 
wind-disk boundary layer where hydrogen transitions from atomic to molecular form, and the temperature plummets
below $T_{\rm bl} \sim 5000$ K; the height of peak magnetization; and the 
base of the active layer.  The height of peak magnetization maintains a relatively uniform aspect ratio $z/R$ -- in contrast
with the thermal scale height -- due to the increasing magnetization in the inner disk.  

Figures \ref{fig:betapeak} and \ref{fig:heights} show that a higher X-ray temperature (5 keV)
reduces the peak magnetization but increases the active column depth.  
For a given $L_X$, there is a reduction in the flux of 
low-energy X-rays that dominate the ionization in the upper disk, but an
increased flux of penetrating X-rays.  
The net effect is to increase the mass-transfer rate by a factor $\sim 2$ 
at 0.1 AU, with only a slight increase at 1 AU.

The active column $\delta\Sigma_{\rm act}$ pushes deeper in the innermost parts of the disk.  
Figure \ref{fig:MdotR} shows the characteristic radial mass transfer rate that each part of the disk
can sustain.  This is constructed by calculating
\ba\label{eq:dmdt}
3\pi\langle \nu \rangle \Sigma_g &=& 3\pi\int_{z_w^-}^{z_w^+} dz
        \biggl[\rho(\nu_{\rm MRI} + \nu_{\rm mix}) - \nn
  &&\quad  {B_RB_\phi\over 6\pi \Omega} \Theta\left({c\Omega R|B_R|\over 4\pi\eta_{\rm O}} - |J_z|\right)\biggr].\nn
\ea
Here $\Sigma_g$ is the total column through the disk, and $\Theta(x) = 1$ (0) for $x > 0$ $(<0)$.  

The largest contribution to the integral (\ref{eq:dmdt}) comes from $\delta\Sigma_g \sim 10$--30 g cm$^{-2}$, where
the main driver of angular momentum transport is MRI turbulence.  
The peak in the vertical profile of $\nu_{\rm MRI}$ shifts to a lower column at
a smaller radius, and merges with the base of the wind-disk boundary layer (Figure \ref{fig:heights}).

The laminar Maxwell stress makes a non-negligible contribution to the 
torque, and we find that it approaches the MRI stress at the innermost radius considered ($10^{-1.25}$ AU).  
The cutoff in this term in Equation (\ref{eq:dmdt}) arises from the low ionization level near the disk midplane.
The toroidal field in the lower
disk is sourced by the shear in the more strongly ionized upper layers of the disk, and diffuses downward.
But when the Ohmic term $4\pi \eta_{\rm O} J_z/c$ in the induction equation is smaller than $({\bf v}\times{\bf B})_z/c$, 
corresponding to a strong drift of the toroidal field in the radial direction, the Maxwell stress $B_RB_\phi/4\pi$ 
cannot be communicated to the neutrals.

\newpage
\subsection{Outstanding Issues}

The stronger magnetizations we find in the inner parts of the PPD has important implications for the
evolution of the disk and the migration of planets, which are explored in Paper II.  
A higher rate of mass transfer forces a strong secular decrease in the surface 
density in the inner disk over a sub-Myr timescale.  This behavior holds over a range 
of imposed field strengths, between $10^{-3}$ and $10^{-2}$ G at $R \sim 1$ AU.

A consequence of our work is that there is a much weaker distinction between PPDs with an optical absorption layer
and `transition' disks with transparent inner cavities, than has been usually supposed.  We find that 
$\Sigma_g$ in the innermost $\sim$ AU of a PPD drops to $\sim 10^2$ g cm$^{-2}$ at a relatively early stage --
even while an absorption layer is sustained in the upper disk by a small abundance of dust that is advected
inward from the outer disk.  

It should be emphasized that obtaining the correct sign of $\partial\Sigma_g/\partial t$ in the inner disk
depends not only on understanding the profile of the imposed magnetic field.  It also depends on a careful 
account of how the ionization rate varies with distance from the protostar.  Our ionization model shows
a stronger inward peak in $\Gamma_i$ at large column than some previous treatments.  

An additional effect involves the stirring of dust from a settled midplane particle layer.
As we show in paper II, this provides a strong downward buffer to $\Sigma_g$ when the ionization rate 
at the midplane rises sufficiently to trigger the MRI.

The scaling of the MRI stress with the linearly amplified toroidal magnetic field remains conjectural.  
Nonetheless, Figure \ref{fig:coll} shows that secular mass depletion in the inner disk is obtained 
for the plausible range of indices $\delta$ in Equation (\ref{eq:nuana}).  This simple analytic model of the stress,
involving a balance between vertical diffusion and linear winding of the entrained magnetic field, reproduces 
well the radial scaling obtained from the vertical disk profiles constructed in this paper.  Softening $\delta$ from 1 to $0.5$ would tend to raise the
MRI-generated stress near the base of the active layer, where $B_\phi^2/8\pi
\lesssim P$, and therefore raise the accretion rate.  From this perspective, our choice of $\delta$ is conservative.

A key question involves the vigor of the MRI turbulence that will result in a situation where
$B^2/8\pi$ approaches the gas pressure.
In a vertical field geometry, such a strong magnetization tends to suppress the MRI \citep{bs13,lesur13}. 

A linear stability analysis, taking into both ambipolar
and Hall drift in a toroidal field geometry, shows that a vigorous instability remains possible in an unstratified
disk for the Elsasser numbers obtained from our ionization calculation (${\rm Am} \sim 10^2-10^3$) and for vertical 
wavenumber $k_zc_g/\Omega \sim 5-10$.  
The inclusion of Hall drift appears to facilitate the instability for a fixed value of Am.
In addition, the orientation of the Poynting flux that is driven by the MRI shows a dramatic difference with the case
of a vertical seed field:  rather than flowing vertically, it is predominantly horizontal.  On this basis, we have
postulated that growing MRI modes will remain trapped in the active layer, and that full-fledged turbulence will be 
triggered by the entrainment of a radial magnetic field from a T-Tauri wind.   

It is worth emphasizing that the mass transfer rate obtained here is weakly sensitive to
the coefficient relating disk magnetization to MRI stress, scaling as $\sim \widetilde\alpha_{\rm MRI}^{1/3}$.  
This turbulence model could be calibrated by direct numerical simulations of a stratified disk with a strong 
seed toroidal magnetic field, ionization levels a order of magnitude higher than those implemented by 
\cite{lesur14}, \cite{bai14} and \cite{gressel15}, and including both ambipolar and Hall drift.


\acknowledgments  We thank the NSERC of Canada for support.

\begin{appendix}
\section{Ionization Rate}\label{s:ionrate2}

Our ionization model follows that of \cite{glassgold1997} with modifications.  X-rays with a thermal
bremsstrahlung spectrum (temperature $T_X$) are emitted by reconnecting magnetic field loops 
at a height $R_X$ above the protostar.  A flux of X-rays in the energy range $(E,E+dE)$,
\be
df_{\rm dir} = f_0 {dE\over E}e^{-E/\kB T_X}; \quad
f_0 \equiv \frac{L_{X}}{8\pi \kB T_{X}}\frac{1}{R^{2}+\left(z-R_{X}\right)^{2}},
\ee
is received at position $(R,z)$ in the disk.  This spectrum is cut off below an energy $E_0 = 1$ keV,
representing absorption by a column $ \sim 1$ g cm$^{-2}$ of stellar wind material.
Note that the ionization rate in the upper disk is proportional to the total X-ray flux, which 
decreases as $\sim R^{-2}$;  by contrast, \cite{glassgold1997} work with the X-ray flux per unit 
area of disk, which scales as $R^{-3}$.

The ionization rate by directly impacting X-rays is
\be\label{eq:gamdir}
d\Gamma_{i,\rm dir}=\frac{dE}{E}e^{-E/\kB T_{X}}\left(1+\frac{E}{0.037~{\rm keV}}\right)\sigma(E)f_0e^{-\tau_r}.
\ee
The term proportional to $E/(0.037~{\rm keV})$ accounts for secondary ionizations by photo-electrons.  
The optical depth along a direct (and nearly radial) ray is
\be\label{eq:taur}
\tau_r(E)=\frac{\left(\sigma(E)+Y_{e}\sigma_{T}\right)}{m_u}\frac{\delta\Sigma_g}{\sin\theta_X}.
\ee
The cross section $\sigma(E)$ per nucleon is obtained by summing over relevant species (H, He, C, N, O, Ne, S)
with volatiles reduced by a factor of 0.01 compared with solar abundances. 
The angle of incidence between X-rays and a surface $z(R)$ is
\be 
\theta_{X}=\arctan\left(\frac{R_{X}-|z|}{R}\right)+\arctan\left(\biggl|{\frac{dz}{dR}}\biggr|\right).
\ee

X-rays are also Compton scattered by atomic electrons towards higher column densities, resulting
in ionization at a rate
\be\label{eq:gamscatt}
d\Gamma_{i,\rm scatt}=\frac{dE}{E}e^{-E/\kB T_{X}}\left(1+\frac{E}{0.037}\right)\sigma(E)f_{z,\rm scatt}.
\ee
The vertical flux of scattered X-rays, radially and then vertically attenuated, is
\ba
f_{z,\rm scatt}(\delta\Sigma_g) &=& \int d\tau_{r}^{'}e^{-\tau_{r}'}{Y_e\sigma_T \over \sigma(E) + Y_e\sigma_T}
\frac{f_0\sin\theta_X}{2}e^{-\sin\theta_X\left(\tau_{r}-\tau_{r}'\right)} \nn
&\simeq& \frac{Y_{e}\sigma_{T}}{\left(\sigma(E)+Y_{e}\sigma_{T}\right)}
\frac{f_0 \sin\theta_X}{2\left(1-\sin\theta_X\right)}\left(e^{-\sin\theta_X\tau_r}-e^{-\tau_r}\right).
\ea
The factor $1/2$ in this expression accounts for the fraction of X-rays that are downscattered, as opposed to upscattered.
Also $\sin\theta_X(\tau_r - \tau_r')$ is the differential vertical optical depth of X-rays downscattered
at radial optical depth $\tau_r'$ and reaching a column $\delta\Sigma_g$ with radial optical depth
$\tau_r$ as given by Equation (\ref{eq:taur}).

We note that $\Gamma_{i,\rm scatt}$, which dominates the ionization rate deep in the active layer of the disk,
decreases with distance from the protostar as $\theta_X/R^2 \sim R^{-3}$, as compared with $\Gamma_{i,\rm dir}
\sim R^{-2}$. This distinguishes our ionization model from that of \cite{baigood2009}, who choose a
$\sim R^{-2.2}$ envelope for the direct and scattered components.  Partly as a result of this choice, they
obtain an active column that increases outward from the protostar.

The net ionization rate $\Gamma_i$ is calculated at each column $\delta\Sigma_g$ by summing the two terms
(\ref{eq:gamdir}) and (\ref{eq:gamscatt}) over $E$.  This procedure must be repeated while solving for
the disk profile, because magnetic pressure support alters $\theta_X \sim z(\delta\Sigma_g)/R$.  The function $\Gamma_i(\delta\Sigma_g)$ is smoothed near the disk midplane, so as to avoid numerical
instabilities; but the magnetic field profile is insensitive to the precise value of $\Gamma_i$ deep in the disk,
where Ohmic diffusion is rapid.

\section{Linear Growth of MRI Modes in a Toroidal Magnetic Field Subject to Hall and Ambipolar Drift}\label{s:nonsym}

A linear stability analysis that includes the effect of both Hall and ambipolar drift on
non-axisymmetric perturbations of a rotating disk has not previously been performed:  existing
treatments either assume axial symmetry \citep{wardle99,kunz04,pandey12}, or
only include the effect of Ohmic resistivity \citep{papaloizou97}.  Here we neglect
the effect of Ohmic drift, partly for reasons of simplicity, and also because it is subdominant
in the mass-transferring layer of a PPD interior to $R \sim 1$ AU.

The background flow is taken to be independent of vertical coordinate $z$, and the rotation profile
is $\Omega(R) \propto R^{-3/2}$.  
The magnetic field is decomposed into a background ${\bf B} = B_\phi \hat\phi + B_z\hat z$ and a linear perturbation
$\delta{\bf B}$.  We neglect any linear growth of $B_\phi$ by the winding of a small seed radial magnetic field, 
so that the background magnetic field is independent of time.  
This allows us to evade the complication of changing $\eta_a$ (which is proportional to $B^2$). 
The peak growth rates we obtain are high enough to justify neglecting the winding up of
the background radial field.

We will not re-derive the time evolution equation for $\delta{\bf B}$ here, but follow the approach
of \cite{balbus01}, who considered the combined effects of Ohmic and Hall drift.   We focus on incompressible perturbations
of a cylindrically symmetric, rotating fluid with angular velocity $\Omega(R) \propto R^{-3/2}$ and 
$\partial P/\partial z = \partial\rho/\partial z = 0$.  The isothermal sound speed $c_g$ introduces a dimensionless
parameter $\sigma = v_{\rm{A}}^2/c_g^2$; and a small parameter $c_g/\Omega R$, which is relevant because we 
only consider modes with vertical wavenumber $k_z \gtrsim (2\pi) \Omega/c_g$. 
We take $c_g/\Omega R = 0.03$.   Hall and ambipolar drift are characterized
by the respective Elsasser numbers $\Lambda_{\rm H}$ and ${\rm Am}$.  

Modes are labeled by wavenumber $[k_R(t),m/R,k_z]$.  Following \cite{balbus92} and \cite{balbus01},
we work in the shearing coordinate system,
\be
R' = R, \quad \phi' = \phi - \Omega(R)t, \quad z' = z,
\ee
and consider the growth or decay of a single Fourier mode $\delta{\bf B} \propto \exp[i(k_R'R' + m\phi' + k_z'z')]$
with ${\bf k}'$ constant.   The wavevector in the background coordinate system is $k_z = k_z'$ and
\be
k_R(t) = k_R' - mt {d\Omega\over dR} = k_R' + {3\over 2} mt\Omega.
\ee

The induction equation reads
\be
{d\delta B_R\over dt} = i({\bf k}\cdot{\bf B})\delta v_R + A_R;\quad\quad {d\delta B_z\over dt} = i({\bf k}\cdot{\bf B})\delta v_z + A_z.
\ee
In our case, the auxiliary vector
\be
{\bf A} = {\bf A}_{\rm H} + {\bf A}_a,
\ee
receives contributions from Hall and ambipolar drift, with Ohmic diffusion neglected:
\be
{\bf A}_{\rm H} = {c({\bf k}\cdot{\bf B}){\bf k}\times\delta{\bf B}\over 4\pi n_e};\quad\quad
{\bf A}_a = -{1\over 4\pi \gamma_i\rho_i \rho}\left[B^2 k^2 \delta{\bf B} - \delta {\bf B}\cdot({\bf k}\times{\bf B})
\,{\bf k}\times{\bf B}\right].\nn
\ee
Here ${\bf k} = (k_R(t),m/R,k_z)$ is the wavevector and
\be
{\bf k}\cdot {\bf B} = {m\over R}B_\phi + k_zB_z
\ee
is independent of time.  The toroidal velocity and magnetic field perturbations are given by
\be
\delta v_\phi = -{R\over m}(k_R\delta v_R + k_z\delta v_z);\quad\quad \delta B_\phi = -{R\over m}(k_R\delta B_R + k_z\delta B_z).
\ee

The perturbation to the toroidal field can thus be eliminated from the toroidal component of the induction equation, 
leaving two coupled equations for $d\delta B_R/dt - A_R$ and $d\delta B_z/dt - A_z$.  We have confirmed that 
these are given by Equations (102) and (103) of \cite{balbus01}.   The new aspects of the problem considered here are 
are contained in the auxilary term ${\bf A}$.

We will consider modes short enough to fit inside a vertical scale height, $k_z \gtrsim 2\pi c_g/\Omega$, so we normalize $k_R$, $k_z$ by
\be
{\bf k}_{R,z} = {\widetilde {\bf k}}_{R,z} {\Omega\over c_g},
\ee
and time by $d\widetilde t = \Omega dt$.   Then
\ba\label{eq:evol}
{d\over d\widetilde t}\left({d\delta B_R\over d\widetilde t} - {A_R\over\Omega}\right) &\;=\;&
 -{2\over m} {\Omega R\over c_g}{\widetilde k_z^2\over \widetilde k^2} \left[\widetilde k_R\left({d\delta B_R\over d\widetilde t} - 
     {A_R\over\Omega}\right) + \widetilde k_z\left({d\delta B_z\over d\widetilde t} - {A_z\over\Omega}\right)\right] \nn
&& -m{c_g\over\Omega R}\left[3{\widetilde k_R\over \widetilde k^2}\left({d\delta B_R\over d\widetilde t} - {A_R\over\Omega}\right)
    + 2{\widetilde k_z\over \widetilde k^2}\left({d\delta B_z\over d\widetilde t} - {A_z\over\Omega}\right)\right] \nn
&& - {v_{{\rm A},\phi}^2\over c_g^2}\left[F_0({\bf k})\,m{c_g\over\Omega R}\right]^2\delta B_R;\nn
{d\over d\widetilde t}\left({d\delta B_z\over d\widetilde t} - {A_z\over\Omega}\right) &\;=\;&
{2\over m} {\Omega R\over c_g}{\widetilde k_R \widetilde k_z\over \widetilde k^2}
\left[\widetilde k_R\left({d\delta B_R\over d\widetilde t} - {A_R\over\Omega}\right) + 
   \widetilde k_z\left({d\delta B_z\over d\widetilde t} - {A_z\over\Omega}\right)\right] \nn
&& -m{c_g\over\Omega R}{\widetilde k_z\over \widetilde k^2}\left({d\delta B_R\over d\widetilde t} 
     - {A_R\over\Omega}\right) - {v_{{\rm A},\phi}^2\over c_g^2}\left[F_0({\bf k})\,m{c_g\over\Omega R}\right]^2\delta B_z.\nn
\ea
The contribution to ${\bf A}$ from ambipolar drift is
\ba
A_{a,R} &\;=\;& -{v_{\rm{A},\phi}^2\over c_g^2} {\Omega\over {\rm Am}} F_0({\bf k}) \biggl[F_1(\delta{\bf B},{\bf k}) + 
{B_z\over B_\phi}{mc_g\over \Omega R} \left(\widetilde k_z\delta B_R - \widetilde k_R\delta B_z\right)\biggr]; \nn
\quad\quad
A_{a,z} &\;=\;& -{v_{\rm{\rm{A},\phi}}^2\over c_g^2} {\Omega\over {\rm Am}} \biggl[F_0({\bf k})F_2(\delta{\bf B},{\bf k}) + 
{B_z\over B_\phi} \widetilde k^2\left({B_z\over B_\phi}\delta B_z
- {\Omega R\over mc_g}\left(\widetilde k_R \delta B_R + \widetilde k_z\delta B_z\right)\right)\biggr], \nn
\ea
and from Hall drift
\be
A_{\rm{H},R} = {B_\phi \over B}{v_{\rm A}^2\over c_g^2} {\Omega\over\Lambda_{\rm H}} F_0({\bf k})\, F_2(\delta{\bf B},{\bf k});\quad\quad
A_{\rm{H},z} = -{B_\phi \over B}{v_{\rm A}^2\over c_g^2} {\Omega\over\Lambda_{\rm H}} F_0({\bf k})\, F_1(\delta{\bf B},{\bf k})
\ee
where
\be
F_0({\bf k})  \equiv 1 + {k_z B_z\over (m/R)B_\phi},
\ee
and
\be
F_1(\delta{\bf B},{\bf k}) = \left[\widetilde k_R^2 + m^2{c_g^2\over (\Omega R)^2}\right]\delta B_R + 
\widetilde k_R\widetilde k_z \delta B_z;\quad\quad
F_2(\delta{\bf B},{\bf k}) = \left[\widetilde k_z^2 + m^2{c_g^2\over (\Omega R)^2}\right]\delta B_z + 
\widetilde k_R\widetilde k_z \delta B_R.
\ee

\begin{figure}[!]
\epsscale{.55}
\plotone{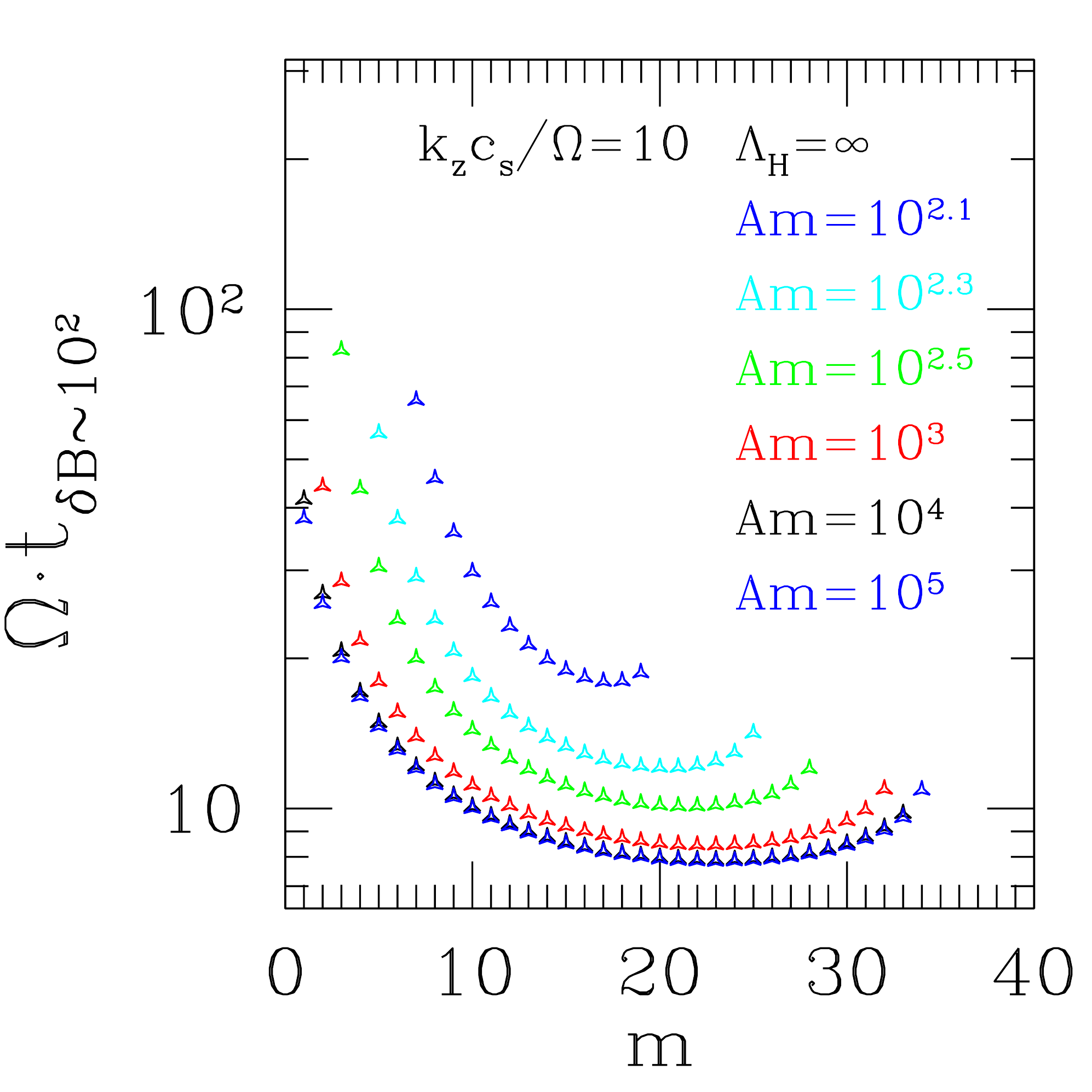}
\caption{Time for a non-axisymmetric mode initialized to $\delta B_R(0) = \delta B_0$, 
$\delta B_z(0) = \delta B_\phi(0) =0$ to grow to an amplitude $|\delta{\bf B}| = 10^2\delta B_0$ in a
rotating fluid with Keplerian shear, strong magnetization ($\sigma = B_\phi^2/4\pi P = 1$) but no 
vertical stratification.  Here the ambipolar Elsasser number takes a range of values, but Hall 
drift is suppressed.  Vertical wavenumber is set to a value that allows the mode to fit within a single scale height
of a stratified disk, $k_z \gtrsim (2\pi)\Omega/c_g$.  Then the small parameter $c_g/\Omega R = 0.03$ 
determines the azimuthal wavenumber $m$ of the fastest-growing modes.   Growth on a timescale 
$\lesssim 10\Omega^{-1}$ is found for $m \sim 20$--30 and ${\rm Am} \gtrsim 200$.  Growth freezes out 
when ${\rm Am} \lesssim 100$; or $m \gtrsim 30$ due to the winding up of the mode by the disk shear.
}
\vskip .1in
\label{fig:growth1}
\end{figure}

\begin{figure}[!]
\epsscale{.55}
\plotone{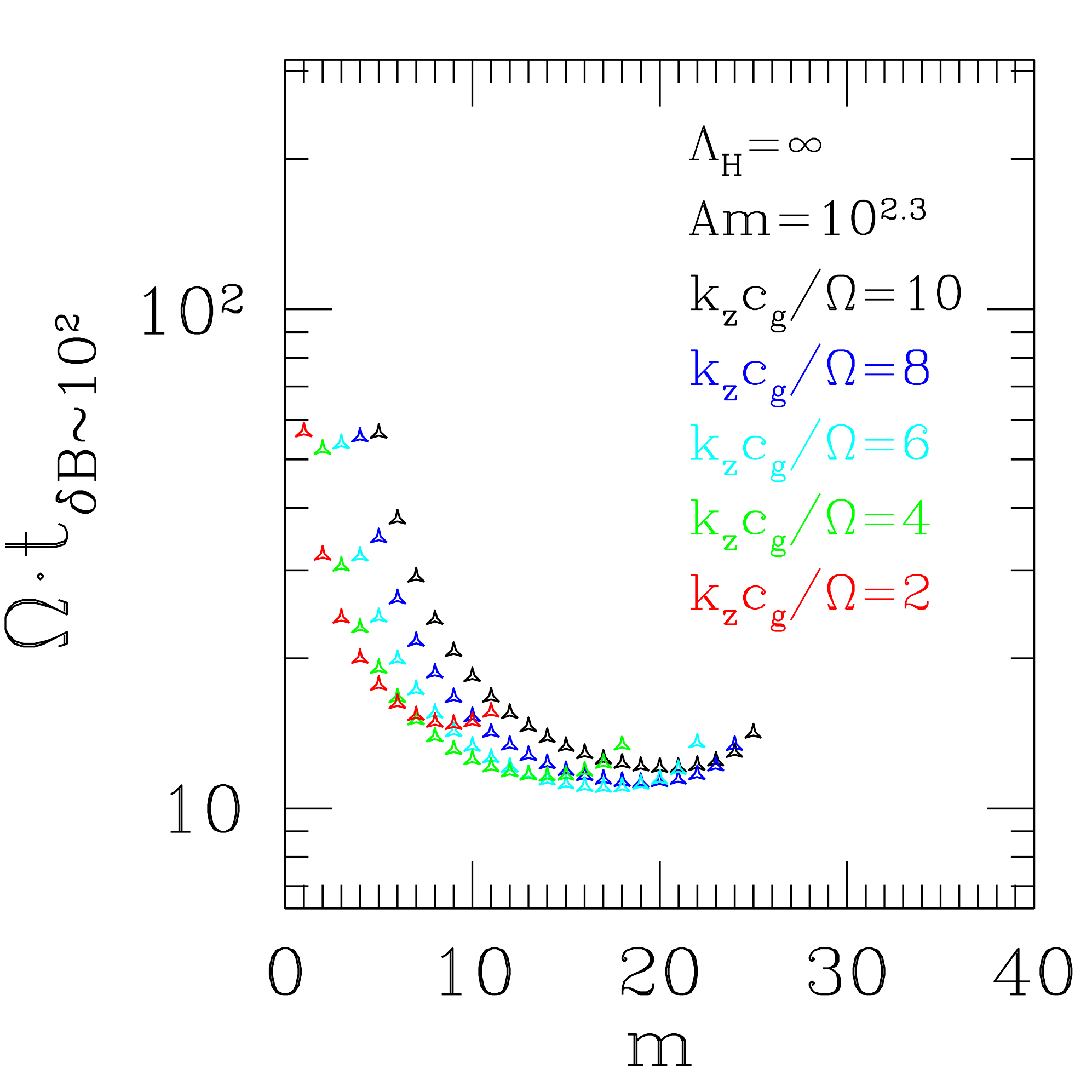}
\caption{Dependence of growth time on vertical wavenumber $k_z$, for the case of
pure ambipolar drift.  Value of ${\rm Am}$ is just above threshold for vigorous
MRI growth.  Background magnetization $B_\phi^2/4\pi P = 1$.}
\vskip .1in
\label{fig:growth2}
\end{figure}

\begin{figure}[!]
\epsscale{1.1}
\plottwo{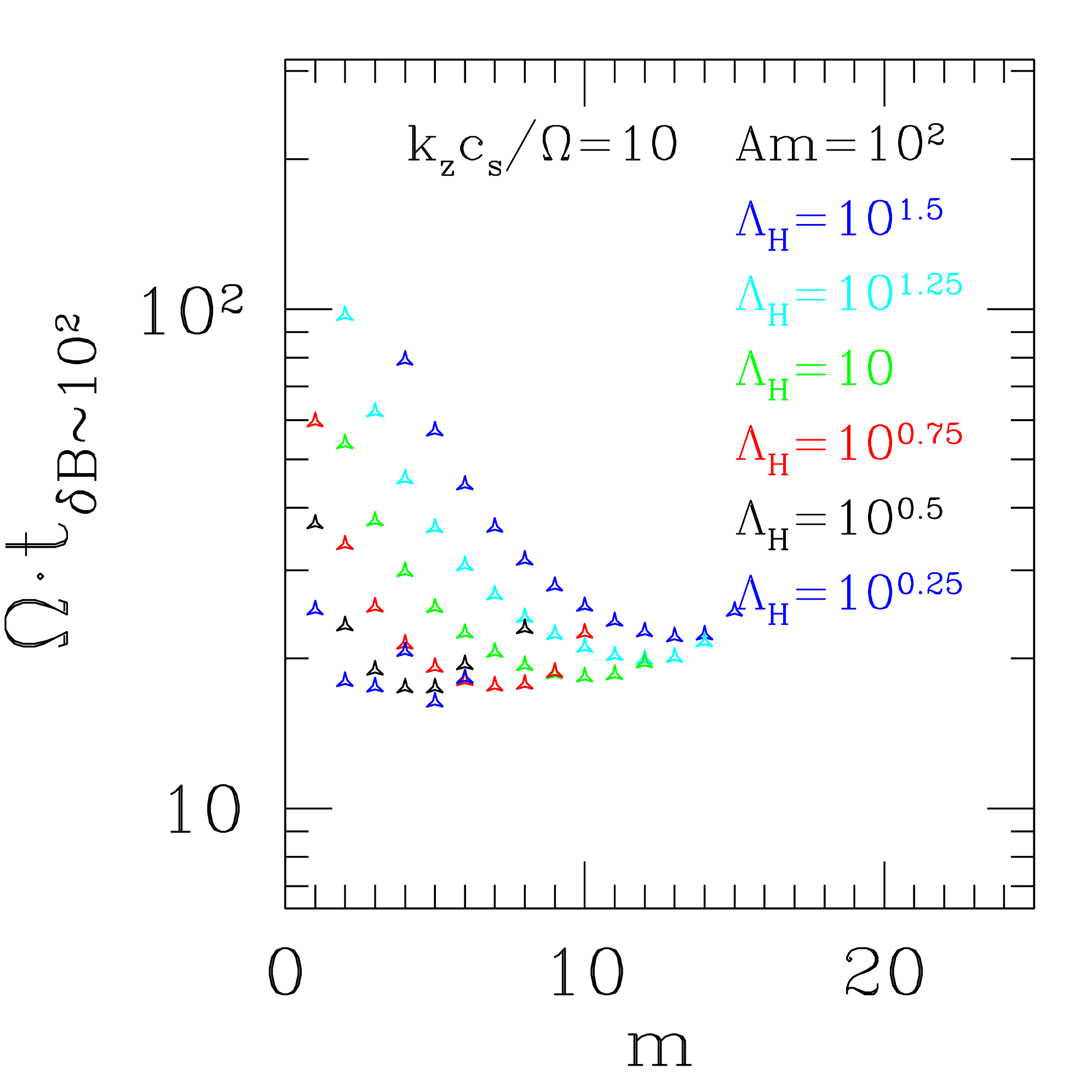}{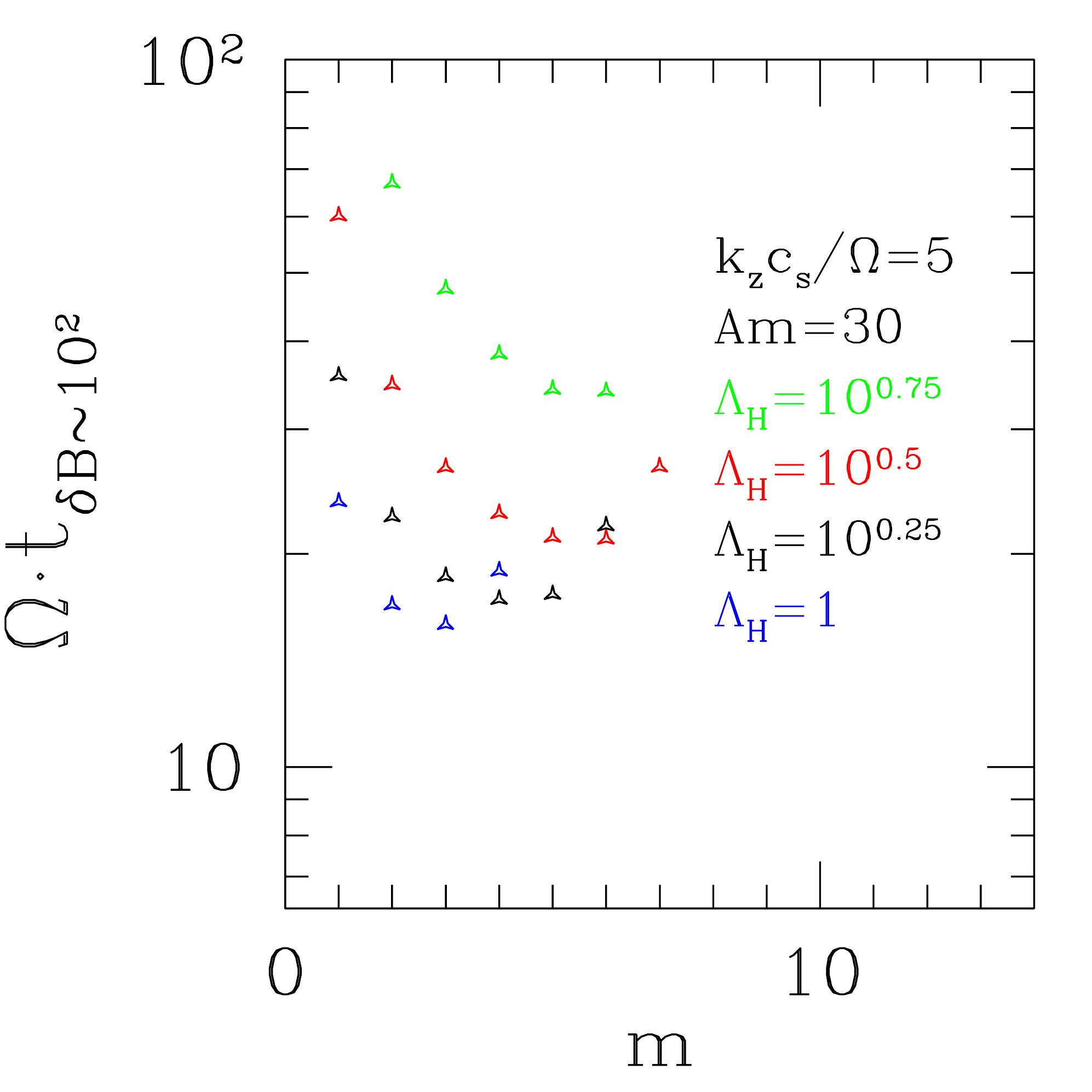}
\caption{Same as Figure \ref{fig:growth1}, but now with Hall drift turned on.
Values of ${\rm Am}$ are below threshold for mode growth when $\Lambda_{\rm H} = \infty$.  
Strong growth appears for $\Lambda_{\rm H} \sim {\rm Am}/10$, peaking at slightly lower values of $m$
than in the ideal MHD regime.}
\vskip .1in
\label{fig:growth3}
\end{figure}

We consider a poloidal wavevector that is large enough, $k_z \sim (3-10)\Omega/c_g$, for the mode 
to fit inside a vertical scale height of a more realistic, stratified disk.
Starting with a perturbation $\delta B_R$ at $t=0$ and invoking 
${\bf k}\cdot \delta{\bf B} = 0$ allows us to set $k_R' = 0$.  Winding up of the mode to
$k_R \sim k_z$ limits the growth at high azimuthal wavenumber $m$.  

Figure \ref{fig:growth1} shows how ambipolar drift can quench the MRI
at high background magnetization $\sigma \equiv B_\phi^2/4\pi P = 1$.  To
begin, we turn off Hall drift.  The criterion for vigorous growth adopted
here is that the mode reach an amplitude $\sim 10^2$ times $\delta B_R(t=0)$
in a finite time.  The corresponding growth time is as short as
$\sim 10\Omega^{-1}$ for $m \sim 20$--30 and ${\rm Am} \gtrsim 200$.  
Growth rapidly becomes less efficient as ${\rm Am}$ is reduced, with significant
growth present down to ${\rm Am} \sim 100$.  We show in Figure \ref{fig:growth2} 
how the choice of vertical wavenumber affects the growth time near the threshold
value of ${\rm Am}$.

The introduction of Hall drift facilitates MRI growth
at lower ${\rm Am}$ (Figure \ref{fig:growth3}).  
Taking $\Lambda_{\rm H} \sim 0.1\,{\rm Am}$, which is realistic for the mass-transferring
layer of a PPD, we find strong growth for ${\rm Am} \sim 30$ and $m \sim 3$--5.  In comparison,
stratified shearing box simulations find strong turbulence is excited in a weaker seed
toroidal field ($\beta_{\phi,0} \gtrsim 10^2$) when the ambipolar number exceeds 
${\rm Am} \gtrsim 10$ \citep{baistone11,simon13}.  

An axisymmetric linear analysis suggests that Hall drift facilitates MRI growth
in an ohmic plasma \citep{wardle12}.  There is also some numerical evidence
that the Hall effect modifies the saturation strength of the MRI-amplified magnetic field 
\citep{sano02}.  However, we are not aware of shearing box calculations which test 
for well-developed MRI turbulence in the regime of strong toroidal magnetization and 
allow for $\Lambda_{\rm H} \sim {\rm Am}/10 \gtrsim 10$, as is implied by the ionization
model adopted here.

\subsection{Constraints on Mode Growth:  Analytic Considerations}\label{s:analytic}

Winding up of a magnetic perturbation by the disk shear eventually shuts off growth, as
$k_R$ rises above $k_z$ \citep{balbus92}.  It is easy to show that Equations (\ref{eq:evol})
do not yield growing modes in the ideal MHD regime as $k_R \rightarrow \infty$:  then
the surviving terms $-({\bf k}\cdot{\bf v}_{\rm A})^2\delta B_{R,z}$ on the right-hand side only
support periodic disturbances.  

A strong MRI depends on the following competition between terms:
\be\label{eq:ineq}
\left|{2\over m}{\Omega R\over c_g} {\widetilde k_R \widetilde k_z^2\over \widetilde k^2}
{d\delta B_{R,z}\over dt}\right| \gtrsim 
{v_{{\rm A},\phi}^2\over c_g^2}\left[F_0({\bf k})\,m{c_g\over\Omega R}\right]^2\delta B_{R,z}.
\ee
The left-hand side peaks when $\widetilde k_R = \widetilde k_z$, at the winding time
\be
\Omega t = {2\over 3} {\Omega R\over m c_g} \widetilde k_z.
\ee

We now allow a finite background $B_z \gtrsim (m c_g/\Omega R)\widetilde k_z B_\phi$.
Then the inequality (\ref{eq:ineq}) reduces to an upper bound on $B_z/B_\phi$
for which strong MRI growth is possible,
\be
{B_z/B_\phi \over 1 + B_z/B_\phi}  \lesssim  
               \left({3\over 2\sigma \widetilde k_z}\right)^{1/2}.
\ee
This result is reflected in Figure \ref{fig:modegrowth}.

We find peak non-axisymmetric mode growth rates remaining near, or below, the 
peak axisymmetric growth rate $3\Omega/4$.  An upper bound on the azimuthal wavenumber
of a mode which experiences such strong growth is easily obtained by imposing
$k_R(t_{\rm grow}) \lesssim k_z$.   Taking $t_{\rm grow} \sim (4/3)\ln(\delta B/\delta B_0)
\Omega^{-1}$, this translates to 
\be
m \lesssim {k_z R\over 2\ln(\delta B/\delta B_0)}
   \sim 30\,\left({k_z\over 10\, h_g^{-1}}\right) \left({h_g\over 0.03 R}\right)^{-1}.
\ee
This is consistent with the results shown in Figures \ref{fig:growth1} and 
\ref{fig:growth2}:  we see that the value of $m$ at which the growth rate peaks
increases with growing $k_z$.  Ambipolar and Hall drift tend to reduce the azimuthal
wavenumber of maximum growth, as expected from the $\sim k^2$ dependence of both
${\bf A}_{\rm A}$ and ${\bf A}_{\rm H}$.  
\end{appendix}




\begin{thebibliography}{} 
\bibitem[Bai(2011)]{bai11} Bai, X.-N.\ 2011, \apj, 739, 50 
\bibitem[Bai(2014)]{bai14} Bai, X.-N.\ 2014, \apj, 791, 137 
\bibitem[Bai \& Goodman(2009)]{baigood2009} Bai, X.-N., \& Goodman, J.\ 2009, \apj, 701, 737 
\bibitem[Bai \& Stone(2011)]{baistone11} Bai, X.-N., \& Stone, J.~M.\ 2011, \apj, 736, 144 
\bibitem[Bai \& Stone(2013)]{bs13} Bai, X.-N., \& Stone, J.~M.\ 2013, \apj, 769, 76 
\bibitem[Balbus \& Hawley(1991)]{balbus91} Balbus, S.~A., \& Hawley, J.~F.\ 1991, \apj, 376, 214 
\bibitem[Balbus \& Hawley(1992)]{balbus92} Balbus, S.~A., \& Hawley, J.~F.\ 1992, \apj, 400, 610 
\bibitem[Balbus \& Terquem(2001)]{balbus01} Balbus, S.~A., \& Terquem, C.\ 2001, \apj, 552, 235 
\bibitem[Bally et al.(2007)]{bally07} Bally, J., Reipurth, B., \& Davis, C. J. \ 2007, in Protostars and Planets V, 
ed. B. Reipurth, D. Jewitt, \& K. Keil (Tucson: Univ. Arizona Press), 215
\bibitem[Begelman \& Pringle(2007)]{begelman07} Begelman, M.~C., \& Pringle, J.~E.\ 2007, \mnras, 375, 1070 
\bibitem[Blaes \& Balbus(1994)]{bb94} Blaes, O.~M., \& Balbus, S.~A.\ 1994, \apj, 421, 163 
\bibitem[Blandford \& Payne(1982)]{bp82} Blandford, R.~D., \& Payne, D.~G.\ 1982, \mnras, 199, 883 
\bibitem[Calvet(1997)]{calvet1997} Calvet, N.\ 1997, Herbig-Haro Flows and the Birth of Stars, IAU Symposium 182, 417, ed. B. Reipurth and C. Bertout 
\bibitem[Chiang \& Goldreich(1997)]{chiang1997} Chiang, E.~I., \& Goldreich, P.\ 1997, \apj, 490, 368 
\bibitem[Cleeves et al.(2013)]{cleeves2013} Cleeves, L.~I., Adams, F.~C., \& Bergin, E.~A.\ 2013, \apj, 772, 5 
\bibitem[D'Alessio et al.(2006)]{dalessio2006} D'Alessio, P., Calvet, N., Hartmann, L., Franco-Hern{\'a}ndez, R., 
\& Serv{\'{\i}}n, H.\ 2006, \apj, 638, 314 
\bibitem[Dalgarno \& McCray(1972)]{dalgarno1972} Dalgarno, A., \& McCray, R.~A.\ 1972, \araa, 10, 375 
\bibitem[Daou et al.(2006)]{daou06} Daou, A.~G., Johns-Krull, C.~M., \& Valenti, J.~A.\ 2006, \aj, 131, 520 
\bibitem[Elmegreen(1978)]{elmegreen1978} Elmegreen, B.~G.\ 1978, Moon and Planets, 19, 261 
\bibitem[Feigelson et al.(2003)]{feigelson03} Feigelson, E.~D., Gaffney, J.~A., III, Garmire, G., Hillenbrand, L.~A., 
\& Townsley, L.\ 2003, \apj, 584, 911 
\bibitem[Flaig et al.(2010)]{flaig10} Flaig, M., Kley, W., \& Kissmann, R.\ 2010, \mnras, 409, 1297 
\bibitem[Flock et al.(2012)]{flock12} Flock, M., Henning, T., \& Klahr, H.\ 2012, \apj, 761, 95 
\bibitem[Fromang et al.(2002)]{fromang02} Fromang, S., Terquem, C., \& Balbus, S.~A.\ 2002, \mnras, 329, 18 
\bibitem[Fromang \& Stone(2009)]{fromang09} Fromang, S., \& Stone, J.~M.\ 2009, \aap, 507, 19 
\bibitem[Furlan et al.(2006)]{furlan2006} Furlan, E., Hartmann, L., Calvet, N., et al.\ 2006, \apjs, 165, 568 
\bibitem[Gammie(1996)]{gammie96} Gammie, C.~F.\ 1996, \apj, 457, 355 
\bibitem[Gammie \& Balbus(1994)]{gammie94} Gammie, C.~F., \& Balbus, S.~A.\ 1994, \mnras, 270, 138 
\bibitem[Glassgold et al.(1997)]{glassgold1997} Glassgold, A.~E., Najita, J., \& Igea, J.\ 1997, \apj, 480, 344 
\bibitem[Glassgold et al.(2004)]{glassgold2004} Glassgold, A.~E., Najita, J., \& Igea, J.\ 2004, \apj, 615, 972 
\bibitem[Goldreich \& Tremaine(1980)]{gt80} Goldreich, P., \& Tremaine, S.\ 1980, \apj, 241, 425 
\bibitem[Gressel et al.(2015)]{gressel15} Gressel, O., Turner, N.~J., Nelson, R.~P., 
     \& McNally, C.~P.\ 2015, \apj, 801, 84 
\bibitem[Guan \& Gammie(2009)]{guan09} Guan, X., \& Gammie, C.~F.\ 2009, \apj, 697, 1901 
\bibitem[Gullbring et al.(1998)]{gullbring1998} Gullbring, E., Hartmann, L., Brice{\~n}o, C., \& Calvet, 
     N.\ 1998, \apj, 492, 323 
\bibitem[Johansen \& Levin(2008)]{jl08} Johansen, A., \& Levin, Y.\ 2008, \aap, 490, 501
\bibitem[Hawley et al.(1995)]{hawley95} Hawley, J.~F., Gammie, C.~F., \& Balbus, S.~A.\ 1995, \apj, 440, 742 
\bibitem[Hayashi(1981)]{hayashi81} Hayashi, C.\ 1981, Progress of Theoretical Physics Supplement, 70, 35 
\bibitem[Hollenbach et al.(2000, p. 401)]{hollenbach2000} Hollenbach, D.~J., Yorke, H.~W., \& Johnstone, D.\ 2000, Tucson: U. Arizona Press 
       Protostars and Planets IV, Tucson: U. Arizona Press 
\bibitem[Ilgner \& Nelson(2006)]{ilgner2006} Ilgner, M., \& Nelson, R.~P.\ 2006, \aap, 445, 205 
\bibitem[Kim et al.(2002)]{kim02} Kim, W.-T., Ostriker, E.~C., \& Stone, J.~M.\ 2002, \apj, 581, 1080 
\bibitem[Koenigl \& Ruden(1993), p. 641]{koenigl1993} Koenigl, A., \& Ruden, S.~P.\ 1993, Protostars and Planets III, 641, Tucson: U. Arizona Press 
\bibitem[Kowal et al.(2003)]{kowal03} Kowal, G., Hanasz, M., \& Otmianowska-Mazur, K.\ 2003, \aap, 404, 533 
\bibitem[Kunz \& Balbus(2004)]{kunz04} Kunz, M.~W., \& Balbus, S.~A.\ 2004, \mnras, 348, 355 
\bibitem[Kunz(2008)]{kunz08} Kunz, M.~W.\ 2008, \mnras, 385, 1494 
\bibitem[Lesur \& Longaretti(2009)]{lesur09} Lesur, G., \& Longaretti, P.-Y.\ 2009, \aap, 504, 309 
\bibitem[Lesur et al.(2013)]{lesur13} Lesur, G., Ferreira, J., \& Ogilvie, G.~I.\ 2013, \aap, 550, AA61 
\bibitem[Lesur et al.(2014)]{lesur14} Lesur, G., Kunz, M.~W., \& Fromang, S.\ 2014, \aap, 566, AA56 
\bibitem[Lin \& Papaloizou(1986)]{lp86} Lin, D.~N.~C., \& Papaloizou, J.\ 1986, \apj, 309, 846 
\bibitem[Lissauer et al.(2014)]{lissauer14} Lissauer, J.~J., Dawson, R.~I., \& Tremaine, S.\ 2014, \nat, 513, 336 
\bibitem[London \& Flannery(1982)]{london82} London, R.~A., \& Flannery, B.~P.\ 1982, \apj, 258, 260 
\bibitem[Masset(2001)]{masset01} Masset, F.~S.\ 2001, \apj, 558, 453 
\bibitem[Masset et al.(2006)]{masset06} Masset, F.~S., Morbidelli, A., Crida, A., \& Ferreira, J.\ 2006, \apj, 642, 478 
\bibitem[Matsuyama et al.(2009)]{matsuyama2009} Matsuyama, I., Johnstone, D., \& Hollenbach, D.\ 2009, \apj, 700, 10 
\bibitem[Matt \& Pudritz(2005)]{matt05} Matt, S., \& Pudritz, R.~E.\ 2005, \apjl, 632, L135 
\bibitem[Miller \& Stone(2000)]{miller00} Miller, K.~A., \& Stone, J.~M.\ 2000, \apj, 534, 398 
\bibitem[Newcomb(1961)]{newcomb61} Newcomb, W.~A.\ 1961, Physics of Fluids, 4, 391 
\bibitem[Pandey \& Wardle(2012)]{pandey12} Pandey, B.~P., \& Wardle, M.\ 2012, \mnras, 423, 222 
\bibitem[Papaloizou \& Terquem(1997)]{papaloizou97} Papaloizou, J.~C.~B., \& Terquem, C.\ 1997, \mnras, 287, 771 
\bibitem[Pariev et al.(2003)]{pariev03} Pariev, V.~I., Blackman, E.~G., \& Boldyrev, S.~A.\ 2003, \aap, 407, 403
\bibitem[Perez-Becker \& Chiang(2011)]{perez11} Perez-Becker, D., \& Chiang, E.\ 2011, \apj, 735, 8
\bibitem[Pessah et al.(2007)]{pessah07} Pessah, M.~E., Chan, C.-k., \& Psaltis, D.\ 2007, \apjl, 668, L51 
\bibitem[Preibisch et al.(2005)]{preibisch05} Preibisch, T., Kim, Y.-C., Favata, F., et al.\ 2005, \apjs, 160, 401 
\bibitem[Pudritz \& Norman(1986)]{pudritz86} Pudritz, R.~E., \& Norman, C.~A.\ 1986, \apj, 301, 571 
\bibitem[Russo \& Thompson(2015)]{russo15} Russo, M., \& Thompson, C.\ 2015, \apj,
in press
\bibitem[Salmeron \& Wardle(2005)]{salmeron05} Salmeron, R., \& Wardle, M.\ 2005, \mnras, 361, 45 
\bibitem[Sano \& Stone(2002)]{sano02} Sano, T., \& Stone, J.~M.\ 2002, \apj, 577, 534 
\bibitem[Sano et al.(2004)]{sano04} Sano, T., Inutsuka, S.-i., Turner, N.~J., \& Stone, J.~M.\ 2004, \apj, 605, 321 
\bibitem[Simon \& Hawley(2009)]{simon09} Simon, J.~B., \& Hawley, J.~F.\ 2009, \apj, 707, 833 
\bibitem[Simon et al.(2013)]{simon13} Simon, J.~B., Bai, X.-N., Stone, J.~M., Armitage, P.~J., \& 
     Beckwith, K.\ 2013, \apj, 764, 66 
\bibitem[Turner et al.(2007)]{turner07} Turner, N.~J., Sano, T., \& Dziourkevitch, N.\ 2007, \apj, 659, 729 
\bibitem[Umebayashi \& Nakano(2009)]{umebayashi2009} Umebayashi, T., \& Nakano, T.\ 2009, \apj, 690, 69 
\bibitem[Valenti \& Johns-Krull(2004)]{valenti04} Valenti, J.~A., \& Johns-Krull, C.~M.\ 2004, \apss, 292, 619 
\bibitem[Ward(1991)]{ward91} Ward, W.~R.\ 1991, Houston, TX: Lunar and Planetary Institute, Abstracts 22, 1463 
\bibitem[Wardle(1999)]{wardle99} Wardle, M.\ 1999, \mnras, 307, 849 

\bibitem[Wardle \& Salmeron(2012)]{wardle12} Wardle, M., \& Salmeron, R.\ 2012, \mnras, 422, 2737 
\end{thebibliography}
\end{document}